\pdfoutput=1
\documentclass[12pt]{article}
\usepackage{graphicx}
\usepackage{latexsym,amsmath,amsfonts,amssymb,color}
\usepackage[latin1]{inputenc}
\usepackage{cite}
\usepackage[american]{babel}
\usepackage{hyperref}
\newcommand{\be}{\begin{equation}}
\newcommand{\ee}{\end{equation}}
\newcommand{\beq}{\begin{equation}}
\newcommand{\eeq}{\end{equation}}
\newcommand{\bea}{\begin{eqnarray}}
\newcommand{\eea}{\end{eqnarray}}

\newcommand{\ba}{\begin{eqnarray}}
\newcommand{\ea}{\end{eqnarray}}
\begin{document}
\baselineskip=15.5pt
\pagestyle{plain}
\setcounter{page}{1}

\def\ie{{\em i.e.},}
\def\eg{{\em e.g.},}
\newcommand{\rc}{\nonumber\\}
\newcommand{\bear}{\begin{eqnarray}}
\newcommand{\eear}{\end{eqnarray}}
\newcommand{\Tr}{\mbox{Tr}}    
\newcommand{\ack}[1]{{\color{red}{\bf Pfft!! #1}}}

\def\a{\alpha}
\def\b{\beta}
\def\c{\gamma}
\def\d{\delta}
\def\eps{\epsilon}           
\def\f{\phi}               
\def\vf{\varphi}  \def\tvf{\hat{\varphi}}
\def\vp{\varphi}
\def\g{\gamma}
\def\h{\eta}
\def\j{\psi}
\def\k{\kappa}                    
\def\l{\lambda}
\def\m{\mu}
\def\n{\nu}
\def\o{\omega}  \def\w{\omega}
\def\p{\pi}
\def\q{\theta}  \def\th{\theta}                  
\def\r{\rho}                                     
\def\s{\sigma}                                   
\def\t{\tau}
\def\u{\upsilon}
\def\x{\xi}
\def\z{\zeta}
\def\pt{\hat{\varphi}}
\def\tt{\hat{\theta}}
\def\lab{\label}
\def\6{\partial}
\def\wg{\wedge}
\def\atanh{{\rm arctanh}}
\def\bpsi{\bar{\psi}}
\def\bt{\bar{\theta}}
\def\bvf{\bar{\varphi}}
\def\W{\Omega}

\numberwithin{equation}{section}

\renewcommand{\theequation}{{\rm\thesection.\arabic{equation}}}


\begin{flushright}
HIP-2019-01/TH
\end{flushright}

\begin{center}

\centerline{\Large {\bf   Gravity dual of a multilayer system}}

\vspace{8mm}

\renewcommand\thefootnote{\mbox{$\fnsymbol{footnote}$}}
Niko Jokela${}^{1,2}$\footnote{niko.jokela@helsinki.fi},
Jos\'e Manuel Pen\'\i n${}^{3,4}$\footnote{jmanpen@gmail.com},\\
Alfonso V. Ramallo${}^{3,4}$\footnote{alfonso@fpaxp1.usc.es},
and Dimitrios Zoakos${}^{5}$\footnote{zoakos@gmail.com}

\vspace{4mm}

${}^1${\small \sl Department of Physics} and ${}^2${\small \sl Helsinki Institute of Physics} \\
{\small \sl P.O.Box 64} \\
{\small \sl FIN-00014 University of Helsinki, Finland} 
\vskip 0.2cm

${}^3${\small \sl Departamento de  F\'\i sica de Part\'\i  culas} 
{\small \sl and} \\
${}^4${\small \sl Instituto Galego de F\'\i sica de Altas Enerx\'\i as (IGFAE)} \\
{\small \sl Universidade de Santiago de Compostela} \\
{\small \sl E-15782 Santiago de Compostela, Spain} 
\vskip 0.2cm
\vskip 0.2cm

${}^5${\small \sl Department of Physics, National and Kapodistrian University of Athens \\ 15784 Athens, Greece
}

\end{center}

\vspace{8mm}
\numberwithin{equation}{section}
\setcounter{footnote}{0}
\renewcommand\thefootnote{\mbox{\arabic{footnote}}}

\begin{abstract}
We construct a gravity dual to a system with multiple $(2+1)$-dimensional layers in a $(3+1)$-dimensional ambient theory.  Following a top-down approach, we generate a geometry corresponding to the intersection of D3- and D5-branes along 2+1 dimensions. The D5-branes create a  codimension one defect in the worldvolume of the D3-branes and are homogeneously distributed along the directions orthogonal to the defect.  We solve the fully backreacted ten-dimensional supergravity equations of motion with smeared D5-brane sources. The solution is supersymmetric, has an intrinsic mass scale, and exhibits anisotropy at short distances in the gauge theory directions. We illustrate the running behavior in several observables, such as Wilson loops, entanglement entropy, and within thermodynamics of probe branes.

\end{abstract}

\newpage
\tableofcontents

\section{Introduction}

The holographic AdS/CFT correspondence relates strongly interacting quantum field theories (QFT) to classical gravity theories in higher dimensions \cite{Maldacena:1997re}. Apart  from being a conceptual breakthrough, this duality has become an important and versatile tool in the study of the possible states of matter (for reviews see \cite{AdS_CFT_reviews}). 

There are basically two types of approaches to implement the holographic idea.  In the so-called bottom-up approach, in order to model a $d$-dimensional QFT,  one considers a gravity system in $d+1$ dimensions.  This is somehow the minimal version of the correspondence and has been very successful in describing many interesting phenomena with models that evade the rigid constraints of string theory. On the contrary, the top-down models use the full machinery of string theory. They employ ten-dimensional gravity solutions of the corresponding equations of motion of type II supergravity. These ten-dimensional solutions are typically more difficult to obtain. However, the extra internal directions encode precious information which is lost in many phenomenological bottom-up approaches. Moreover, knowing the brane configuration which gives rise to the supergravity background allows to determine its field theory dual.  Actually,  one can  engineer such holographic duals from the  corresponding brane setup. 
\begin{figure}[ht]
\center
 \includegraphics[width=0.30\textwidth]{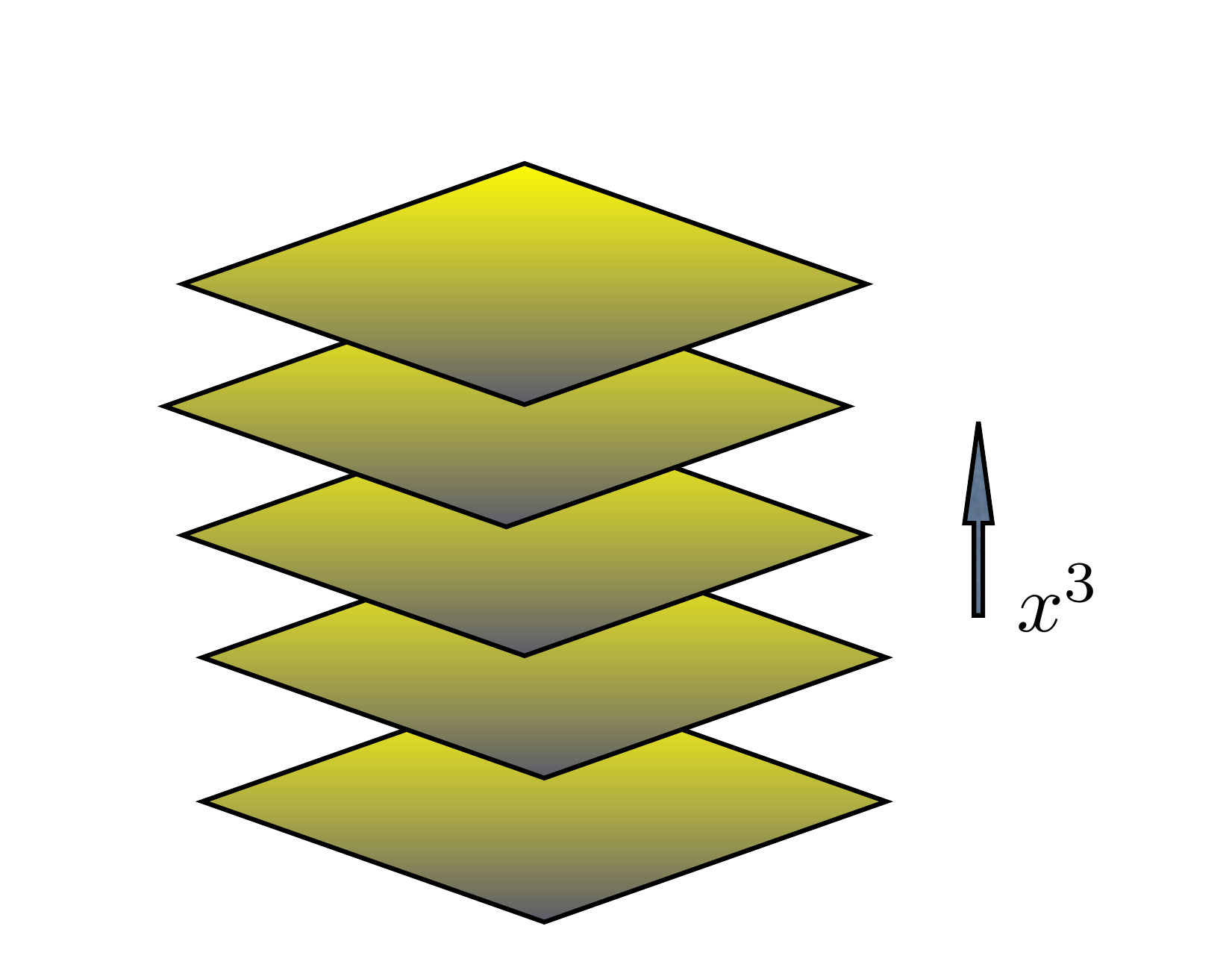}
  \caption{Our system  has a bulk $(3+1)$-dimensional theory together with multiple $(2+1)$-dimensional layers. The direction of the coordinate $x^3$ is perpendicular to the layers. }
\label{Multiple_layers}
\end{figure}

In this paper we will employ string theory techniques to find gravity duals of systems composed by multiple parallel $(2+1)$-dimensional layers in a ($3+1$)-dimensional ambient theory (see Fig.~\ref{Multiple_layers}). The ambient 4d theory will be ${\cal N}=4$ super Yang-Mills theory with gauge group $SU(N_c)$, which is realized in a stack of coincident $N_c$ color branes. The multiple layers are obtained by adding parallel D5-branes sharing two spatial directions with the stack of color branes, according to the array:
\beq
\begin{array}{cccccccccccl}
 &0&1&2&3& 4& 5&6 &7&8&9 &  \\
(N_c)\,\,D3: &\times & \times &\times &\times &\_ &\_ & \_&\_ &\_ &\_ &      \\
(N_f)\,\,D5: &\times &\times&\times&\_&\times&\times&\times&\_&\_&\_ &
\end{array}
\label{D3D5intersection}
\eeq
The field theory corresponding to the array (\ref{D3D5intersection}) is well-known \cite{Karch:2002sh,DeWolfe:2001pq,Erdmenger:2002ex,Skenderis:2002vf}. It consists of a supersymmetric  theory with matter hypermultiplets (flavors) living on the $(2+1)$-dimensional defect and coupled to the ambient ${\cal N}=4$ theory. To construct a multilayer structure of the type shown in Fig.~\ref{Multiple_layers} we consider a continuous distribution of D5-branes along the third direction in (\ref{D3D5intersection}). In the absence of D5-branes, the geometry generated by the D3-branes is $AdS_5\times {\mathbb S}^5$. We want to obtain a supergravity solution which includes the backreaction of the  flavor D5-branes. This backreaction can be obtained by using the techniques developed to study unquenched holographic flavor (see \cite{Nunez:2010sf} for a review and references).  In some cases the solutions found with these techniques are analytic and preserve some amount of  supersymmetry.   In this approach  the D5-branes of the array (\ref{D3D5intersection}) are smeared both in the cartesian direction orthogonal to the defect and on the internal directions. The dual gravity background can be obtained by solving the supergravity equations of motion with D5-brane sources. These backgrounds are dual to anisotropic systems, since there is a distinct  field theory direction, the direction orthogonal to the defect. Let us also comment on the novelty of our approach. There are several previous holographic works which provide anisotropy, see, {\emph{e.g.}}, \cite{Azeyanagi:2009pr,Kiritsis:2011zq,Mateos:2011tv,Ammon:2012qs,Jain:2014vka,Cheng:2014qia,Banks:2015aca,Roychowdhury:2015cva,Gursoy:2016ofp,Giataganas:2017koz,Itsios:2018hff,Gursoy:2018ydr}.  The main difference between our model and other anisotropic models in the literature is that in our case the anisotropy is produced by the presence of dynamical objects (the D5-branes of the multiple layers) and not by fluxes or fields depending anisotropically on the coordinates.

The D3- and D5-branes of the array (\ref{D3D5intersection}) can be separated in the directions 789 transverse to both types of branes. When this separation is zero the mass of the hypermultiplets  living on the defect vanishes, \ie\ we have massless flavors. This D3-D5  massless flavor case was considered in \cite{Conde:2016hbg}, where an analytic supersymmetric solution was found that displays a Lifshitz-like anisotropic scaling invariance.  The non-zero temperature generalization of the scaling solution was found and studied in \cite{Penin:2017lqt}. In this paper, we study the massive flavor case of the D3-D5 intersection (\ref{D3D5intersection}). The solutions we present here  preserve the same supersymmetry as the massless flavor case, but do not possess the scaling invariance of the latter.

The gravity duals of theories with massive flavors are running solutions which naturally represent a renormalization group flow. This flow is generated by changing the quark mass (see \cite{Bea:2013jxa} for an example of these massive flavored backgrounds in the ABJM theory). When the mass of the  quarks is very large, the flavors decouple and we expect to recover the unflavored solution ($AdS_5\times {\mathbb S}^5$ in our case). On the other hand, when the flavors are massless we obtain the anisotropic scaling solution of \cite{Conde:2016hbg}. For a finite non-vanishing value of the quark mass we expect to get a background interpolating between these two solutions:  Unflavored in the IR and massless flavored in the UV.  Once we have the background at our disposal we can study the effects of the flow on different observables. In general, we expect to obtain the results corresponding to the isotropic  $AdS_5\times {\mathbb S}^5$ solution in the IR and to be able to tune the amount of anisotropy by changing some of the parameters of our solutions. In our analysis of several observables we will find that, indeed, the UV behavior is determined by the scaling solution of \cite{Conde:2016hbg}, whereas the long distance IR behavior depends on a free parameter of our geometry. 

The rest of this paper is organized as follows.  In Sec.~\ref{ansatz} we present our ansatz for the metric, for the dilaton, and for the forms. All functions of the model depend on a master function, which in turn satisfies a second order differential equation. This equation follows from supersymmetry analysis detailed in App.~\ref{Background_details}. A crucial ingredient entering the equations is the so-called profile function, which encodes the distribution of sources along the holographic coordinate. To determine this function for D5-brane sources one needs to analyze in detail the embeddings of the D5-branes and their kappa symmetry. This analysis is deferred to App.~\ref{profile_details}.

In Sec.~\ref{integration} we tackle the problem of integrating the master equation. This task requires the redefinition of some of the functions and a change of variables. In its final form our solution depends on a constant parameter which characterizes the IR deformation of the metric. In Sec.~\ref{Wilson_loops} we begin our study of the observables in our background. In this section we study the Wilson loops and the potentials for quark-antiquark pairs, when these particles are in the same layer or separated in the direction orthogonal to the layers. In Sec.~\ref{entanglement_entropy} we do a similar analysis for the entanglement entropy of slabs.

In Sec.~\ref{D5_probe} we explore our supergravity solution with a probe D5-brane with a worldvolume gauge field dual to a chemical potential.  We analyze the zero temperature thermodynamics of the probe and, in particular, the UV-IR flow of the speed of sound.  This is not the only interesting configuration of probe branes with non-zero worldvolume gauge field. An interested reader is invited to App.~\ref{Higgs} where we consider D5-branes with worldvolume flux along the internal directions. This internal flux induces a bending of the probe brane along the direction $x^3$ in the array (\ref{D3D5intersection}), which can be interpreted as a recombination of the flavor D5-branes and the color D3-branes, realizing the Higgs branch of the theory \cite{Arean:2006vg,Arean:2007nh}. Finally, in Sec.~\ref{conclusions} we summarize our results and discuss some research lines for the future.

\section{Supergravity ansatz}
\label{ansatz}

In this section we review the supergravity setup of \cite{Conde:2016hbg}, corresponding to the array (\ref{D3D5intersection}) of D3- and D5-branes. More details are given in App.~\ref{Background_details}.  The D3-branes are color branes which generate an $AdS_5\times 
{\mathbb S}^5$ space, whereas the flavor D5-branes create a codimension one  defect in the $(3+1)$-dimensional gauge theory and, when the backreaction is included, the original $AdS_5\times 
{\mathbb S}^5$  metric gets deformed.  The D5-branes are homogeneously distributed in the internal space in such a way that some amount of supersymmetry is preserved. When the ${\mathbb S}^5$ space  is represented as a $U(1)$ bundle over ${\mathbb C}{\mathbb P}^2$, the deformation of the five-sphere depends on a single radial function, which measures the relative squashing between the fiber and the base of the deformed ${\mathbb S}^5$. 
Choosing a convenient radial coordinate $\zeta$ (with boundary corresponding to $\zeta=\infty$),  the ten-dimensional Einstein frame metric takes the form:
\bear
&&ds^2_{10}\,=\,h^{-{1\over 2}}\,\big[-(dx^0)^2+(dx^1)^2+(dx^2)^2\,+\,e^{-2\phi}\,(dx^3)^2\big]\rc\rc
&&\qquad\qquad\qquad\qquad
+h^{{1\over 2}}\,\Big[\zeta^2 e^{-2f}\,d\zeta^2\,+\,\zeta^2\,
ds^2_{{\mathbb C}{\mathbb P}^2}\,+\,e^{2f}\,(d\tau+A)^2\Big]\,\,,
\label{metric_ansatz_zeta}
\eear
where $\phi$ is the dilaton of type IIB supergravity, $h$ is the warp factor, and $f$ is the squashing function. These functions are assumed to depend only on  $\zeta$. Moreover, $A$ is a one-form on ${\mathbb C}{\mathbb P}^2$ which implements the non-trivial $U(1)$ bundle. The Minkowski directions $x^1$ and $x^2$ are parallel to the defect, whereas $x^3$ is orthogonal to it.

Besides the metric and the dilaton, the type IIB supergravity solution contains a 
RR five-form $F_5$ and a RR three-form $F_3$. The former is self-dual and given by the standard ansatz in terms of the dilaton $\phi$ and warp factor $h$:
\beq
F_5\,=\,\partial_{\zeta}\,\big(e^{-\phi}\,h^{-1}\big)\,\big(1+*\big)\,d^4x\wedge d\zeta\,\,.
\label{F5_sol}
\eeq
Clearly, $d\,F_5=0$, since the D3-branes have been replaced by a flux in the supergravity solution. 
On the contrary, the D5-branes are dynamical and are governed by the standard DBI+WZ action which, in particular, contains the term:
\beq
S_{WZ}\,=\,T_5\,\sum^{N_f}\,\int_{{\cal M}_6}\,\hat C_{6}\,\,,
\label{SWZ_D5}
\eeq
where $1/T_5\,=\,(2\pi)^5\,g_s\alpha'^3$ and $ C_{6}$ is the six-form potential for $F_7\,=\,-e^{\phi} *F_3$  (the hat over $C_{6}$ denotes its pullback to the D5-brane worldvolume ${\cal M}_6$).  Therefore, $S_{WZ}$ contributes to the equation of motion of $C_{6}$ or, equivalently, to the Bianchi identity of $F_3$. Indeed, let us write the six-dimensional integral in (\ref{SWZ_D5}) as a ten-dimensional integral:
\beq
\sum^{N_f}\,\int_{{\cal M}_6}\,\hat C_{6}\,=
\,
\int_{{\cal M}_{10}}\,\Xi\wedge C_{6}\,\,,
\eeq
where $\Xi$ is a four-form with support on the worldvolume of the D5's and with legs along the directions orthogonal to ${\cal M}_6$, which is just the RR  D5-brane charge distribution. The equation of motion for $ C_{6}$  is:
\beq
dF_3\,=\,2\,\kappa_{10}^2\,T_5\,\Xi \ ,
\label{Bianchi_F3}
\eeq
where  $2\,{\kappa}_{10}^2\,=\,(2\pi)^7\,g_s^2\alpha'^4$. In the smearing approach $\Xi$ does not contain Dirac $\delta$-function singularities. Its form can be obtained once the ansatz of $F_3$ compatible with supersymmetry is fixed.  This has been done in \cite{Conde:2016hbg}, a result which we now review.  

The  ${\mathbb C}{\mathbb P}^2$  manifold is a K\"ahler-Einstein space endowed with a K\"ahler two-form $J=dA/2$, which can be canonically written as $J=e^1\wedge e^2+e^3\wedge e^4$, where $e^1,\ldots,e^4$ are vielbein one-forms of ${\mathbb C}{\mathbb P}^2$, whose explicit coordinate expression can be found in App.~\ref{Background_details}. Let us introduce the complex two-form $\hat\Omega_2$ as:
\beq
\hat\Omega_2\,=\,e^{3 i\tau}\,(e^1+i e^2)\wedge (e^3+i e^4) \ .
\label{hat_Omega_2}
\eeq
Then,  $F_3$ is given by:
\beq
F_3\,=\,Q_f\,p(\zeta)\,dx^3\wedge {\rm Im }\,\hat\Omega_2\,\,,
\label{F3_ansatz}
\eeq
where $Q_f$ is a constant and $p(\zeta)$ is an arbitrary function of the holographic coordinate $\zeta$.  By computing the exterior derivative of $F_3$ we get its modified Bianchi identity:
\beq
dF_3\,=\,-Q_f\,\big[ 3 \,p(\zeta)\,dx^3\wedge {\rm Re}\,\hat\Omega_2\wedge (d\tau+A)\,+\,
p'(\zeta)\,dx^3\,\wedge d\zeta\wedge {\rm Im}\,\hat\Omega_2\big]\,\,.
\label{dF_3}
\eeq
Comparing (\ref{dF_3}) and (\ref{Bianchi_F3}) we can extract the D5-brane charge distribution $\Xi$ which, in what follows, we will refer to as the smearing form.  Clearly, $\Xi$ does not depend on the $x^3$ coordinate, although it contains $dx^3$ in its expression. This means that we are continuously distributing our D5-branes along $x^3$, giving rise to a system of multiple $(2+1)$-dimensional parallel layers.  Moreover, the function $p(\zeta)$ introduces a profile of the charge distribution in the holographic coordinate. Notice that the D3- and D5-branes in the array (\ref{D3D5intersection}) can be separated in the 789 directions. In principle, we could have an arbitrary distribution of D5-branes in these transverse coordinates, which is  reflected in the fact that the profile function $p(\zeta)$ is arbitrary. However, for a stack of flavor D5-branes with the same quark mass the function $p(\zeta)$ has a well-defined form (see App.~\ref{profile_details}) and $Q_f$ is related to the density of smeared branes along the direction $x^3$. As shown in App.~\ref{profile_details}, if we distribute $N_f$ D5-branes along a distance $L_3$ in the third cartesian direction, then  $Q_f={4\pi\,g_s\,\alpha'\, N_f\over 9\sqrt{3}\,L_3}$ (see (\ref{Q_f-N_f-L3})). 

The preservation of two supercharges for our ansatz imposes a system of first-order differential equations in the radial variable. These BPS equations are reviewed in App.~\ref{Background_details}, where it is shown that they can be reduced to a single second-order  differential equation for a master function $W(\zeta)$. This equation is:
\beq
{d\over d\zeta}\Big(\zeta\,{d W\over d\zeta}\Big)\,+\,6\,{d W\over d\zeta}\,=\,
-{6\,Q_f\,p(\zeta)\over \zeta^2\,\sqrt{W}}\,\,.
\label{Master_W}
\eeq
From $W$ we can reconstruct the full solution. The squashing function $f(\zeta)$  is given by:
\beq
e^{2f}\,=\,{6\,\zeta^2\,W\over 6\,W\,+\,\zeta\,{dW\over d\zeta}}\,\,,
\label{g_f_W}
\eeq
while the dilaton is:
\beq
e^{-\phi}\,=\,W\,+\,{1\over 6}\,\zeta\,{d W\over d\zeta}\,\,.
\label{dilaton_W}
\eeq
A nice way of measuring the deformation of the metric (\ref{metric_ansatz_zeta}) with respect to the $AdS_5\times {\mathbb S}^5$ geometry is obtained by considering a squashing factor, which we define as follows
\beq
q\,=\,{e^f\over \zeta}\,\,.
\label{q_def}
\eeq
It is clear from our ansatz  (\ref{metric_ansatz_zeta})  that $q$ represents the relative size of the $U(1)$ fiber with respect to the 
${\mathbb C}{\mathbb P}^2$ base. In terms of the master function $W$, $q$ is given by:
\beq
q\,=\,{\sqrt{6\,W}\over \sqrt{6\,W+\zeta\,{dW\over d\zeta}}}\,\,.
\label{q_W}
\eeq

Once $f$ and $\phi$ are known,  the warp factor $h$ can be obtained by integrating the following first-order differential equation:
\beq
{dh\over d\zeta}\,+\,Q_f\,{e^{{3\phi\over 2}-f}\,p\over \zeta}\,h\,=\,-{Q_c\over \zeta^3}\,e^{-2f}\,\,,
\label{warp_factor_eq}
\eeq
where $Q_c$ is related to the number $N_c$ of D3-branes as $Q_c=16\,\pi g_s \alpha'{}^{\,2}\,N_c$.  In what follows we will study several solutions of the master equation (\ref{Master_W}).

\subsection{Unflavored solution}\label{unflavored}

Let us consider the master equation in the case in which there are no flavor brane sources. It turns out that 
the general solution of (\ref{Master_W})  can be analytically found in this case. Indeed, when $Q_f=0$ 
 the master equation (\ref{Master_W}) can be trivially integrated once as:
\beq
\zeta\,{d W\over d\zeta}\,+\,6\,W\,={\rm constant}\,\,.
\eeq
A further integration gives:
\beq
W\,=\,C\,\Big(1\,-\,{b^6\over \zeta^6}\Big)\,\,,
\label{W_inside}
\eeq
where $C$ and $b$ are constants. Plugging this expression of $W$ into the right-hand-side of  (\ref{dilaton_W}) one readily verifies that the dilaton is constant and given by:
\beq
e^{-\phi}\,=\,C\,\,.
\eeq
Moreover, the function $f$ and the squashing function $q$ are given by:
\beq
e^{2f}\,=\,\zeta^2\,\Big(1\,-\,{b^6\over \zeta^6}\Big)\,\,,
\qquad\qquad\qquad
q\,=\,\Big(1\,-\,{b^6\over \zeta^6}\Big)^{{1\over 2}}\,\,,
\eeq
and the warp factor $h$ can be obtained by integrating (\ref{warp_factor_eq}).
This solution coincides with the general unflavored one found  in \cite{Conde:2016hbg}.  For $b=0$ this geometry is just
$AdS_5\times {\mathbb S}^5$. When $b\not=0$  the solution approaches $AdS_5\times {\mathbb S}^5$ in the UV. If we take $b$ to be real and positive, then the minimal value of $\zeta$ is $\zeta=b$ and the metric has a blown-up ${\mathbb C}{\mathbb P}^2$  cycle at $\zeta=b$. It was argued in \cite{Benvenuti:2005qb} that this $b\not=0$ background is dual to the superconformal ${\cal N}=4$ field theory deformed by the VEV of a dimension 6 operator.

\subsection{Massless flavored solution}
\label{massless_flavors}
Let us now consider the massless flavor case with $Q_f\not=0$ and $p=1$. In this case it is possible to find a special solution of (\ref{Master_W}):
\beq
W\,=\,A\,\zeta^{\alpha}\,\,.
\eeq
Indeed, by plugging this ansatz into the master equation we readily get that the exponent $\alpha$ is given by:
\beq
\alpha\,=\,-{2\over 3}\,\,.
\eeq
Similarly, we  can obtain the value of the constant $A$. The final formula for $W$ is:
\beq
W\,=\,{9\over 8}\,\big(\sqrt{2}\,Q_f\big)^{{2\over 3}}\,\zeta^{-{2\over 3}}\,\,.
\label{W_massless_flavored}
\eeq
Using this expression in (\ref{g_f_W}) and (\ref{dilaton_W}) we arrive at the following values of $f$ and $\phi$:
\beq
e^{2f}\,=\,{9\over 8}\,\zeta^2\,\,,
\qquad\qquad\qquad
e^{-\phi}\,=\,\big(\sqrt{2}\,Q_f\big)^{{2\over 3}}\,\zeta^{-{2\over 3}}\,\,.
\label{f_phi_zeta}
\eeq
It is straightforward to verify that this solution coincides with the one found in \cite{Conde:2016hbg} for massless flavors.\footnote{The relation between $\zeta$ and the  radial variable $r$ used in \cite{Conde:2016hbg} and in the ansatz (\ref{10d_metric_explicit})  is
$\zeta\,=\,{3\over 2\sqrt{2}}\,r$.}
The warp factor for this solution is:
\beq
h\,=\,{\bar R^4\over \zeta^4}\,\,,
\label{h_masless_flavored}
\eeq
where $\bar R$ is the same as in \cite{Conde:2016hbg}, 
\beq
\bar R^4\,=\,{4\over 15}\,Q_c\,\,.
\eeq
Since the profile function $p$ is constant, this solution represents massless smeared flavors extending all the way down to  $\zeta=0$. As shown in \cite{Conde:2016hbg}, the background corresponding to the master function (\ref{W_massless_flavored}) is invariant under a set of anisotropic scale transformations in which the $x^3$ coordinate transforms with an anomalous scaling dimension.  Moreover, the squashing factor (\ref{q_def}) $q$ is constant and equal to ${3\over 2\sqrt{2}}\approx 1.06$, see (\ref{f_phi_zeta}). The purpose of this paper is to find solutions corresponding to massive flavors, which should interpolate between the unflavored solution of subsection \ref{unflavored} at the IR and the scaling solution studied in this subsection at the UV. We start to discuss these solutions in the next subsection.

\begin{figure}[ht]
\center
 \includegraphics[width=0.35\textwidth]{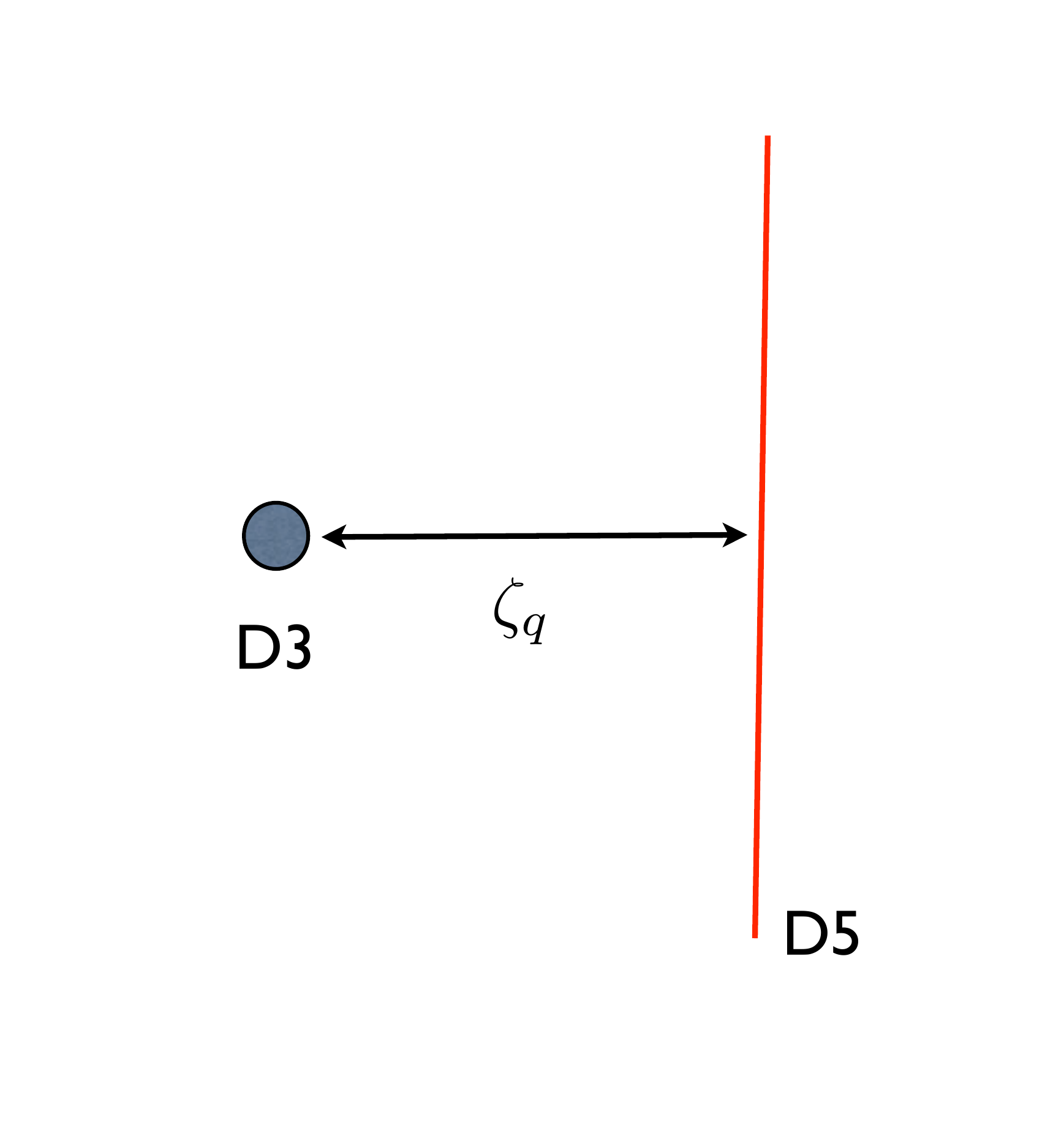}
 \qquad\qquad
 \includegraphics[width=0.50\textwidth]{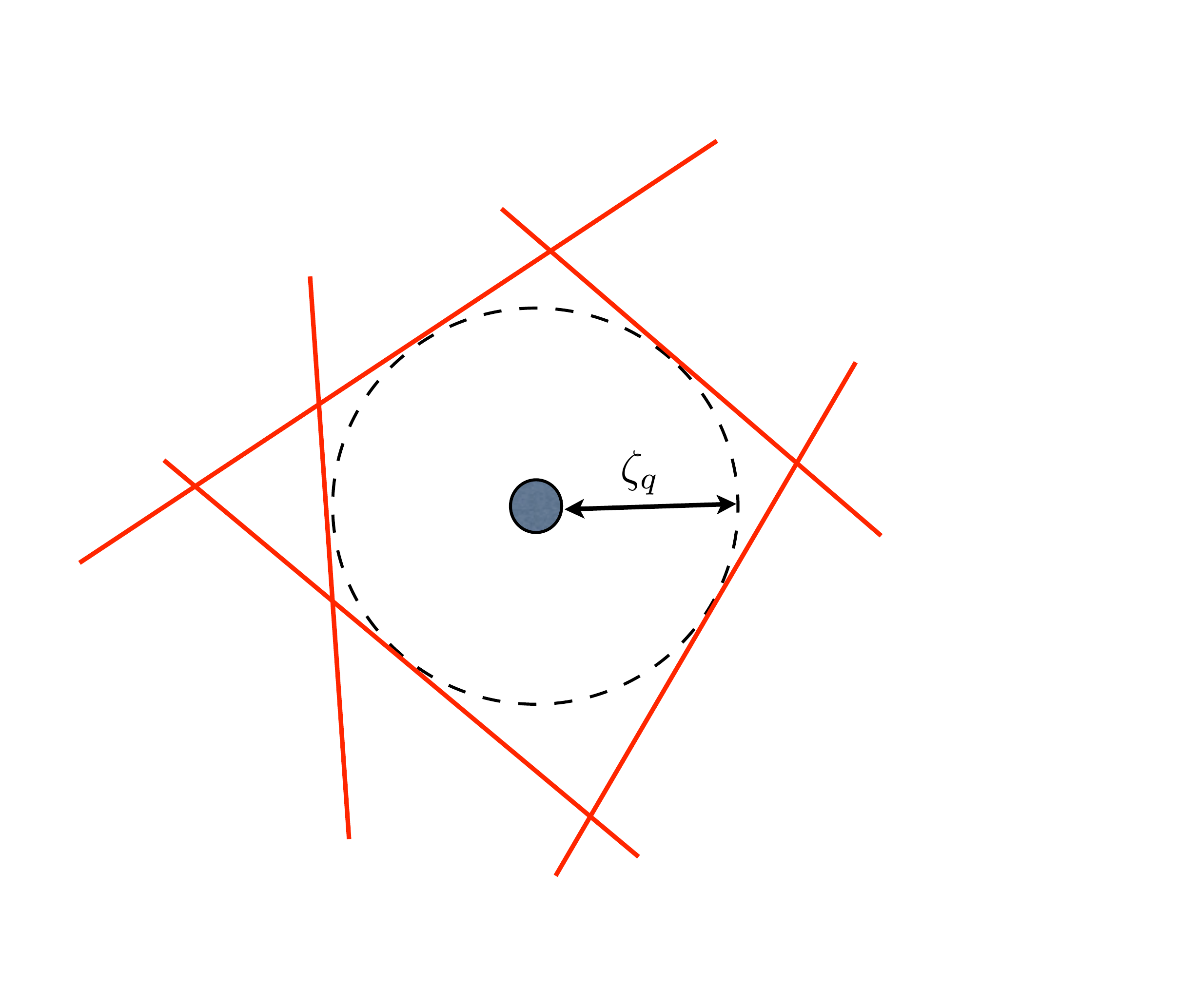}
  \caption{On the left we depict a localized embedding of a D5-brane with a separation $\zeta_q$ from the stack of  color D3-branes. On the right, several D5-branes with different orientations and the same distance $\zeta_q$ generate a cavity $\zeta\le \zeta_q$ which does not contain flavor sources. }
\label{Cavity}
\end{figure}

\subsection{Massive flavors}
\label{massive_flavors}

In the holographic approach the mass of the fundamentals is related to the distance between the color and flavor branes (in our case D3's and D5's, respectively). When these two sets of branes are separated, the fundamentals are massive and there is an  IR region of the geometry which is not occupied by the the flavor brane and, as a consequence, the D5-brane charge density vanishes there. In the smearing setup  we have many D5-branes with different orientations in the internal space which, if they correspond to  flavors with the same mass,  should have the same separation from the D3-branes. As illustrated in Fig.~\ref{Cavity}, the sourceless region has a $\zeta$ coordinate less or equal to some value $\zeta_q$, a region we will call cavity in the following.  The profile function $p(\zeta)$ vanishes inside the cavity and should approach the value appropriate for massless flavors, {\emph{i.e.}}, $p\to 1$ as $\zeta\to\infty$. To determine the explicit form of the function $p(\zeta)$ one has to specify the set of source D5-branes of our smeared distribution. This is done in detail in App.~\ref{profile_details}. The final result for the profile function is:
\beq
p(\zeta)\,=\,\Bigg[1-\Big({\zeta_q\over \zeta}\Big)^2\Bigg]^{{1\over 2}}\,
\Bigg[1\,+\,{1\over 2}\,\Big({\zeta_q\over \zeta}\big)^2\Bigg]\,\,\Theta(\zeta-\zeta_q)\,\,.
\label{p_zeta_explicit}
\eeq
Notice that $p(\zeta)$ is continuous at $\zeta=\zeta_q$ and asymptotically $p(\zeta\to\infty)=1$. The master function $W$ for the profile (\ref{p_zeta_explicit}) must be obtained numerically. However, we can expand (\ref{Master_W}) in a series expansion about any radial coordinate $\zeta$. Let us start in the neighborhood of $\zeta=\infty$. Indeed, the profile function (\ref{p_zeta_explicit}) can be expanded in powers of ${\zeta_q\over \zeta}$ as:
\beq
p(\zeta)= 1\,-\,{3\over 8}\,{\zeta_q^4\over \zeta^4}\,-\,{1\over 8}\,{\zeta_q^6\over \zeta^6}\,-\,{9\over 128}
\, {\zeta_q^8\over \zeta^8}\,+\,{\cal O}\Big( {\zeta_q^{10}\over \zeta^{10}}\Big)\,\,.
\eeq
Plugging this expansion into the master equation (\ref{Master_W}) and integrating order by order, we get the following solution for the master function $W(\zeta)$:
\beq
W = {9\over 8}\,\Big[{\sqrt{2}\,Q_f\over \zeta}\Big]^{{2\over 3}}\,\Bigg(1\,-\,{1\over 6}\,
{\zeta_q^4\over \zeta^4}\,+\,{1\over 6}\,{\zeta_q^6\over \zeta^6}\,+\,{35\over 2304}\,{\zeta_q^8\over \zeta^8}\,+\,
{\cal O}\Big( {\zeta_q^{10}\over \zeta^{10}}\Big)\,\Bigg)\,\,.
\label{W_UV_zeta}
\eeq
The asymptotic  UV expansions of the different functions of the background are easily obtained from (\ref{W_UV_zeta}). We have collected these expansions in App.~\ref{Background_details}. 

Another useful expansion is the neighborhood where the sources set in, {\emph{i.e.}}, at the edge of the cavity. We have hence perturbatively solved the master equation for $\zeta\ge \zeta_q$ and $\zeta-\zeta_q$ small; details are relegated in App.~\ref{Background_details}.

Recall that inside the cavity, \ie\ for $\zeta\le \zeta_q$, we have the analytic solution for the master function. This solution was written in (\ref{W_inside}) and depends on two parameters $C$ and $b$. We want to extend this solution for 
$\zeta\ge \zeta_q$ and determine  the values of $C$ and $b$  for which the function $W$ behaves as in (\ref{W_UV_zeta}) for $\zeta\to \infty$. This matching has to be done numerically, for which we implemented the shooting technique accompanied with a convenient change of variables.  We will discuss this in the next section.

\section{Integration of the master equation}\label{integration}

Let us introduce a new holographic variable $x$, related to $\zeta$ as follows:
\beq
x\,\equiv\,{\zeta-b\over \zeta_q}\,\,.
\eeq
Clearly, for our interpolating solutions $x\ge 0$. In this new variable the cavity (\ie\ the region without flavor branes) corresponds to $0\le x\le x_q$, where $x_q$  is the edge of the cavity given by:
\beq
x_q\,=\,1\,-\,{b\over \zeta_q}\,\,.
\label{x_q_b_zeta_q}
\eeq
Notice that $x_q$ depends on the ratio between the deformation parameter $b$ in the sourceless region and the location of the boundary of this unflavored region in the $\zeta$ variable. Moreover,  as $b\le \zeta_q$, we have that:
\beq
x_q\le 1\,\,.
\eeq
Let us write the background in terms of the new variables. It is convenient to absorb the constant $C$ appearing in the master function (\ref{W_inside}) inside the cavity. Accordingly, we define $\hat W$ as:
\beq
\hat W\,=\,{W\over C}\,\,.
\label{hat_W_def}
\eeq
In the region without flavor sources this function is given by:
\beq
\hat W(x)\,=\,1\,-\,{(1-x_q)^6\over (x+1-x_q)^6}\,\,,
\qquad\qquad
0\,\le\,x\,\le\,x_q\,\,.
\label{hat_W_inside}
\eeq
\begin{figure}[ht]
\center
 \includegraphics[width=0.60\textwidth]{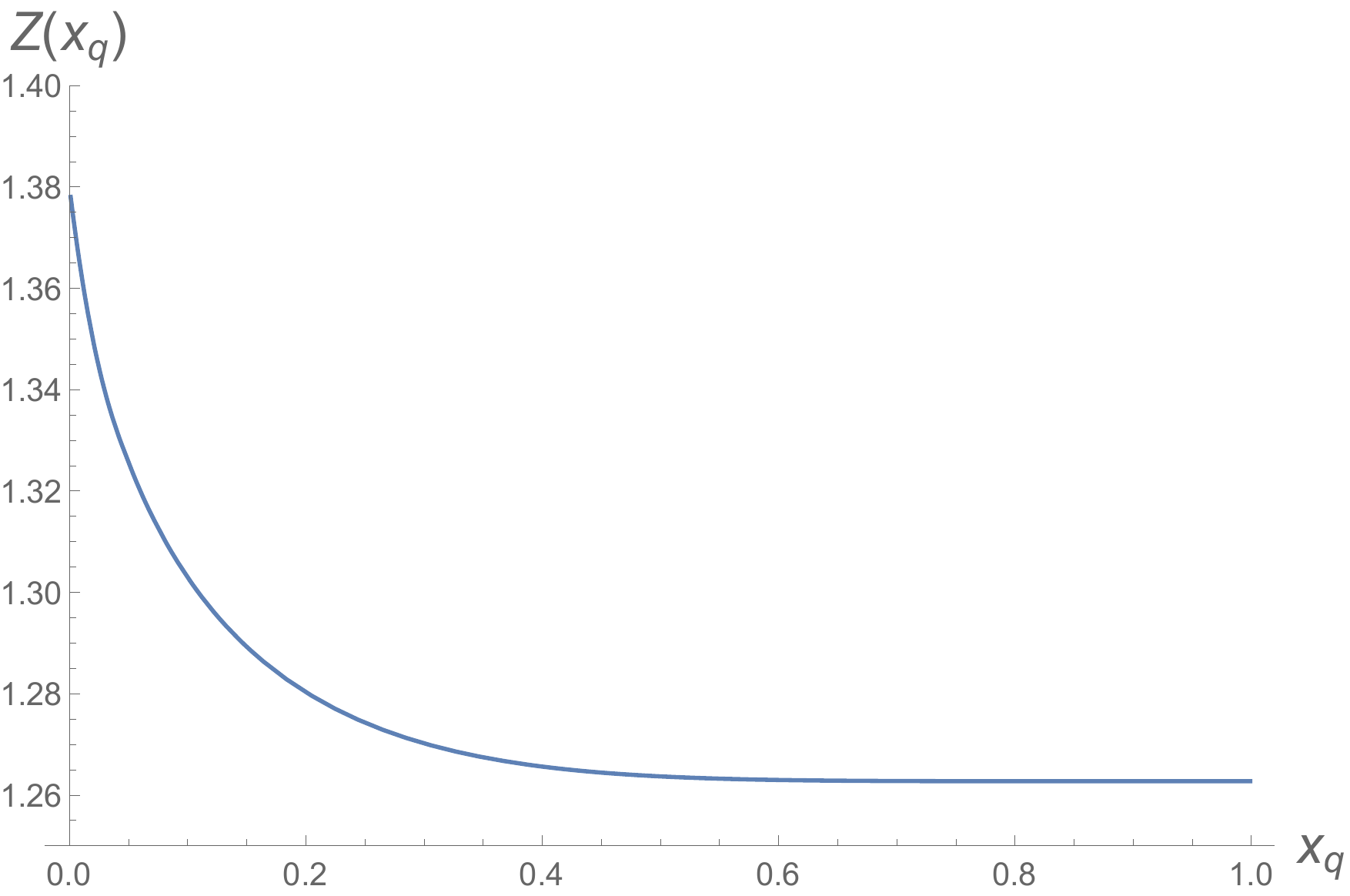}
  \caption{Plot of the  function $Z(x_q)=C^{{3\over 2}}\,\zeta_q/Q_f$, obtained by the shooting method of the master equation (\ref{master_eq_x}). This function has as asymptotics $Z(x_q=0)\approx 1.38$ and $Z(x_q=1)\approx 1.26$. Notice that $Z(x_q)$ is almost constant in the interval $0.5\le x_q\le 1$. }
\label{zetaq_xq}
\end{figure}

This function vanishes at the origin $x=0$ and $\hat W$ and its derivative take the following values at the edge of the cavity:
\beq
\hat W(x=x_q)\,=\,1\,-\,(1-x_q)^6\,\,,
\qquad\qquad
{d\hat W\over dx}\Big|_{x=x_q}\,=\,6(1-x_q)^6\,\,.
\label{W_Wprime_border}
\eeq
Notice that, for $x_q=1$, \ie\ in the case in which the cavity is largest, $\hat W(x)=1$ inside the cavity and the solution becomes (isotropic) AdS in this region. This implies that $x_q$ measures the amount of anisotropy of our solution.

The profile function in terms of the $x$ variable is:
\beq
p(x)\,=\,\theta(x-x_q)\,
{\sqrt{x-x_q}\,\sqrt{x-x_q+2}\over 2\,(x+1-x_q)^3}\,
\Big[1\,+\,2(x+1-x_q)^2\Big]\,\,.
\eeq
Moreover,  the master equation for $\hat W$ becomes:
\beq
{d\over dx}\,\Big[(x+1-x_q)\,{d\hat W\over dx}\Big]\,+\,6\,{d\hat W\over dx}\,=\,-
{6\,Q_f\over C^{{3\over 2}}\,\zeta_q}\,{p(x)\over \sqrt{\hat W}}\,\,.
\label{master_eq_x}
\eeq
This equation is solved analytically  by (\ref{hat_W_inside}) inside the cavity, where $0\le x\le x_q$.  For $x\ge x_q$ we can solve (\ref{master_eq_x}) numerically by imposing the initial conditions (\ref{W_Wprime_border})
and by requiring that as $x\to\infty$ it asymptotically behaves as the massless flavored solution,
\beq
\hat W\,\approx\,{9\over 8}\,\Big[{\sqrt{2}\,Q_f\over C^{{3\over 2}}\,\zeta_q}\Big]^{{2\over 3}}\,\,
x^{-{2\over 3}}\,\,,
\qquad\qquad
(x\to\infty)\,\,.
\label{asymp_hat_W}
\eeq
Notice that (\ref{master_eq_x}) depends on two parameters $x_q$ and $C^{{3\over 2}}\,\zeta_q/Q_f$. In order to have a solution with the asymptotic behavior (\ref{asymp_hat_W}) these two parameters must satisfy a relation, which can be determined numerically. This relation is
\beq
\zeta_q\,=\,{Q_f\over C^{{3\over 2}}}\,Z(x_q)\,\,,
\eeq
where the function $Z(x_q)$ is determinend numerically for $0\le x_q\le 1$. Our results for this function have been plotted in Fig.~\ref{zetaq_xq}. We observe that $Z(x_q)$ is a decreasing function of $x_q$ and, for given values of $Q_f$ and the constant $C$, $\zeta_q$ reaches its maximum at $x_q=0$. Notice also that $x_q$ gives the size of the cavity in the $x$ variable. This size should be related to the quark mass $m_q$.  In our holographic setup $m_q$ can be determined by evaluating the Nambu-Goto action of a fundamental string extended along the holographic direction, from the origin of the space to the tip of the flavor brane. In the $\zeta$ variable we have:
\beq
m_q\,=\,{1\over 2\pi\,\alpha'}\,\int_b^{\zeta_q}\,d\zeta\,e^{{\phi\over 2}}\,
\sqrt{-\det g_2}\,\,,
\eeq
where the $e^{{\phi\over 2}}$ factor is due to the fact that we are working in the Einstein frame. By using the metric ansatz (\ref{metric_ansatz_zeta}), as well as (\ref{g_f_W}) and (\ref{dilaton_W}), we obtain $m_q$ in terms of an integral involving the master function $W$:
\beq
m_q\,=\,{1\over 2\pi\,\alpha'}\,\int_b^{\zeta_q}\,
{d\zeta\over \sqrt{W}}\,=\,{1\over 2\pi\,\alpha'\,\sqrt{C}}\,
\int_b^{\zeta_q}\,{d\zeta\over \sqrt{1-{b^6\over \zeta^6}}}\,\,.
\eeq
Putting $\alpha'=1$ from now on and writing the result in terms of $x_q$, we get:
\beq
m_q\,=\,{Q_f\over 2\pi\,C^2}\,Z(x_q)\,
\Bigg[(x_q-1)\,{\sqrt{\pi}\,\Gamma\big({5\over 6}\big)\over \Gamma\big({1\over 3}\big)}\,+\,
F\big(-{1\over 6}\,,\,{1\over 2}\,;\,{5\over 6}\,;\,(1-x_q)^6\big)\Bigg]\,\,.
\label{m_q_C_x_q}
\eeq
\begin{figure}[ht]
\center
 \includegraphics[width=0.50\textwidth]{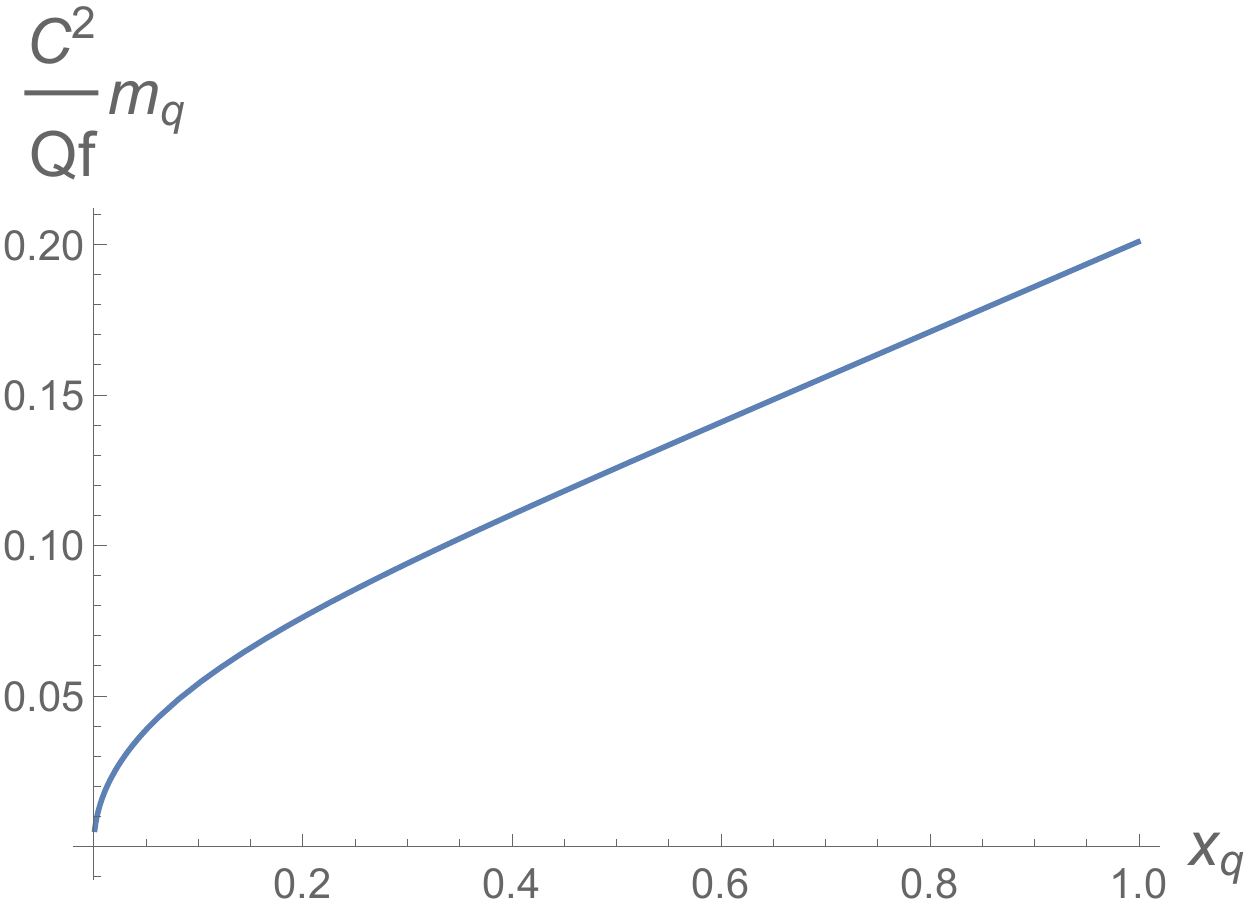}
  \caption{Quark mass $m_q$, rescaled by $C^2/Q_f$,   as a function of $x_q$.  }
\label{mq_xq}
\end{figure}

In Fig.~\ref{mq_xq} we plot $C^2\,m_q/Q_f$ as a function of $x_q$. We observe in this plot that $m_q$ grows with $x_q$ and that $m_q=0$ for $x_q=0$, \ie\  when the cavity has zero size. Moreover, we can increase the quark mass by decreasing $C$ ($C\to 0$ corresponds to $m_q\to\infty$ for $x_q\not=0$). The precise relation between $m_q$ and the constant $C$ depends on the value of $x_q$. For $x_q\sim 0,1$ one can expand the right-hand side of (\ref{m_q_C_x_q}) and find:
\beq
m_q\approx {Q_f\over 2\pi C^2}\,\sqrt{{2\over 3}}\,Z(0)\,\sqrt{x_q}\,\,,\qquad (x_q\to 0)\,\,,
\qquad\qquad
m_q\approx {Q_f\over 2\pi C^2}\,Z(1)\,\,,\qquad (x_q\to 1)\,\,.
\eeq

All the functions of the background can be written in terms of $Z$ and $\hat W$. For example, the squashing factor $q$ is:
\beq
q\,=\,{\sqrt{\hat W}\over \sqrt{\hat W\,+\,{1+x-x_q\over 6}\,{d\hat W\over dx}}}\,\,,
\label{q_hat_W}
\eeq
whereas $f$, $g$, and the dilaton $\phi$ are:
\bear
 e^{2g} & = & {Q_f^2\over C^3}\,Z^2(x_q)\,(1+x-x_q)^2\nonumber \\
e^{2f} & = & {Q_f^2\over C^3}\,Z^2(x_q)\,\,{(1+x-x_q)^2\,\hat W\over
 \hat W+ {1+x-x_q\over 6}\,{d\hat W\over dx}}\,=\,
 {Q_f^2\over C^3}\,Z^2(x_q)\,\,(1+x-x_q)^2\,q^2(x)\nonumber \\
 e^{-\phi} & = & C\,\Big[ \hat W+ {1+x-x_q\over 6}\,{d\hat W\over dx}\Big]\,\,=\,\,
 C\,{ \hat W\over q^2(x)}\,\,,
\eear
where, in the second step, we have written these functions in terms of $q(x)$.

\begin{figure}[ht]
\center
\includegraphics[width=0.70\textwidth]{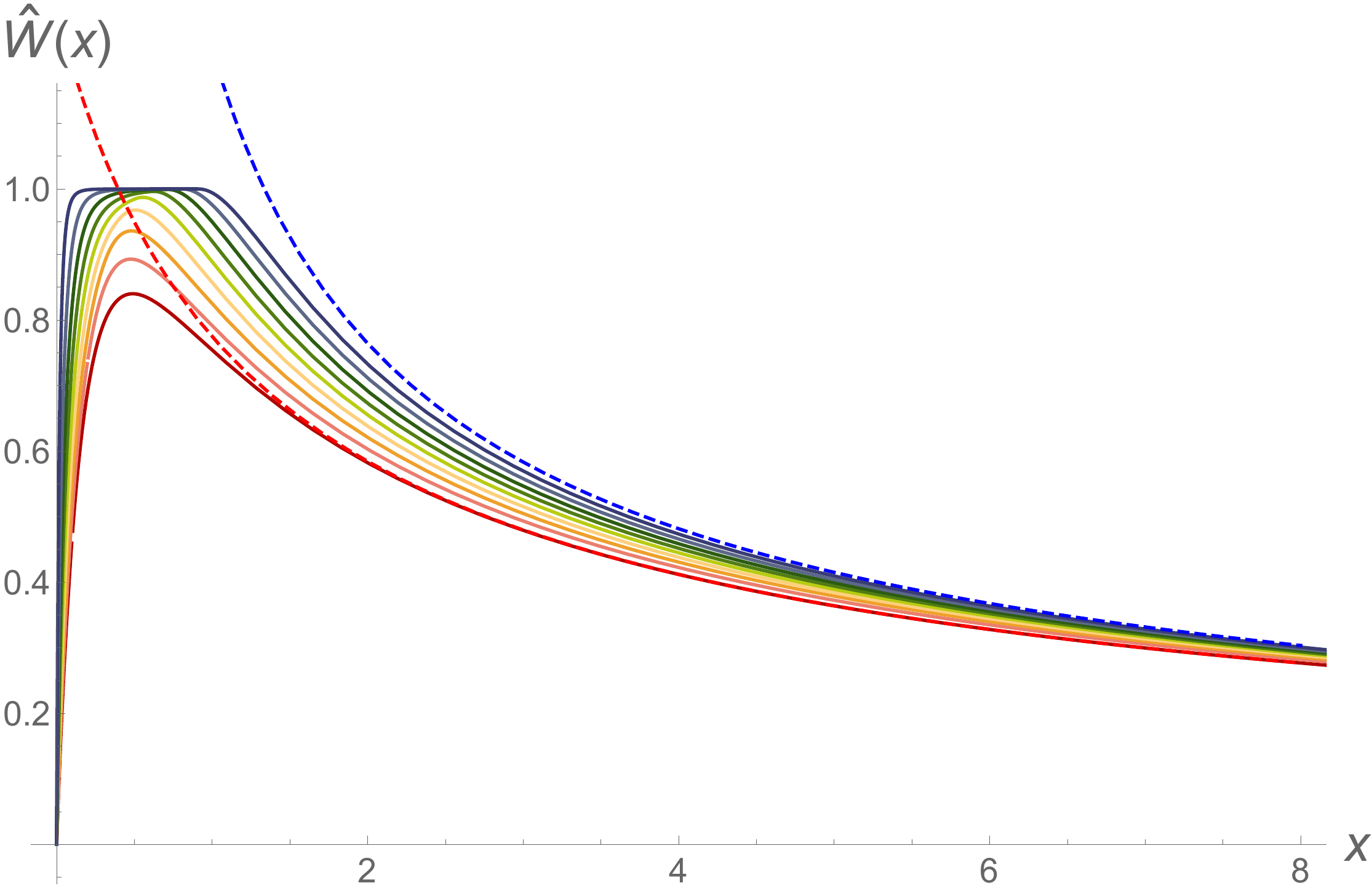}
\caption{Plot of $\hat W$ versus the holographic variable $x$ for different values of $x_q$ in the range $0.1\le x_q\le 0.9$. The continuous bottom dark red (top dark blue) curve corresponds to $x_q=0.1$ ($x_q=0.9$). The  continuous curves between them correspond to $x_q=i/10$ for $i=2,\ldots,8$ (bottom-up).  The dashed curves are the leading terms in the UV expansion (\ref{W_UV_x}) for $x_q=0.1$ (red) and $x_q=0.9$ (blue).}
\label{W_x}
\end{figure}

In Fig.~\ref{W_x} we plot the function $\hat W$ for different values of $x_q$.  We notice that, as $x_q\to 1$, $\hat W$ is almost constant and equal to 1 in the interior of the cavity, where it is given by (\ref{hat_W_inside}). Moreover, outside the cavity, \ie\ for $x\ge x_q$ it rapidly approaches the asymptotic expression (\ref{W_UV_zeta}), which can be rewritten in terms of $x$ and $x_q$ as:
\bear
&&\hat W = {9\over 8}\,\Big[{\sqrt{2}\over Z(x_q)\,(x+1-x_q)}\Big]^{{2\over 3}}\,\Bigg(1-{1\over 6}{1\over (x+1-x_q)^4}+\,
{1\over 6}{1\over (x+1-x_q)^6}\rc\rc
&&\qquad\qquad\qquad\qquad\qquad\qquad\qquad
+{35\over 2304}\,{1\over (x+1-x_q)^8}\,+\,\ldots\Bigg)\,\,.
\label{W_UV_x}
\eear
The UV expansion of $f$, $\phi$, $h$, and $q$  in the $x$ variable can be obtained from (\ref{f_UV_zeta})-(\ref{q_UV_zeta}) by substituting $\zeta\,=\,\zeta_q(x+1-x_q)$.

The comparison of the numerical values for $x\ge x_q$ and those given by (\ref{W_UV_x}) shows that the agreement with the UV expansion is better if $x_q$ is small, since in this case the background is closer to the massless scaling solutions (for $x_q=0$ there is no cavity). On the other hand, for $x_q$ close to one the flavor effects are smaller in the IR,  since $\hat W$ is almost constant and equal to one inside the cavity. Indeed, from (\ref{x_q_b_zeta_q}) we get that $x_q\sim 1$ implies  that the IR deformation parameter $b$ is small.  This  is consistent with the fact that, for a given value of $C$,  the quark mass is maximal in this case (see figure \ref{mq_xq}).

Let us now write the metric in the $x$ variable. First of all, we define the rescaled warp factor and cartesian coordinates as:
\beq
\hat h\,=\,{Q_f^4\over C^6}\,h\,\,,
\qquad
\hat x^{\mu}\,=\,{Q_f\over C^{{3\over 2}}}\,\,x^{\mu}\,\,,
\qquad
(\mu\,=\,0,1,2)\,\,,
\qquad
\hat x^{3}\,=\,{Q_f\over C^{{1\over 2}}}\,\,x^{3}\,\,.
\label{hatted_def}
\eeq
Then, in terms of these hatted variables, we have:
\bear
&&ds_{10}^2\,=\,\hat h^{-{1\over 2}}\,
\Big[-(d\hat x^0)^2\,+\,(d \hat x^1)^2\,+\,(d\hat x^2)^2\,+\,
{\hat W^2\over q^4}\,(d \hat x^3)^2\,\Big]\,
\qquad\qquad\qquad
\rc\rc
&&
\qquad\qquad
+Z^2(x_q)\,\hat h^{{1\over 2}}\,\Big[
{(dx)^2\over q^2}\,+\,(1+x-x_q)^2\,
\Big(ds^2_{{\mathbb C \mathbb P}^2}+q^2\,
(d\tau+A)^2\Big)\Big]\,\,\,.
\eear
We have only one free parameter, $x_q$, corresponding to a family of geometries. Actually, in our unquenched flavored background one would expect the geometry to depend on two quantities, the amount of flavors and their mass. We have rescaled out the quark mass with the definition of the hatted quantities in (\ref{hatted_def}) and therefore $x_q$ will serve us to parametrize the amount of flavors. 

It is interesting to write down the relation between the hatted and unhatted cartesian coordinates in terms of the quark mass $m_q$. Taking into account that $C\sim m_q^{-{1\over 2}}$, we get for the longitudinal ($x^1\,x^2$) and transverse ($x^3$) directions:
\beq
\hat x_{\parallel}\sim m_q^{{3\over 4}}\,x_{\parallel}\,\,,
\qquad\qquad
\hat x_{\perp}\sim m_q^{{1\over 4}}\,x_{\perp}\,\,.
\label{hatted_distances}
\eeq
Therefore, for fixed distances $x_{\parallel}$ and $x_{\perp}$, taking $m_q\to 0$ is equivalent to considering the UV 
$\hat x_{\parallel}, \hat x_{\perp}\to 0$ region in the hatted variables, whereas taking large $m_q$ amounts to zooming in the IR region of large  $x_{\parallel}$ and $x_{\perp}$.

\begin{figure}[ht]
\center
\includegraphics[width=0.60\textwidth]{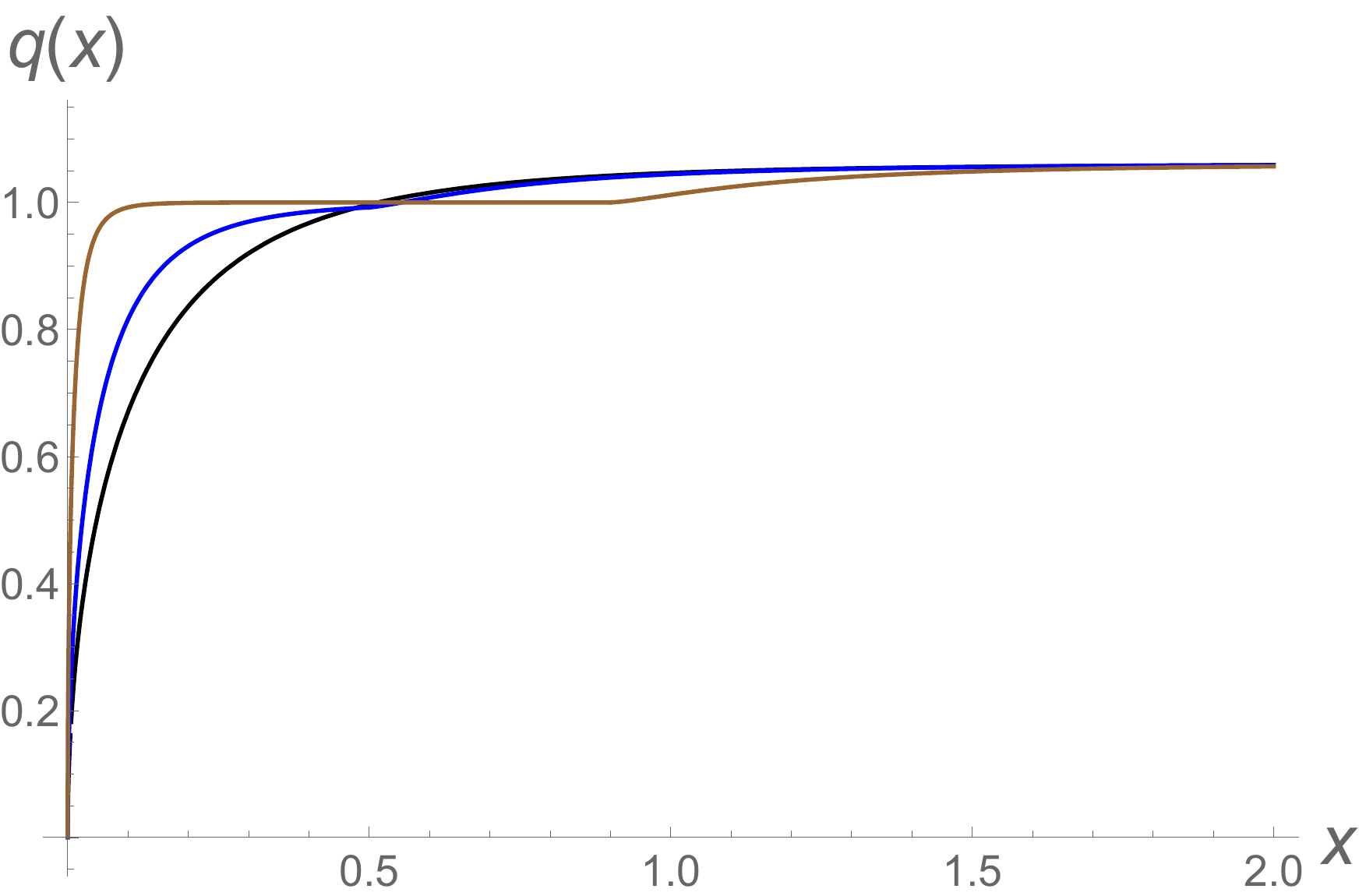}
\caption{The squashing function $q(x)$ for $x_q=0.005$ (black), $x_q=0.5$ (blue) and $x_q=0.9$ (brown). As 
$x\rightarrow 0$ the squashing factor goes to zero for all the values of $x_q$, except for the case $x_q=1$. In that 
particular case $q(x)$ remains finite and equal to one inside the cavity.}
\label{q_x}
\end{figure}

For fixed quark mass, $x_q$ is the parameter that controls the flavor effects in the IR. 
To illustrate how the flavor  branes deform the metric as we move in the holographic direction, we have plotted in Fig.~\ref{q_x} the squashing function $q(x)$ for different values of $x_q$.  For $x_q\sim 1$ the function $q(x)$ is nearly constant and equal to one in the sourceless region $x\le x_q$ and grows monotonically for $x\ge x_q$, until it reaches its asymptotic UV value $q=3/2\sqrt{2}$. It follows from these results  and those represented in Fig.~\ref{W_x} that the geometry becomes more and more isotropic inside the cavity as we increase the parameter $x_q$.

In the following sections we will study our gravity dual by computing several observables with the purpose of exploring their change as we move from the UV to the IR as we vary the size of the cavity $x_q$.

\section{Wilson loops}\label{Wilson_loops}

In holography, the potential energy between a ``quark" and an ``antiquark" is obtained from the solution of the equation of motion of a fundamental string hanging from the UV boundary \cite{Maldacena:1998im,Rey:1998ik}. These equations are derived from the Nambu-Goto action:
\beq
S\,=\,{1\over 2\pi}\,\,\int d\tau d\sigma\,e^{{\phi\over 2}}\,\sqrt{-\det g_2}\,\,,
\label{NG_action_general}
\eeq
where $g_2$ is the induced metric on the worldsheet of the string.  We will calculate the $q\bar q$ potential in two cases. First we will consider the  intralayer case,  in which the quark and the antiquark are in the same layer and have the same value of the coordinate $x^3$. After this we will consider the interlayer configuration, where the quarks are separated in the anisotropic direction.

\subsection{Intralayer potential}
We take $(t, \hat x^1)$ as worldvolume coordinates of a fundamental string and we will consider an ansatz with $x=x(\hat x^1)$ with the other cartesian coordinates being constant. The induced metric on the two-dimensional worldsheet is: 
\beq
ds_2^2\,=\,-\hat h^{-{1\over 2}}\,(d\hat x^0)^2\,+\,\hat h^{-{1\over 2}}\,\Big(1\,+\,
{\hat h\,Z^2_q\over q^2}\,(x')^2\,\Big)\,(d\hat x^1)^2\,\,,
\eeq
where  the prime denotes derivative with respect to $\hat x^1$ and we have denoted $Z_q\equiv Z(x_q)$. The Nambu-Goto action of the string is:
\beq
{S\over \hat T}\,=\,{1\over 2\pi}\,\int d\hat x^1\,e^{{\phi\over 2}}\,\sqrt{-g_2}\,\equiv
\int d\hat x^1\, L\,\,,
\eeq
where $\hat T\,=\,\int d\hat x^0$ and the Lagrangian density $L$ is:
\beq
L\,=\,{1\over 2\pi\,\sqrt{C}}\,
{\hat h^{-{1\over 2}}\,q\over \sqrt{\hat W}}\,
\sqrt{1\,+\,Z_q^2\,{\hat h\over q^2}\,(x')^2}\,\,.
\eeq
As $L$ does not depend explicitly on the holographic coordinate $x$, one has the following first integral:
\beq
x'\,{\partial L\over \partial x'}\,-\,L\,=\,{\rm constant}\,\,,
\eeq
from which we  get:
\beq
{q\over \sqrt{\hat h\,\hat W}}\,{1\over \sqrt{1\,+\,Z_q^2\,{\hat h\over q^2}\,(x')^2}}\,=\,
{q_0\over \sqrt{\hat h_0\,\hat W_0}}\,\,,
\eeq
where $q_0$, $\hat h_0$, and $\hat W_0$ are the values of the functions 
$q$, $\hat h$, and $\hat W$, respectively, at the turning point $x=x_0$. From this relation we obtain  
$x'$ as:
\beq
x'\,=\,\pm\,{q\over Z_q\,\sqrt{\hat h}}\,\sqrt{{\hat h_0\,\hat W_0\,q^2\over 
\hat h\,\hat W\,q_0^2}\,-\,1}\,\,,
\eeq
which can be straightforwardly integrated to give:
\beq
\hat x^1(x)\,=\,\pm Z_q\,\int_{x_0}^{x}\,
{\sqrt{\hat  h(\bar x)}\over q(\bar x)}\,
{d\bar x\over 
\sqrt{{\hat h_0\,\hat W_0\,q^2(\bar x)\over 
\hat h(\bar x)\,\hat W(\bar x)\,q_0^2}\,-\,1}}\,\,.
\eeq
The (hatted) quark-antiquark distance at the boundary is:
\beq
\hat d_{\parallel}\,=\,2\,Z_q
\int_{x_0}^{\infty}\,
{\sqrt{\hat  h( x)}\over q( x)}\,
{d x\over 
\sqrt{{\hat h_0\,\hat W_0\,q^2( x)\over 
\hat h( x)\,\hat W( x)\,q_0^2}\,-\,1}}\,\,.
\label{hat_d_parallel}
\eeq

Let us now calculate the on-shell action of the fundamental string. Plugging the solution into the action, we get:
\beq
{S_{on-shell}\over \hat T}\,=\,{Z_q\over \pi\sqrt{C}}\,\,
\int_{x_0}^{x_{max}}\,{dx\over \sqrt{\hat W}\,
\sqrt{1\,-\,{q_0^2\,\hat h\,\hat W\over q^2\,\hat h_0\,\hat W_0}}}\,\,.
\eeq
As usual, this on-shell action is divergent and must be regulated. We do it by subtracting the action of two straight fundamental strings stretched from the origin $x=0$ to $x=x_{max}$:
\beq
{S^{reg}_{on-shell}\over \hat T}\,=\,{S_{on-shell}\over \hat T}\,-\,
2\,{Z_q\over 2\pi}\,\int_{0}^{x_{max}}\,dx\,{e^{{\phi\over 2}}\over q}\,=\,
{S_{on-shell}\over \hat T}\,-\,{Z_q\over \pi\sqrt{C}}\,
\int_{0}^{x_{max}}\,{dx\over \sqrt{\hat W}}\,\,.
\eeq
The quark-antiquark potential is then given by:
\beq
V_{q\bar q}\,=\,{{S^{reg}_{on-shell}\over  T}}\,=\,{Q_f\over C^{{3\over 2}}}\,
{S^{reg}_{on-shell}\over \hat T}\,\,,
\eeq
where we have used the relation between $T$ and $\hat T$:
\beq
\hat T\,=\,{Q_f\over C^{{3\over 2}}}\,T\,\,.
\eeq
More explicitly:
\beq
V_{q\bar q}\,=\,{Z_q\,Q_f\over \pi\,C^2}\,
\Bigg[\int_{x_0}^{\infty}\,{dx\over \sqrt{\hat W}}\,\Bigg(
{1\over \sqrt{1\,-\,{q_0^2\,\hat h\,\hat W\over q^2\,\hat h_0\,\hat W_0}}}-1\Bigg)\,-\,
\int_0^{x_0}{dx\over \sqrt{\hat W}}\Bigg]\,\,.
\label{intralayer_pot}
\eeq

\begin{figure}[ht]
\center
 \includegraphics[width=0.45\textwidth]{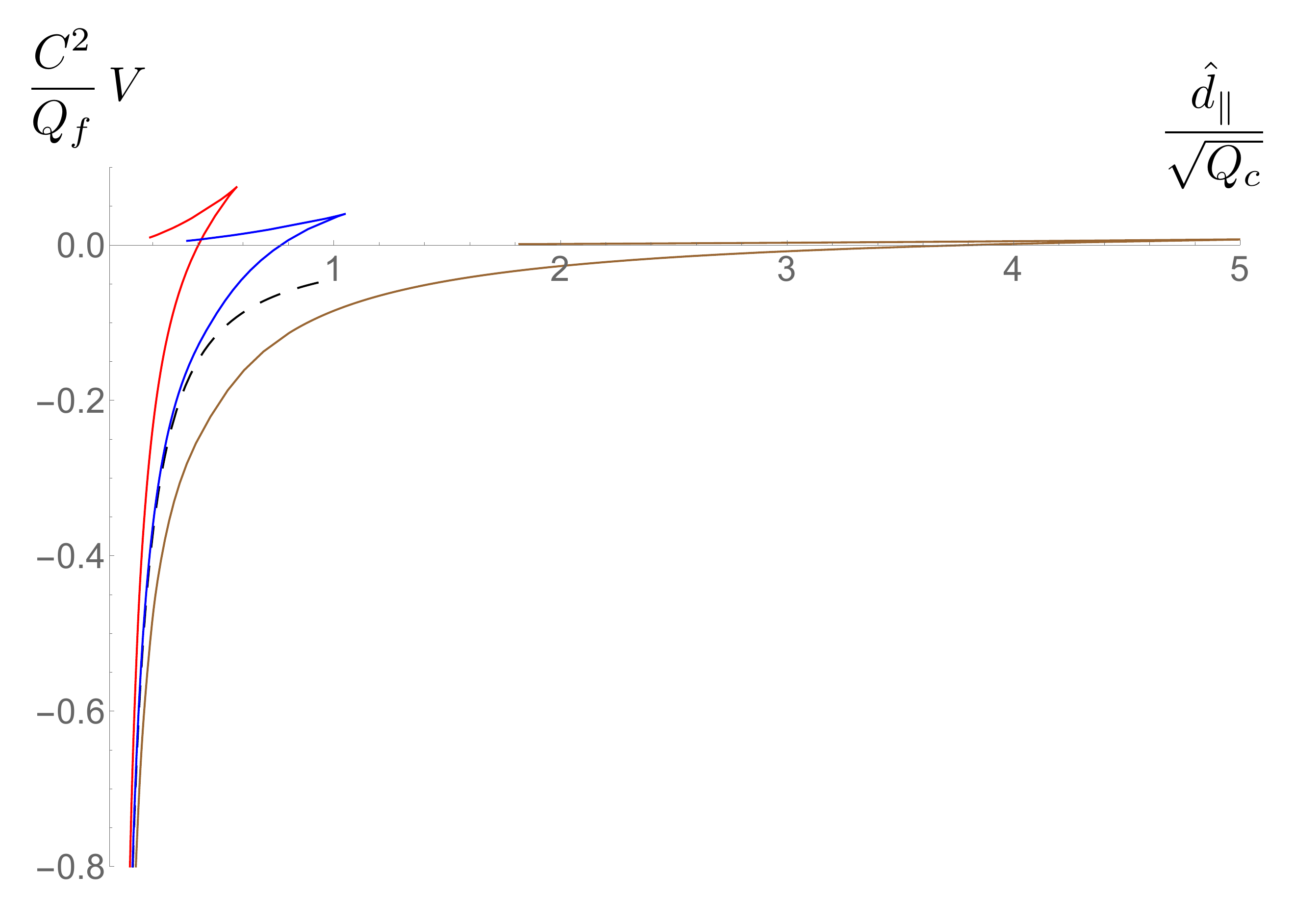}
   \caption{We depict the $\bar q q$ intralayer potential (\ref{intralayer_pot}) versus $\hat d_{\parallel}\propto m_q^{{3\over 4}}\, d_{\parallel}$ for $x_q=0.1$ (red),  $x_q=0.5$  (blue), and $x_q=0.9$ (brown). The dashed curve is the UV limit (\ref{intralayer_pot_UV}). Notice that the maximal separation for $\bar q q$ increases with $x_q$. }
\label{V_parallel}
\end{figure}

We have numerically evaluated the potential $V_{q\bar q}$ as a function of $\hat d_{\parallel}$. The results are presented in Fig.~\ref{V_parallel} for different  values of $x_q$. We notice that all curves become coincident for small $\hat d_{\parallel}$. As $\hat d_{\parallel}\sim m_q^{{3\over 4}}\,d_{\parallel}$ (see eq. (\ref{hatted_distances})),  one expects to recover the massless scaling solution in the UV domain $\hat d_{\parallel}\to 0$.  Indeed, we prove below that $V_{q\bar q}\sim \hat d_{\parallel}^{-{4\over 3}}$ in the UV region of small $\hat d_{\parallel}$. This behavior matches the one obtained numerically, as shown in Fig.~\ref{V_parallel}.  As we move towards the IR by decreasing the turning point coordinate $x_0$ and increasing  
$\hat d_{\parallel}$, we obtain that there is a maximal value  of $\hat d_{\parallel}$ (corresponding to a minimal value $x_0^{\min}<x_q$  of the turning point coordinate). For $x_0<x_0^{\min}$ the dominant configuration is a disconnected one, in which the two ends of the string go straight from the boundary to the origin. This behavior has been obtained previously in backgrounds dual to unquenched flavors \cite{Bigazzi:2008gd,Bigazzi:2008zt,Bigazzi:2008qq,Ramallo:2008ew} (in other types of backgrounds, see also \cite{Avramis:2006nv,Avramis:2007mv,Avramis:2006em,Avramis:2007wb,Bea:2017iqt}). 
Indeed, dynamical quarks produce string breaking and a maximal length due to pair creation. In our case this breaking is not produced in the scaling solution with $m_q=0$.  Moreover, the critical distance at which the string breaks grows with $x_q$, as is also evident in Fig.~\ref{V_parallel}, and becomes very large when $x_q\sim 1$.  This is easy to understand since the breaking occurs  when the string penetrates deeply in the cavity, whose size is maximal when $x_q\sim 1$, and the integrals (\ref{hat_d_parallel}) and (\ref{intralayer_pot}) get their main contribution from the sourceless region inside the cavity. For large enough values of $\hat d_{\parallel}$ the dominant configuration is the disconnected one with  zero energy, which means that the external quarks are completely screened by dynamical quarks popping out  from the vacuum. 
A recent interesting work \cite{Gursoy:2018ydr}, in a seemingly unrelated context of holographic QCD, parallels our findings. The authors of \cite{Gursoy:2018ydr} demonstrated that large amounts of anisotropy will completely screen the interactions between quarks and anti-quarks, while in the absence of anisotropy the model would otherwise be confining.

\subsubsection{UV limit}
Let us  now evaluate   the potential in the UV limit in which $d_{\parallel}$ is small (large $x_0$) and our embedding is close to the boundary. In this limit we can use that,   at leading order in the UV,  the squashing  function $q$ is constant and that $\hat W\sim x^{-{1\over 3}}$ and $\hat h\sim x^{-4}$ (see  eqs. (\ref{W_UV_x}), (\ref{h_UV_zeta}), and (\ref{q_UV_zeta})). 
We will use these values to calculate the integrals for $\hat d_{\parallel}$ and $V_{q\bar q}$ in (\ref{hat_d_parallel}) and (\ref{intralayer_pot}). At leading order we get the following relation between $\hat d_{\parallel}$  and $x_0$:
\beq
\hat d_{\parallel}\approx
{8\,\sqrt{2}\over 3\sqrt{15}}\,{\sqrt{Q_c}\over Z_q}\,\sqrt{\pi}\,
{\Gamma\Big({5\over 7}\Big)\over \Gamma\Big({3\over 14}\Big)}\,{1\over x_0}\,\,.
\label{d_parallel_UV}
\eeq
Similarly, we approximate  the potential $V_{q\bar q}$  by the following integral:
\beq
V_{q\bar q}\approx{2\sqrt{2}\over 3\,\cdot 2^{{1\over 6}}\pi}\,{Z_q^{{4\over 3}}\,Q_f\over C^2}\,
x_0^{{4\over 3}}\,\,\Bigg[
\int_1^{\infty} dz\,z^{{1\over 3}}\,
\Bigg({z^{{7\over 3}}\over \sqrt{z^{{14\over 3}}-1}}\,-\,1\Bigg)\,-\,{3\over 4}\Bigg]\,\,.
\eeq
Performing the integral, we get:
\beq
V_{q\bar q}\approx -{1\over 2^{{1\over 6}}\sqrt{2\pi}}\,
{Z_q^{{4\over 3}}\,Q_f\over C^2}\,
{\Gamma\Big({5\over 7}\Big)\over \Gamma\Big({3\over 14}\Big)}\,\,x_0^{{4\over 3}}\,\,.
\eeq
By using the relation (\ref{d_parallel_UV}) we can eliminate $x_0$ in favor of the $q\bar q$ distance $\hat d_{\parallel}$. After some calculation we get:
\beq
{C^2\over Q_f}\,V_{q\bar q}\approx -\beta_{\parallel} \,\,\Bigg({\sqrt{Q_c}\over \hat d_{\parallel}}\Bigg)^{{4\over 3}}\,\,,
\qquad\qquad
\beta_{\parallel}= {16\pi^{{1\over 6}}\over 9\,\cdot 5^{{2\over 3}}}
\Bigg({\Gamma\Big({5\over 7}\Big)\over \Gamma\Big({3\over 14}\Big)}\Bigg)^{{7\over 3}}\,\,,
\label{intralayer_pot_UV}
\eeq
which coincides with the  result found in \cite{Penin:2017lqt} for the massless scaling background. In Fig.~\ref{V_parallel} we show that  the potential (\ref{intralayer_pot_UV}) does indeed coincide with the numerical results in the UV 
domain $\hat d_{\parallel}\to 0$ for all values of the parameter $x_q$.

\subsection{Interlayer potential}

We now consider a Wilson loop that extends in the $x^3$ direction with $x^1$ and $x^2$ constant, which corresponds to two fundamentals located at  different layers. Accordingly, we take  $(t, \hat x^3)$ as worldvolume coordinates and consider an ansatz in which $x=x(\hat x^3)$. The two-dimensional induced metric is now:
\beq
ds_2^2 = -\hat h^{-{1\over 2}}\,(d\hat x^0)^2\,+\,{\hat h^{-{1\over 2}}\over q^4}\,
\big[\hat W^2\,+\,Z_q^2\,\hat h\,q^2\,(x')^2\,\big]\,(d\hat x^3)^2\,\,,
\eeq
where the prime now denotes derivative with respect to $\hat x^3$. The Nambu-Goto action is:
\beq
{S\over \hat T}\,=\,{1\over 2\pi}\,\int d\hat x^3\,e^{{\phi\over 2}}\,\sqrt{-g_2}\,\equiv
\int d\hat x^3\, L\,\,,
\eeq
with
\beq
L\,=\,{1\over 2\pi\,\sqrt{C}}\,
{\hat h^{-{1\over 2}}\over q\, \sqrt{\hat W}}\,
\sqrt{\hat W^2\,+\,Z_q^2\,\hat h\, q^2\,(x')^2}\,\,.
\eeq
The first integral derived from $L$ is:
\beq
{\sqrt{\hat W}\over q\,\sqrt{\hat h}}\,\,{1\over
\sqrt{1\,+\,{Z_q^2\,q^2\, \hat h\over \hat W^2}\,(x')^2}}\,=\,
{\sqrt{\hat W_0}\over q_0\,\sqrt{\hat h_0}}\,\,,
\eeq
where again $q_0$, $\hat h_0$, and $\hat W_0$ are the values of the functions 
$q$, $\hat h$, and $\hat W$, respectively, at the turning point $x=x_0$. Then:
\beq
x'\,=\,\pm\,{\hat W\over Z_q\,q\sqrt{\hat h}}\,\sqrt{{\hat h_0\,q^2_0\,\hat W\over 
\hat h\,q^2\,\hat W_0}\,-\,1}\,\,.
\eeq
Integrating this equation we get:
\beq
\hat x^3(x)\,=\,\pm Z_q\,\int_{x_0}^{x}\,
{q(\bar x)\,\sqrt{\hat  h(\bar x)}\over \hat W(\bar x) }\,
{d\bar x\over 
\sqrt{{\hat h_0\,q^2_0\,\hat W(\bar x)\over 
\hat h(\bar x)\,q^2(\bar x)\,\hat W_0}\,-\,1}
}\,\,.
\eeq
Therefore, the quark-antiquark distance along the direction transverse to the  layers is:
\beq
\hat d_{\perp}\,=\,2\,Z_q
\int_{x_0}^{\infty}\,
{q(x)\,\sqrt{\hat  h( x)}\over \hat W( x)}\,
{d x\over 
\sqrt{{\hat h_0\,q^2_0\,\hat W( x)\over 
\hat h( x)\,q^2( x)\,\hat W_0}\,-\,1}
}\,\,.
\eeq
The unregulated on-shell action for this case is:
\beq
{S_{on-shell}\over \hat T}\,=\,{Z_q\over \pi\sqrt{C}}\,\,
\int_{x_0}^{x_{max}}\,{dx\over \sqrt{\hat W}\,
\sqrt{1\,-\,{q^2\,\hat h\,\hat W_0\over q^2_0\,\hat h_0\,\hat W}}}\,\,.
\eeq
Proceeding as in the intralayer case to regulate this action, we arrive at the following 
quark-antiquark potential:
\beq
V_{q\bar q}\,=\,{Z_q\,Q_f\over \pi\,C^2}\,
\Bigg[\int_{x_0}^{\infty}\,{dx\over \sqrt{\hat W}}\,\Bigg(
{1\over \sqrt{1\,-\,{q^2\,\hat h\,\hat W_0\over q^2_0\,\hat h_0\,\hat W}}}-1\Bigg)\,-\,
\int_0^{x_0}{dx\over \sqrt{\hat W}}\Bigg]\,\,.
\label{intrelayer_pot}
\eeq

\begin{figure}[ht]
\center
 \includegraphics[width=0.50\textwidth]{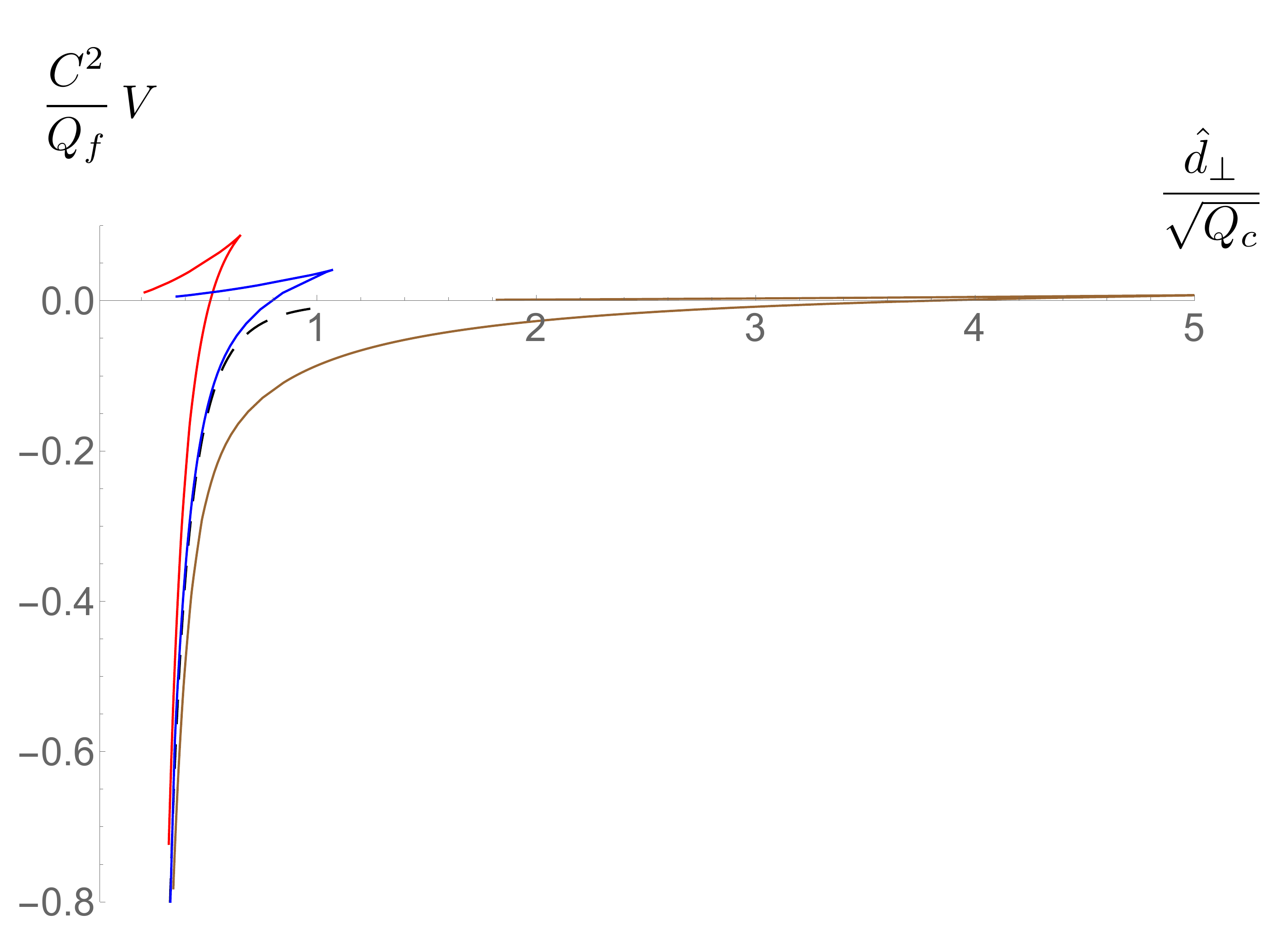}
 \,\,
 \includegraphics[width=0.47\textwidth]{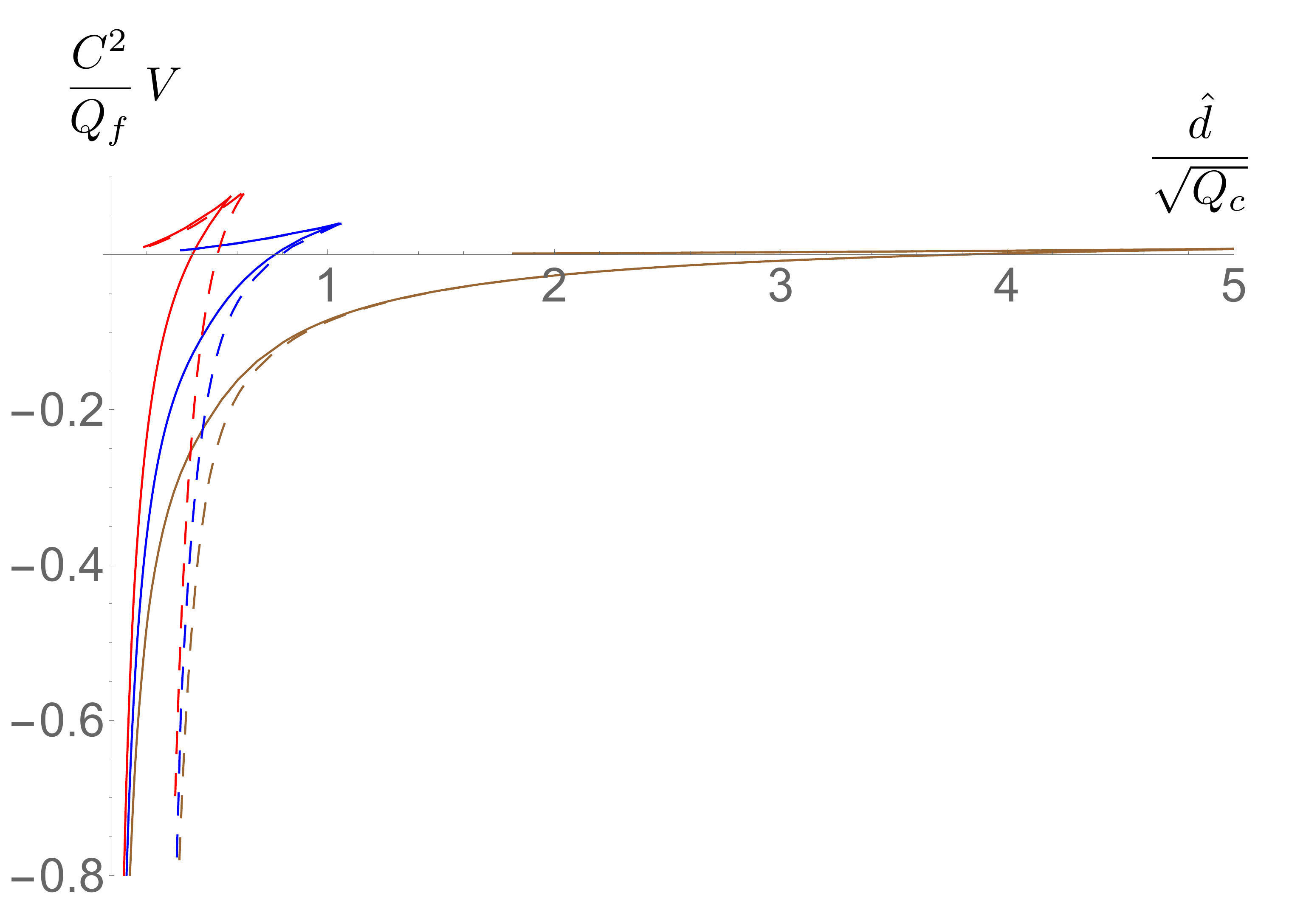}
  \caption{On the left we plot the $\bar q q$ interlayer potential (\ref{intrelayer_pot}) versus $\hat d_{\perp}\propto m_q^{{1\over 4}}\, d_{\perp}$ for $x_q=0.1$ (red),  $x_q=0.5$  (blue), and $x_q=0.9$ (brown).  The dashed curve is the UV potential (\ref{intrelayer_pot_UV}). On the right we compare the intralayer and interlayer potentials   for $x_q=0.1$ (red),  $x_q=0.5$  (blue), and $x_q=0.9$ (brown). The continuous (dashed) curves correspond to the intralayer (interlayer) potentials.   }
\label{V_perp}
\end{figure}

The numerical results for the interlayer potential have been plotted in Fig.~\ref{V_perp}. They are qualitatively similar to the intralayer case of Fig.~\ref{V_parallel}. In the UV region $\hat d_{\perp}\propto m_q^{{1\over 4}}\, d_{\perp}\to 0$, the potential decays as $V_{q\bar q}\sim d_{\perp}^{-4}$, in agreement with our analytic calculation of Sec.~\ref{interlayer_pot_UV}. In this case there is also a maximal length which increases with $x_q$. We have also compared the intralayer and interlayer potentials for the same value of $x_q$. These potentials are plotted together on the right panel of Fig.~\ref{V_perp},  where we notice that they have very different  behavior in the UV but they become very similar in the IR (cf. also \cite{Gursoy:2018ydr}). This IR similarity increases as $x_q$ approaches its maximal value $x_q=1$, which is consistent with the fact that for large values of $x_q$ the isotropic unflavored limit is rapidly attained in the IR.

\subsubsection{UV limit}
\label{interlayer_pot_UV}

Let us consider the UV limit in which $\hat d_{\perp}$ is small and we have the following approximate relation between $d_{\perp}$ and $x_0$:
\beq
\hat d_{\perp}\approx
{8\,\cdot 2^{{1\over 6}}\over \sqrt{15}}\,{\sqrt{Q_c}\over Z_q^{{1\over 3}}}\,\sqrt{\pi}\,
{\Gamma\Big({3\over 5}\Big)\over \Gamma\Big({1\over 10}\Big)}\,{1\over x_0^{{1\over 3}}}\,\,.
\eeq
In this limit the potential can be approximated as:
\beq
V_{q\bar q}\approx{2\sqrt{2}\over 3\,\cdot 2^{{1\over 6}}\pi}\,{Z_q^{{4\over 3}}\,Q_f\over C^2}\,
x_0^{{4\over 3}}\,\,\Bigg[
\int_1^{\infty} dz\,z^{{1\over 3}}\,
\Bigg({z^{{5\over 3}}\over \sqrt{z^{{10\over 3}}-1}}\,-\,1\Bigg)\,-\,{3\over 4}\Bigg]\,\,.
\eeq
The integral can be computed analytically and yields the following result for the potential:
\beq
V_{q\bar q}\approx -{1\over 2^{{1\over 6}}\sqrt{2\pi}}\,
{Z_q^{{4\over 3}}\,Q_f\over C^2}\,
{\Gamma\Big({3\over 5}\Big)\over \Gamma\Big({1\over 10}\Big)}\,\,x_0^{{4\over 3}}\,\,.
\eeq
In terms  of $\hat d_{\perp}$ we get:
\beq
{C^2\over Q_f}\,
V_{q\bar q}\approx -\beta_{\perp}\,\, 
\Bigg({\sqrt{Q_c}\over \hat d_{\perp}}\Bigg)^{4}\,\,,
\qquad\qquad
\beta_{\perp}= {2^{12}\,\pi^{{3\over 2}}\over 3^2\,\cdot 5^{2}}
\Bigg({\Gamma\Big({3\over 5}\Big)\over \Gamma\Big({1\over 10}\Big)}\Bigg)^{5}\,\,,
\label{intrelayer_pot_UV}
\eeq
which is equivalent to the  result  obtained in \cite{Penin:2017lqt} for the massless background.
As illustrated in Fig.~\ref{V_perp}, the potential (\ref{intrelayer_pot_UV}) nicely matches the numerical results.

\section{Entanglement entropy}
\label{entanglement_entropy}

The entanglement entropy of a region and its complement is a good measure of the quantum correlations of the system. In holography the entanglement entropy is obtained by minimizing an area functional for an eight-dimensional surface embedded in ten-dimensional spacetime \cite{Ryu:2006bv,Ryu:2006ef}.  Let $A$ be a spatial region in the gauge theory. The holographic entanglement entropy between $A$ and its complement is:
\beq
S_{A}\,=\,{1\over 4 G_{10}}\,\int_{\Sigma} d^8\xi\,\sqrt{\det g_8}\,\,,
\eeq
where $\Sigma$ is the eight-dimensional spatial surface whose boundary is $A$ and minimizes 
$S_{A}$,   $G_{10}$ is the ten-dimensional Newton constant ($G_{10}=8\pi^6$ in our units) and $g_8$ is the induced metric on $\Sigma$ in the Einstein frame. In this section we will apply this prescription when $A$ is a slab of infinite extent in the two cartesian directions and having a finite width in the remaining cartesian coordinate. Clearly, there are two  cases to study, namely parallel and transverse slabs, which we analyze separately. We note that again we find striking similarity with the results in \cite{Gursoy:2018ydr}.

\subsection{Parallel slab}
First we consider the case in which $A$ is an infinite slab with a finite width parallel to the layers, namely:
\beq
A\,=\,\Big\{-{\hat l_{\parallel}\over 2}\,\le\,\hat x^1\,\le {\hat l_{\parallel}\over 2}\,, \,-\infty< \hat x^2, \hat x^3<+\infty\Big\}\,\,.
\eeq
We will parameterize $\Sigma$ by a function $x\,=\,x(\hat x^1)$. The eight-dimensional induced metric on $\Sigma$ is:
\bear
&&ds^2_8\,=\,\hat h^{-{1\over 2}}\,\Big[1\,+\,Z_q^2\,{\hat h\over q^2}\,(x')^2\Big]\,(d\hat x^1)^2\,+\,
\hat h^{-{1\over 2}}\,\Big[(d\hat x^2)^2\,+\,{\hat W^2\over q^4}\,(d\hat x^3)^2\Big]\,\rc\rc
&&\qquad\qquad\qquad\qquad
+\hat h^{{1\over 2}}\,Z_q^2\,(1+x-x_q)^2\, \Big(ds^2_{{\mathbb C \mathbb P}^2}+q^2\,
(d\tau+A)^2\Big)\,\,,
\eear
where the prime now denotes derivative with respect to $\hat x^1$. If we integrate over all the coordinates except $x$, we get:
\beq
{S_{\parallel}\over \hat L_2\hat L_3}\,=\,
{Z_q^5\over 32\,\pi^3}\,\int d\hat x^1\,(1+x-x_q)^5\,{\hat h^{{1\over 2}}\hat W\over q}\,
\sqrt{1+Z_q^2\,{\hat h\over q^2}\,(x')^2}\,\,.
\eeq
The first integral derived from $S_{\parallel}$ is:
\beq
{\hat h^{{1\over 2}}\hat W\over q}\,{(1+x-x_q)^5\over
\sqrt{1+Z_q^2\,{\hat h\over q^2}\,(x')^2}}\,=\,
{\hat h^{{1\over 2}}_0\hat W_0 \over q_0}\,(1+x_0-x_q)^5\,\,,
\eeq
where the subscript nought denotes that the corresponding quantity is evaluated at the minimal value $x_0$ of $x$. From this last equation we get:
\beq
 x'\,=\,\pm\,{q\over Z_q\,\sqrt{\hat h}}\,
\sqrt{{(1+x-x_q)^{10}\over (1+x_0-x_q)^{10}}\,
{\hat h\,\hat W^2\,q_0^2\over \hat h_0\,\hat W^2_0\,q^2}\,-\,1}\,\,,
\eeq
which can be integrated to give  $\hat l_{\parallel}$:
\beq
\hat l_{\parallel}\,=\,2Z_q\,\int_{x_0}^{\infty}\,
{\sqrt{\hat h}\over q}\,{dx\over 
\sqrt{{(1+x-x_q)^{10}\over (1+x_0-x_q)^{10}}\,
{\hat h\,\hat W^2\,q_0^2\over \hat h_0\,\hat W^2_0\,q^2}\,-\,1}}\,\,.
\label{hat_l_parallel}
\eeq
The entanglement entropy for this configuration is given by the divergent integral:
\beq
{S_{\parallel}\over \hat L_2\hat L_3}\,=\,{Z_q^6\over 16\,\pi^3}\,
\int_{x_0}^{x_{max}}\,{\hat h\,\hat W\over q^2}\,
{(1+x-x_q)^5\over 
\sqrt{1-
{(1+x_0-x_q)^{10}\over (1+x-x_q)^{10}}\,
{\hat h_0\,\hat W^2_0\,q^2\over \hat h\,\hat W^2\,q^2_0}}}\,dx\,\,.
\eeq
We will regularize $S_{\parallel}$ by subtracting the entropy of a configuration that we call the flat surface in the following. To conform with the homology constraint, the surface is not disconnected, but it is connected at the bottom $x=0$. The full flat surface consists of three constant pieces: two straight $\hat x^1=\pm l_{\parallel}/2$ ones and a horizontal one $x=const.=0$, each individually being solutions to the equation of motion. It turns out that the contribution of the horizontal surface to the area integral vanishes. The entropy for the flat embedding is thus
\beq
{S_{\parallel}^{flat}\over \hat L_2\hat L_3}\,=\,{Z_q^6\over 16\,\pi^3}\,
\int_{0}^{x_{max}}\,{\hat h\,\hat W\over q^2}\,
(1+x-x_q)^5\,dx\,\,.
\eeq
Thus, we define the finite entanglement entropy as:
\beq
{S_{\parallel}^{finite}\over \hat L_2\hat L_3}\,=
{S_{\parallel}\,-\,S_{\parallel}^{flat}\over \hat L_2\hat L_3}\,\,.
\label{reg_S_parallel}
\eeq

\begin{figure}[ht]
\center
 \includegraphics[width=0.55\textwidth]{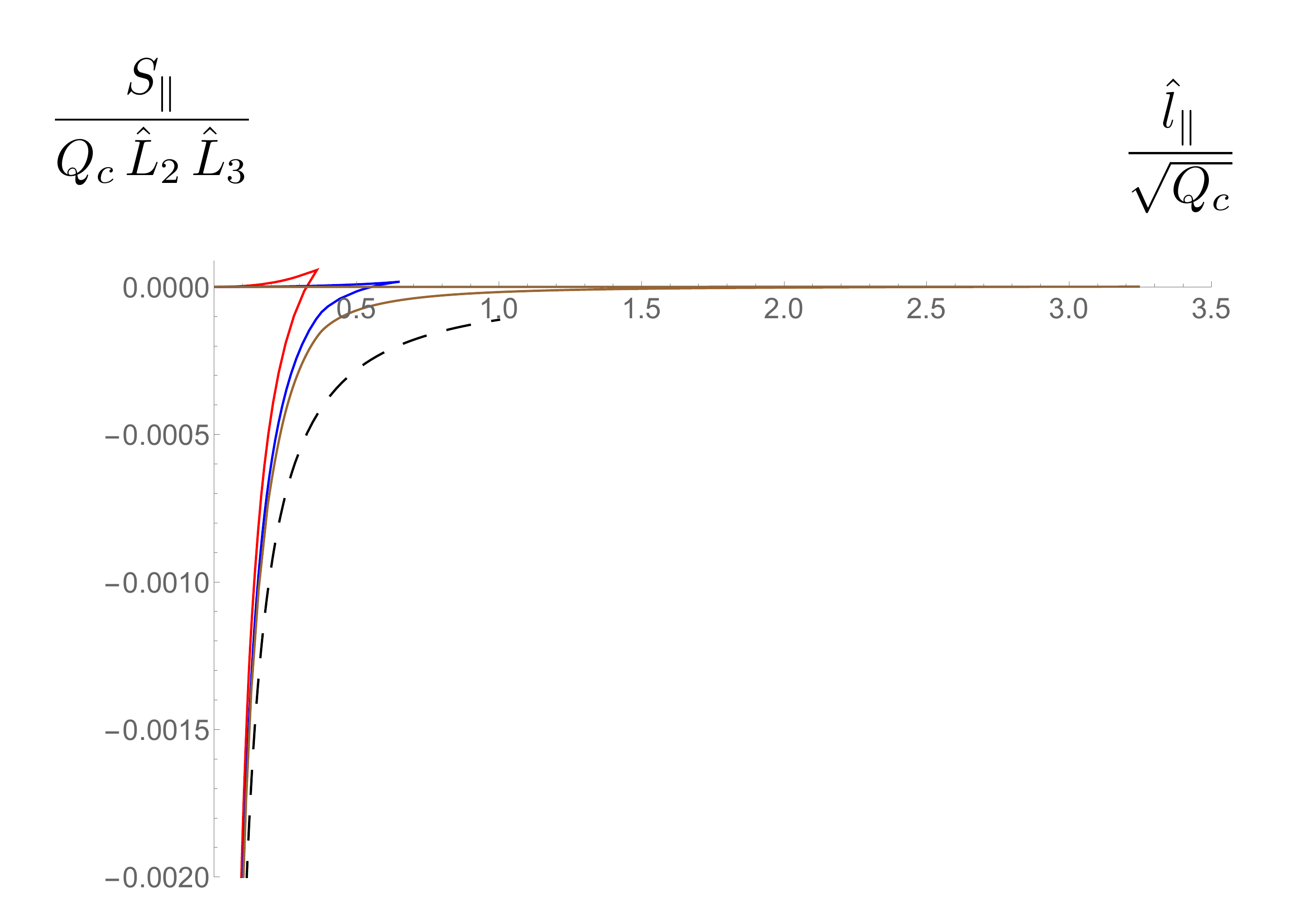}
 \caption{Plot of the entanglement entropy for a parallel slab as a function of its rescaled width. The continuous lines are the results of the numerical integration of (\ref{finite_S_parallel}) and (\ref{hat_l_parallel})  for $x_q=0.1$ (red),  $x_q=0.5$  (blue) and $x_q=0.9$ (brown). The dashed curve is the UV result  (\ref{S_parallel_UV}). }
\label{S_parallel}
\end{figure}

After some calculations, we get:
\bear
&&{S_{\parallel}^{finite}\over \hat L_2\hat L_3}\,=\,-{Z_q^6\over 16\,\pi^3}\,
\Bigg(\int_{x_0}^{\infty}{\hat h\,\hat W\over q^2}\,
(1+x-x_q)^5\,\Bigg[\,1\,-\,
{1\over 
\sqrt{1-
{(1+x_0-x_q)^{10}\over (1+x-x_q)^{10}}\,
{\hat h_0\,\hat W^2_0\,q^2\over \hat h\,\hat W^2\,q^2_0}}}\,\,\Bigg]\,dx\,\rc\rc
&&\qquad\qquad\qquad\qquad\qquad\qquad\qquad
+\int_{0}^{x_0}{\hat h\,\hat W\over q^2}\,
(1+x-x_q)^5\,dx\,\Bigg)\,\,.
\label{finite_S_parallel}
\eear

The numerical results for $S_{\parallel}$ versus $\hat l_{\parallel}$ are presented in Fig.~\ref{S_parallel} for several values of the parameter $x_q$. In this plot we notice that $S_{\parallel}$ becomes positive when $\hat l_{\parallel}$ is large enough. According to our regularization procedure (\ref{reg_S_parallel}) this means that the disconnected surface is dominant with respect to the connected one when $\hat l_{\parallel}>\hat l_{\parallel}^c$, where $\hat l_{\parallel}^c$ is a critical length which grows with $x_q$.  On the contrary, for small $\hat l_{\parallel}$ the entropy behaves as 
 $S_{\parallel}\sim\hat  l_{\parallel}^{-{4\over 3}}$, with a coefficient independent of $x_q$. As pointed out in \cite{Penin:2017lqt}, this universal behavior is the one corresponding to an effective D2-brane and can be obtained analytically,  as we show in the next subsection. 
A similar behavior of the EE has previously been obtained in backgrounds dual to confining theories and unquenched flavors, see \cite{Klebanov:2007ws,Kol:2014nqa,Georgiou:2015pia}.

\subsubsection{UV limit}

When $\hat l_{\parallel}$ is small, the minimal value $x_0$ of $x$ is large and we can use the expansion of the functions of the background valid for large $x$. At leading order, we get:
\beq
\hat l_{\parallel}\approx{8\,\sqrt{2}\over 3\sqrt{15}}\,
{\sqrt{Q_c}\over Z_q}\,
\int_{x_0}^{\infty}{dx\over x^2\sqrt{(x/x_0)^{{14\over 3}}\,-\,1}}\,=\,
{8\,\sqrt{2\pi}\over 3\sqrt{15}}\,
{\Gamma\Big({5\over 7}\Big)\over \Gamma\Big({3\over 14}\Big)}\,
{\sqrt{Q_c}\over Z_q}\,{1\over x_0}\,\,.
\eeq
Similarly, the regulated entanglement entropy for the parallel slab is:
\beq
{S_{\parallel}^{finite}\over \hat L_2\hat L_3}\approx -\,
{Z_q^{{4\over 3}}\,Q_c\over 15\cdot 2^{{5\over 3}}\,\pi^3}\,x_0^{{4\over 3}}\,
\Bigg[{3\over 4}\,-\,
\int_1^{\infty} dz\,z^{{1\over 3}}\,
\Bigg({z^{{7\over 3}}\over \sqrt{z^{{14\over 3}}-1}}\,-\,1\Bigg)\Bigg]\,\,,
\eeq
which can be integrated analytically with the result:
\beq
{S_{\parallel}^{finite}\over \hat L_2\hat L_3}\approx -\,
{Z_q^{{4\over 3}}\,Q_c\over 5\cdot 2^{{11\over 3}}\,\pi^{{5\over 2}}}\,
{\Gamma\Big({5\over 7}\Big)\over \Gamma\Big({3\over 14}\Big)}\,x_0^{{4\over 3}}\,\,.
\eeq
If we eliminate $x_0$ and write the entanglement entropy in terms of $\hat l_{\parallel}$, we get:
\beq
{S_{\parallel}^{finite}\over Q_c\, \hat L_2\,\hat L_3}\approx -\gamma_{\parallel}\,
\Bigg(
{\sqrt{Q_c}\over \hat l_{\parallel}}\Bigg)^{{4\over 3}}\,\,,
\qquad\qquad
\gamma_{\parallel}\,=\,{2\over 45\cdot 5^{{2\over 3}}\,\pi^{{11\over 6}}}
\Bigg({\Gamma\Big({5\over 7}\Big)\over \Gamma\Big({3\over 14}\Big)}\Bigg)^{{7\over 3}}\,\,.
\label{S_parallel_UV}
\eeq
which is  the same result as in  \cite{Penin:2017lqt} and, as shown in Fig.~\ref{S_parallel}, matches  perfectly the numerical results  when $\hat l_{\parallel}$ is small.

\subsection{Transverse slab}

We now consider a slab with finite width in the direction of $\hat x^3$. The region $A$ in this case is:
\beq
A\,=\,\Big\{-\infty< \hat x^1, \hat x^2<+\infty\,,\,-{\hat l_{\perp}\over 2}\,\le\,\hat x^3\,\le {\hat l_{\perp}\over 2}\Big\}
\,\,,
\eeq
and the induced metric on $\Sigma$ becomes:
\bear
&&ds^2_8\,=\,\hat h^{-{1\over 2}}\,
\Big[(d\hat x^1)^2\,+\,(d\hat x^2)^2\Big]+\hat h^{-{1\over 2}}\,
{\hat W^2\over q^4}\,
\Big[1\,+\,Z_q^2\,{\hat h\,q^2\over \hat W^2}\,(x')^2\Big]\,(d\hat x^3)^2\,\rc\rc
&&\qquad\qquad\qquad\qquad
+\hat h^{{1\over 2}}\,Z_q^2\,(1+x-x_q)^2\, \Big(ds^2_{{\mathbb C \mathbb P}^2}+q^2\,
(d\tau+A)^2\Big)\,\,,
\eear
where now $x=x(\hat x^3)$. The transverse entropy functional is:
\beq
{S_{\perp}\over \hat L_1\,\hat L_2}\,=\,
{Z_q^5\over 32\,\pi^3}\,\int d\hat x^3\,(1+x-x_q)^5\,{\hat h^{{1\over 2}}\hat W\over q}\,
\sqrt{1+Z_q^2\,{\hat h\,q^2\over \hat W^2}\,(x')^2}\,\,.
\eeq
Now the first integral is:
\beq
{\hat h^{{1\over 2}}\hat W\over q}\,{(1+x-x_q)^5\over
\sqrt{1+Z_q^2\,{\hat h\,q^2\over \hat W^2}\,(x')^2}}\,=\,
{\hat h^{{1\over 2}}_0\hat W_0 \over q_0}\,(1+x_0-x_q)^5\,\,,
\eeq
from which it follows that:
\beq
 x'\,=\,\pm\,{\hat W\over Z_q\,q\,\sqrt{\hat h}}\,
\sqrt{{(1+x-x_q)^{10}\over (1+x_0-x_q)^{10}}\,
{\hat h\,\hat W^2\,q_0^2\over \hat h_0\,\hat W^2_0\,q^2}\,-\,1}\,\,.
\eeq
Therefore, the transverse length $\hat l_{\perp}$ is:
\beq
\hat l_{\perp}\,=\,2Z_q\,\int_{x_0}^{\infty}\,
{q\,\sqrt{\hat h}\over \hat W}\,{dx\over 
\sqrt{{(1+x-x_q)^{10}\over (1+x_0-x_q)^{10}}\,
{\hat h\,\hat W^2\,q_0^2\over \hat h_0\,\hat W^2_0\,q^2}\,-\,1}}\,\,.
\label{l_perp}
\eeq
The entropy functional evaluated on the minimal surface for this configuration is:
\beq
{S_{\perp}\over \hat L_1\hat L_2}\,=\,{Z_q^6\over 16\,\pi^3}\,
\int_{x_0}^{x_{max}}\,\hat h\,
{(1+x-x_q)^5\over 
\sqrt{1-
{(1+x_0-x_q)^{10}\over (1+x-x_q)^{10}}\,
{\hat h_0\,\hat W^2_0\,q^2\over \hat h\,\hat W^2\,q^2_0}}}\,dx\,\,,
\eeq
whose divergent part is:
\beq
{S_{\perp}^{div}\over \hat L_1\hat L_2}\,=\,
{Z_q^6\over 16\,\pi^3}\,
\int_{0}^{x_{max}}\,\hat h\,
(1+x-x_q)^5\,dx\,\,.
\eeq
We define $S_{\perp}^{finite}$ as:
\beq
{S_{\perp}^{finite}\over \hat L_1\,\hat L_2}\,=\,
{S_{\perp}\,-\,S_{\perp}^{div}\over \hat L_1\,\hat L_2}\,\,. 
\eeq
\begin{figure}[ht]
\center
 \includegraphics[width=0.45\textwidth]{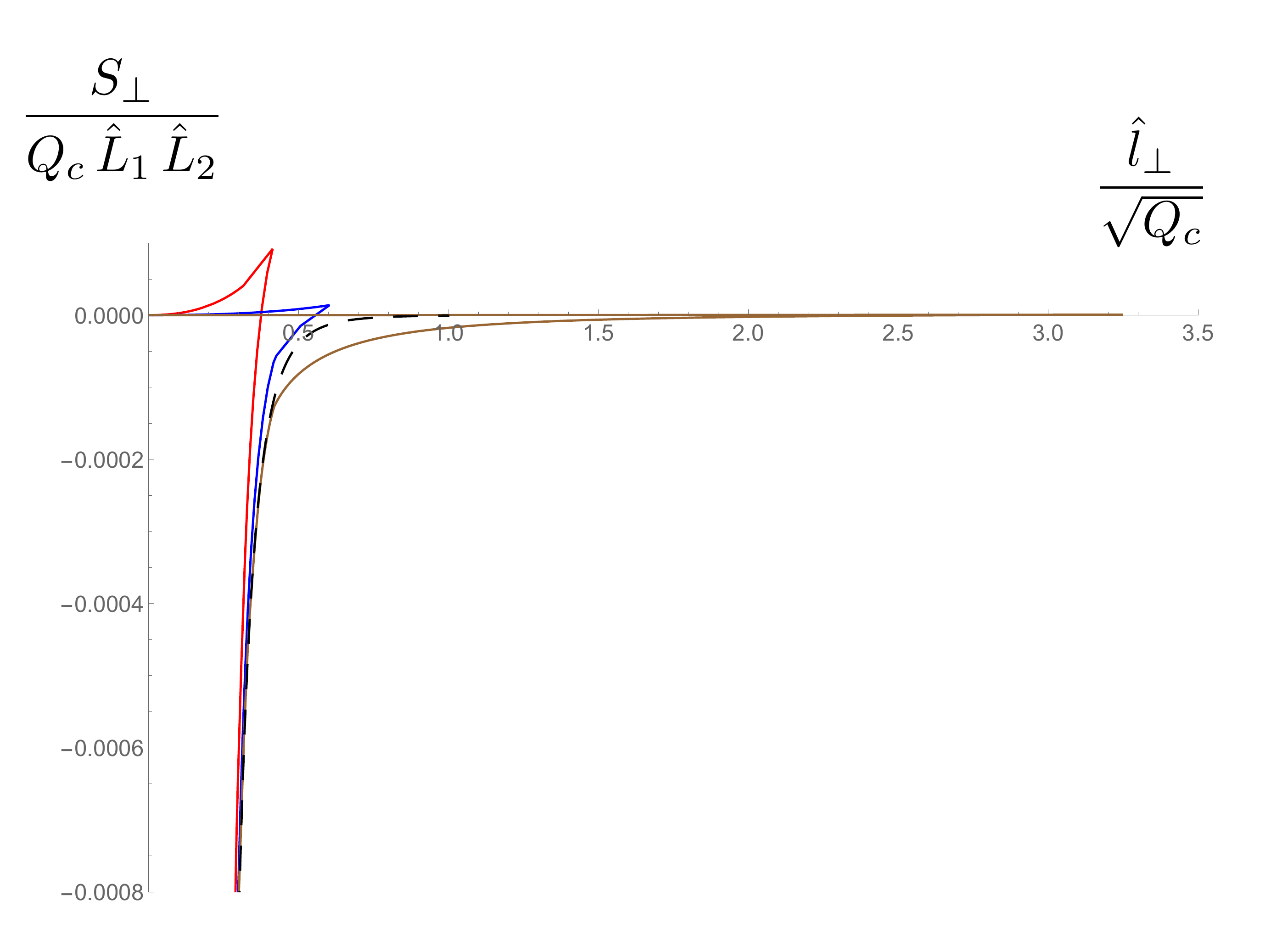}
 \qquad\qquad
 \includegraphics[width=0.40\textwidth]{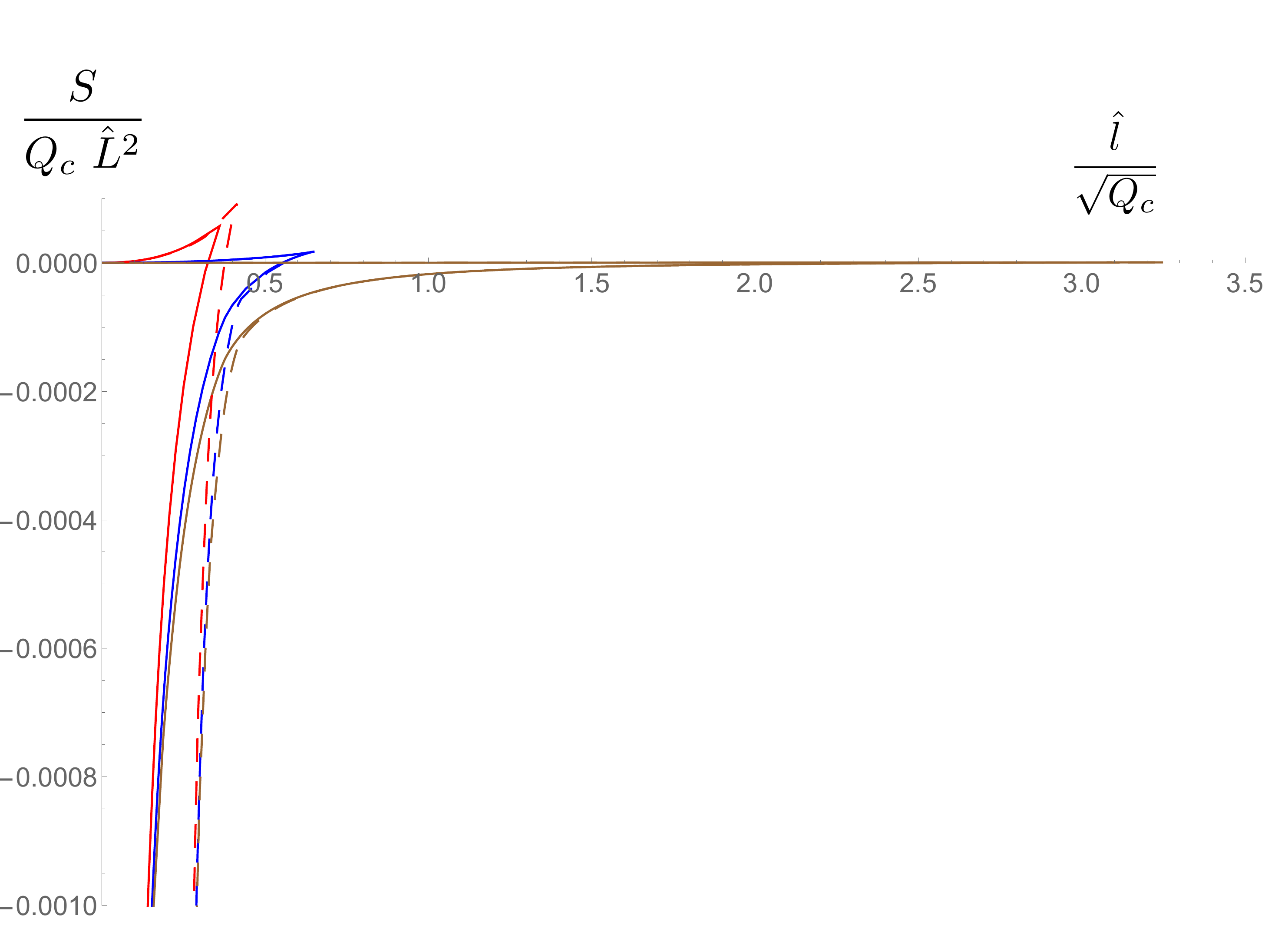}
  \caption{ On the left we depict the entropy of a transverse slab as a function of $\hat l_{\perp}$. 
  The dashed curve is the UV result (\ref{S_perp_UV}). On the right we compare ${S_{\parallel}\over \hat Q_c\,\hat L_2\,\hat L_3}$ and ${S_{\perp}\over \hat Q_c\,\hat L_1\,\hat L_2}$ as functions of their corresponding rescaled width. In both panels curves with the same color correspond to the same value of $x_q$: red for $x_q=0.1$, blue for $x_q=0.5$, and brown for  $x_q=0.9$.}
 \label{S_perp}
\end{figure}
After some calculations one can demonstrate that:
\bear
&&{S_{\perp}^{finite}\over \hat L_1\,\hat L_2}\,=\,-{Z_q^6\over 16\,\pi^3}\,\Bigg(
\int_{x_0}^{\infty}
\hat h\,(1+x-x_q)^5\Bigg[1-{1\over 
\sqrt{1-{(1+x_0-x_q)^{10}\over (1+x-x_q)^{10}}\,
{\hat h_0\,\hat W^2_0\,q^2\over \hat h\,\hat W^2\,q^2_0}}}\,\Bigg]\rc\rc
&&\qquad\qquad\qquad\qquad\qquad\qquad\qquad
+\int_{0}^{x_0}\hat h\,(1+x-x_q)^5\,dx\,\Bigg)\,\,.
\label{S_perp_finite}
\eear

In Fig.~\ref{S_perp} we plot $S_{\perp}$ as a function of $\hat l_{\perp}$ obtained by the numerical computation of the integrals in (\ref{S_perp_finite}) and (\ref{l_perp}). For small $\hat l_{\perp}$ the curves for different values of $x_q$ coincide and behave as $S_{\perp}\sim \hat l_{\perp}^{-6}$. This behavior is found analytically in the next subsection. For large $\hat l_{\perp}$ the dominant configuration is the disconnected one and  $S_{\perp}$ becomes positive. We have also compared in Fig.~\ref{S_perp} the parallel and transverse entanglement entropies. In the UV the difference is significant, but at larger distances the two entanglement entropies become very similar. This similarity is more and more pronounced as $x_q\to 1$.

\subsubsection{UV limit}

We now evaluate the entropy in the limit in which $x_0$ is very large and $\hat l_{\perp}$ is very small. Using the UV expansion at leading order of the different functions of the background, we get:
\beq
\hat l_{\perp}\approx{8\cdot 2^{{1\over 6}}\over 3\,\sqrt{15}}\,
{\sqrt{Q_c}\over Z_q^{{1\over 3}}}\,
\int_{x_0}^{\infty}{dx\over x^{{4\over 3}}\sqrt{(x/x_0)^{{14\over 3}}\,-\,1}}\,=\,
{8\cdot 2^{{1\over 6}}\over \sqrt{15}}\,\sqrt{\pi}\,
{\Gamma\Big({4\over 7}\Big)\over \Gamma\Big({1\over 14}\Big)}
{\sqrt{Q_c}\over Z_q^{{1\over 3}}}\,{1\over x_0^{{1\over 3}}}\,\,,
\eeq
whereas $S_{\perp}^{finite}$ is:
\beq
{S_{\perp}^{finite}\over \hat L_1\,\hat L_2}\approx
\,-{Z_q^2\,Q_c\over 60\,\pi^3}\,x_0^2
\Bigg[{1\over 2}\,-\,\int_{1}^{\infty}
dz\,z\,\Bigg({z^{{7\over 3}}\over \sqrt{z^{{14\over 3}}-1}}\,-\,1\Bigg)\Bigg]\,\,,
\eeq
which, after performing the integral becomes:
\beq
{S_{\perp}^{finite}\over \hat L_1\,\hat L_2}\approx
-{Z_q^2\,Q_c\over 120\,\pi^{{5\over 2}}}\,
{\Gamma\Big({4\over 7}\Big)\over \Gamma\Big({1\over 14}\Big)}\,x_0^2\,\,.
\eeq
Eliminating $x_0$ in favor of $\hat l_{\perp}$  we reproduce the result of \cite{Penin:2017lqt}:
\beq
{S_{\perp}^{finite}\over Q_c \,\hat L_1\,\hat L_2}\approx-\gamma_{\perp}\,\Bigg(
{\sqrt{Q_c}\over\hat l_{\perp}}\Bigg)^6\,\,,
\qquad\qquad
\gamma_{\perp}\,=\,\Big({16\over 15}\Big)^4\,\sqrt{\pi}
\Bigg({\Gamma\Big({4\over 7}\Big)\over \Gamma\Big({1\over 14}\Big)}\Bigg)^{7}\,\,.
\label{S_perp_UV}
\eeq

\subsection{Flow of mutual information}
The mutual information of two entangling regions $A_1$ and $A_2$ is a measure of the information shared by these two domains and is defined as:
\beq
I(A_1, A_2)\,=\,S(A_1)\,+\,S(A_2)\,-\,S(A_1\cup A_2)\,\,.
\eeq
We will analyze the evolution with the intrinsic scale of the background of $I(A_1, A_2)$ when $A_1$ and $A_2$ are two  slabs of equal length $l$ parallel to each other which are  separated by a distance $s$.  In holography there are two possible surfaces contributing to the entanglement entropy of two slabs \cite{Headrick:2010zt,Ben-Ami:2014gsa}. One of these  configurations, which has zero mutual information, dominates when the slab separation is large.  Below some critical separation the second configuration is dominant and $I(A_1, A_2)$ becomes positive. In reduced units, the critical separation $\hat s$ for two strips of length $\hat l$ is determined by the vanishing of the mutual information:
\beq
2\,S(\hat l)\,=\,S(2\hat l+\hat s)\,+\,S(\hat s)\,\,.
\label{critical_s}
\eeq
Let us analyze (\ref{critical_s}) for our background in the UV region, where  both $\hat l$ and $\hat s$ are small and the entanglement entropy has the scaling behavior written in (\ref{S_parallel_UV}) ((\ref{S_perp_UV})) for parallel (transverse) slabs. The resulting equation takes the form:
\beq
{1\over \big(2+{\hat s\over \hat l}\big)^a}\,+\,
{1\over \big({\hat s\over \hat l}\big)^a}\,=\,2\,\,,
\label{s_over_l}
\eeq
where $a={4\over 3}$ ($a=6$) for parallel (transverse) slabs. The solution of (\ref{s_over_l}) in these two cases is:
\beq
 \Bigg({\hat s\over \hat l}\Bigg)_{\parallel}\,\approx\,0.663\,\,,
 \qquad\qquad
  \Bigg({\hat s\over \hat l}\Bigg)_{\perp}\,\approx\,0.891\,\,.
  \label{s_l_UV}
 \eeq
 Let us  next study the critical point in the long distance IR region, where we expect to approach 
 the conformal isotropic behavior of $AdS_5\times {\mathbb S}^5$. In this last case the critical separation is also determined by (\ref{s_over_l}) with $a=2$ and, thus:
\beq
\Bigg({\hat s\over \hat l}\Bigg)_{AdS_5\times {\mathbb S}^5}\,=\,\sqrt{3}-1\approx 0.732 \ .
\label{s_l_AdS}
\eeq
Interestingly, viewed as an inverse it equals the metallic mean $\hat l/\hat s = \sigma_{p=1,q=1/2} = \frac{\sqrt 3+1}{2}$ \cite{Spinadel,Balasubramanian:2018qqx}.
Outside the UV region the critical $\hat s/\hat l$ depends on $\hat l$. We have numerically verified that this flow behaves as  expected. For parallel (transverse) slabs $\hat s/\hat l$  grows (decreases) from its UV value (\ref{s_l_UV}) as $\hat l$ is increased and it actually becomes very close to the conformal isotropic value (\ref{s_l_AdS}) for large $\hat l$ and $x_q$ close to one (see \cite{Balasubramanian:2018qqx} for another recent example of a holographic flow of mutual information). We note that for very large distances $\hat l$, the dominant phase is that for  flat embeddings and the full phase diagram resembles that of \cite{Ben-Ami:2014gsa}.

\section{Thermodynamics of a massless probe brane} \label{D5_probe}
 
In this section we test our background with a probe D5-brane, embedded as in the array (\ref{D3D5intersection}),  in which we switch on  a worldvolume gauge field $A=A_0\,dx^0$ dual to a chemical potential  $\mu$.   Our goal is to study the zero-temperature thermodynamics of the probe and its evolution as we move from the UV to the IR. For simplicity we will consider massless embeddings in which  the probe reaches the IR end  of the space. These type of embeddings have been analyzed in detail in App.~\ref{eom_probe}, where we check that the equations of motion of the probe are satisfied if the worldvolume gauge potential $A_0$ satisfies (\ref{first_integral_At}). In this section  we will work directly in the radial coordinate $\zeta$, for which the Lagrangian density takes the form:
\beq
\tilde {\cal L}\,=\,e^{g-f}\,{\cal L}\,=\,
-{\cal T}\,{\zeta^2\over \sqrt{W}}\,
\Big[\sqrt{1-W\,\Big({\partial A_0\over \partial \zeta}\Big)^{\,2}}\,-\,1\Big]\,\,,
\label{cal_L_At_zeta}
\eeq
where $ {\cal L}$ is the Lagrangian density (\ref{cal_L_At}). In (\ref{cal_L_At_zeta})  ${\cal T}$  is the constant  defined in  (\ref{cal_T})  and we  have written $\tilde {\cal L}$ in terms of the master function $W$. The chemical potential is just the value of $A_0$ at the UV boundary. In our variables:
\beq
\mu\,=\,d\,\int_b^{\infty}\,
{d\zeta\over \sqrt{W}}\,{1\over \sqrt{d^2+\zeta^4}}\,\,,
\label{chemical_pot_massivebck}
\eeq
where $d$ is the integration constant of (\ref{first_integral_At}) which, as we check below, is proportional to the charge density. 
The grand potential $\Omega$  is given by minus the on-shell action, which in our case is finite and there is no need of regularizing it. This is due to the cancelation of the divergences between the DBI and WZ terms.  Removing the Minkowski volume factor, we get:
\beq
\Omega\,=\,{\cal N}\,\int_b^{\infty}\,d\zeta\,{\zeta^2\over \sqrt{W}}\,
\Big[{\zeta^2\over \sqrt{d^2+\zeta^4}}\,-\,1\,\Big]\,\,,
\label{grand_pot_massivebck}
\eeq
where ${\cal N}={16\pi^2\over 3\sqrt{3}}$.  The charge density $\rho$ can be written as:
\beq
\rho\,=\,-{\partial \Omega\over \partial \mu}\,=\,-
{{\partial \Omega\over \partial d}\over {\partial \mu\over \partial d}}\,\,.
\eeq
From (\ref{chemical_pot_massivebck}) and (\ref{grand_pot_massivebck}) we can compute the derivatives with respect to $d$ that are needed to calculate $\rho$:
\bear
&&{\partial \mu\over \partial d}\,=\,
\int_b^{\infty}\,d\zeta\,{\zeta^4\over \sqrt{W}}\,
{1\over (d^2+\zeta^4)^{{3\over 2}}}\rc\rc
&&{\partial \Omega\over \partial d}\,=\,-{\cal N}\,d\,\int_b^{\infty}\,d\zeta\,{\zeta^4\over \sqrt{W}}\,
{1\over (d^2+\zeta^4)^{{3\over 2}}}\,\,.
\eear
Clearly, one has:
\beq
{\partial \Omega\over \partial d}\,=\,-{\cal N}\,d\,
{\partial \mu\over \partial d}\,\,,
\eeq
and we get that, indeed, $\rho$ is related to $d$ as expected:
\beq
\rho\,=\,{\cal N}\,d\,\,.
\eeq
The energy density $\epsilon$ is given by:
\beq
\epsilon\,=\,\Omega\,+\mu\rho\,\,.
\label{epsilon_def}
\eeq
Plugging the values of $\Omega$ and $\rho$ into (\ref{epsilon_def}), we get:
\beq
\epsilon\,=\,{\cal N}\,
\int_b^{\infty}\,
{d\zeta\over \sqrt{W}}\,\zeta^2\,\Bigg[\sqrt{1+{d^2\over \zeta^4}}
\,-\,1\Bigg]\,\,.
\eeq
Therefore:
\beq
{\partial \epsilon\over \partial d}\,=\,{\cal N}\,d\,\int_b^{\infty}\,{d\zeta\,\over \sqrt{W}}\,
{1\over (d^2+\zeta^4)^{{1\over 2}}}\,=\,{\cal N}\,\mu\,\,,
\eeq
as expected. Taking into account that $p=-\Omega$, the speed of sound $u_s$ can be obtained as:
\beq
u_s^2\,=\,-{{\partial \Omega\over \partial d}\over {\partial \epsilon\over \partial d}}\,=\,-
{1\over {\cal N}\,\mu}\,{\partial \Omega\over \partial d}\,\,.
\eeq
Thus, we get the following expression of $u_s^2$:
\beq
u_s^2\,=\,{d\over \mu}\,\int_b^{\infty}\,d\zeta\,{\zeta^4\over \sqrt{W}}\,
{1\over (d^2+\zeta^4)^{{3\over 2}}}\,\,.
\label{us_int_mu}
\eeq
More explicitly, plugging into (\ref{us_int_mu}) the expression (\ref{chemical_pot_massivebck}) of $\mu$, we obtain that
$u_s^2$ is given by the following ratio of two integrals over the holographic coordinate:
\beq
u_s^2\,=\,{
\int_b^{\infty}\,{d\zeta\over \sqrt{W}}\,
{\zeta^4\over (d^2+\zeta^4)^{{3\over 2}}}
\over \int_b^{\infty}\,
{d\zeta\over \sqrt{W}}\,{1\over \sqrt{d^2+\zeta^4}}}\,\,.
\label{us_zeta_variable}
\eeq
In the unflavored ($W=1$, $b=0$) and massless flavored ($W\propto \zeta^{-{2\over 3}}$, $b=0$) 
cases  we get the following  $d$-independent results:
\beq
u_s^2({\rm unflavored})\,=\,{1\over 2}\,\,,
\qquad\qquad
u_s^2({\rm massless\,\, flavored})\,=\,{2\over 3}\,\,.
\label{speeds_of_sound_analytic}
\eeq
The unflavored result $u_s^2=1/2$ is the one expected for a conformal worldvolume theory in $2+1$ dimensions. 
In general, one should get a value depending on $d$, which interpolates between these two values.  In order to facilitate the numerical calculations, let us rewrite these results in $x$ coordinate. Recall that $\zeta$ and $x$ are related as:
\beq
\zeta\,=\,{Q_f\over C^{{3\over 2}}}\,Z_q\,\big(x+1-x_q)\,\,.
\eeq
It turns out that $Q_f$ and $C$ can be scaled out from our formulas. First of all, we define  the rescaled density and chemical potential as:
\beq
\hat d\,\equiv\,{C^3\over Q_f^2}\,d\,\,,
\qquad\qquad
\hat \mu\,\equiv\,{C^2\over Q_f}\,\mu\,\,.
\eeq
Then, we have:
\beq
\hat\mu\,=\,\hat d\,Z_q\int_0^{\infty}\,
{dx\over \sqrt{\hat W}}\,
{1\over \big[\hat d^2\,+\,Z_q^4\,(x+1-x_q)^4\big]^{{1\over 2}}}\,\,,
\eeq
where $\hat W$ was defined in (\ref{hat_W_def}). 
Moreover, $\Omega$ and $\epsilon$ can be recast as:
\bear
\Omega & = & \hat{\cal N}\,Z_q^3\,\int_0^{\infty}\,{dx\over \sqrt{\hat W}}\,\,
(x+1-x_q)^2\,\Bigg[
{(x+1-x_q)^2 \, Z_q^2 \over \sqrt{\hat d^2\,+\,Z_q^4\,(x+1-x_q)^4}}\,-\,1\Bigg]\rc\rc\rc
\epsilon & = & \hat{\cal N}\,Z_q^2\int_0^{\infty}\,{dx\over \sqrt{\hat W}}\,
(x+1-x_q)^2\,\Bigg[\sqrt{1+{\hat d^2\over Z_q^4}\,{1\over (x+1-x_q)^4}}\,-\,1\Bigg]\,\,,
\eear
where $\hat{\cal N}$ is defined as:
\beq
\hat{\cal N}\,\equiv\,
{Q_f^3\over C^5}\,{\cal N}\,\,.
\eeq
The speed of sound can either be written as:
\beq
u_s^2\,=\,Z_q^5\,{\hat d\over \hat\mu}\,
\int_0^{\infty}\,{dx\over \sqrt{\hat W}}\,
{(x+1-x_q)^4\over \big[\hat d^2\,+\,Z_q^4\,(x+1-x_q)^4\big]^{{3\over 2}}}\,\,,
\eeq
or if we define the  integrals $I_1$ and $I_2$ as:
\bear
I_1 & = & \int_0^{\infty}\,{dx\over \sqrt{\hat W(x)}}\,{(x+1-x_q)^4\over \big[\hat d^2\,+\,Z_q^4\,(x+1-x_q)^4\big]^{{3\over 2}}}\rc\rc
I_2 & = & \int_0^{\infty}\,{dx\over \sqrt{\hat W(x)}}\,
{1\over \big[\hat d^2\,+\,Z_q^4\,(x+1-x_q)^4\big]^{{1\over 2}}}\,\,,
\label{I1_I_2_def}
\eear
then $u_s$ is obtained from the ratio between $I_1$ and $I_2$:
\beq
 u_s^2 \,=\,Z_q^4\,{I_1\over I_2}\,\,.
 \label{us_I1_I2}
 \eeq
Recall that we can relate the quark mass of the background $m_q$ to the constant $C$, namely 
$C\sim \sqrt{Q_f}/\sqrt{m_q}$. Using this relation we get that $\hat d\sim d/( \sqrt{Q_f}\,m_q^{{3\over 2}})$. Therefore, the UV limit $m_q\to 0$ corresponds to taking $\hat d\to\infty$. Accordingly, we should recover in this large $\hat d$ limit the massless flavored result of (\ref{speeds_of_sound_analytic}):
\beq
 u_s^2 (\hat d\to \infty)\to {2\over 3}\,\,,
\qquad\qquad ({\rm UV\,\,limit}). 
\label{naive_UV_us}
\eeq
The UV limit written above can also be analytically verified.  Indeed, let us introduce a new integration variable $y$, related to $x$ as:
\beq
 x\,=\,x_q\,-\,1\,+{\hat d^{{1\over 2}}\over Z_q}\,y\,\equiv\,x(y)\,\,.
 \eeq
The minimal value of $y$, corresponding to $x=0$, is $y=y_q$, where:
\beq
 y_q\,=\,{1-x_q\over \hat d^{{1\over 2}}}\,Z_q\,\,.
\eeq
It is now straightforward to write $I_1$ and $I_2$ as:
\bear
 I_1 & = & {1\over Z_q^5\,\hat d^{{1\over 2}}}\,
 \int_{y_q}^{\infty}\,{y^4\,dy\over \sqrt{\hat W(x(y))}\,(1+y^4)^{{3\over 2}}}\rc\rc
 I_2 & = & {1\over Z_q\,\hat d^{{1\over 2}}}\,
  \int_{y_q}^{\infty}\,{dy\over \sqrt{\hat W(x(y))}\,(1+y^4)^{{3\over 2}}}\,\,.
\eear
When $\hat d\to\infty$ the lower limit of the integrals becomes $y_q=0$. Moreover, the argument of the function $\hat W$ is large and one can use its UV asymptotic expression:
\beq
 \hat W(x(y))\,\approx\,{9\over 8}\,\Bigg[{\sqrt{2}\over Z_q}\Bigg]^{{2\over 3}}\,
 \Bigg[x_q-1+{\hat d^{{1\over 2}}\over Z_q}\,y\Bigg]^{-{2\over 3}}\approx
 {9\over 8}\,\Bigg[{\sqrt{2}\over \hat d^{{1\over 2}} }\Bigg]^{{2\over 3}}\,y^{-{2\over 3}}\,\,,
 \qquad\qquad
 (\hat d \to \infty)\,\,.
 \eeq
Then for large $\hat d$:
 \bear
 I_1 & \approx & {1\over Z_q^5\,\hat d^{{1\over 2}}}\,
\int_{0}^{\infty}\,dy\,{y^{{13\over 3}}\over (1+y^4)^{{3\over 2}}}\,=\,
{1\over Z_q^5\,\hat d^{{1\over 2}}}\,
{\Gamma\big({1\over 6}\big)\,\Gamma\big({1\over 3}\big)\over
4\,\sqrt{\pi}}\rc\rc
I_2 & \approx & {1\over Z_q\,\hat d^{{1\over 2}}}\,
\int_{0}^{\infty}\,dy\,{y^{{1\over 3}}\over (1+y^4)^{{3\over 2}}}\,=\,
{1\over Z_q\,\hat d^{{1\over 2}}}\,
{\Gamma\big({1\over 6}\big)\,\Gamma\big({1\over 3}\big)\over
6\,\sqrt{\pi}}\,\,,
\eear
and therefore we get the expected UV result:
\beq
u_s^2\,\approx {2\over 3}\,\,,
\qquad\qquad
(\hat d\to\infty)\,\,.
\label{UV_estimate_us}
\eeq

\begin{figure}[ht]
\center
 \includegraphics[width=0.50\textwidth]{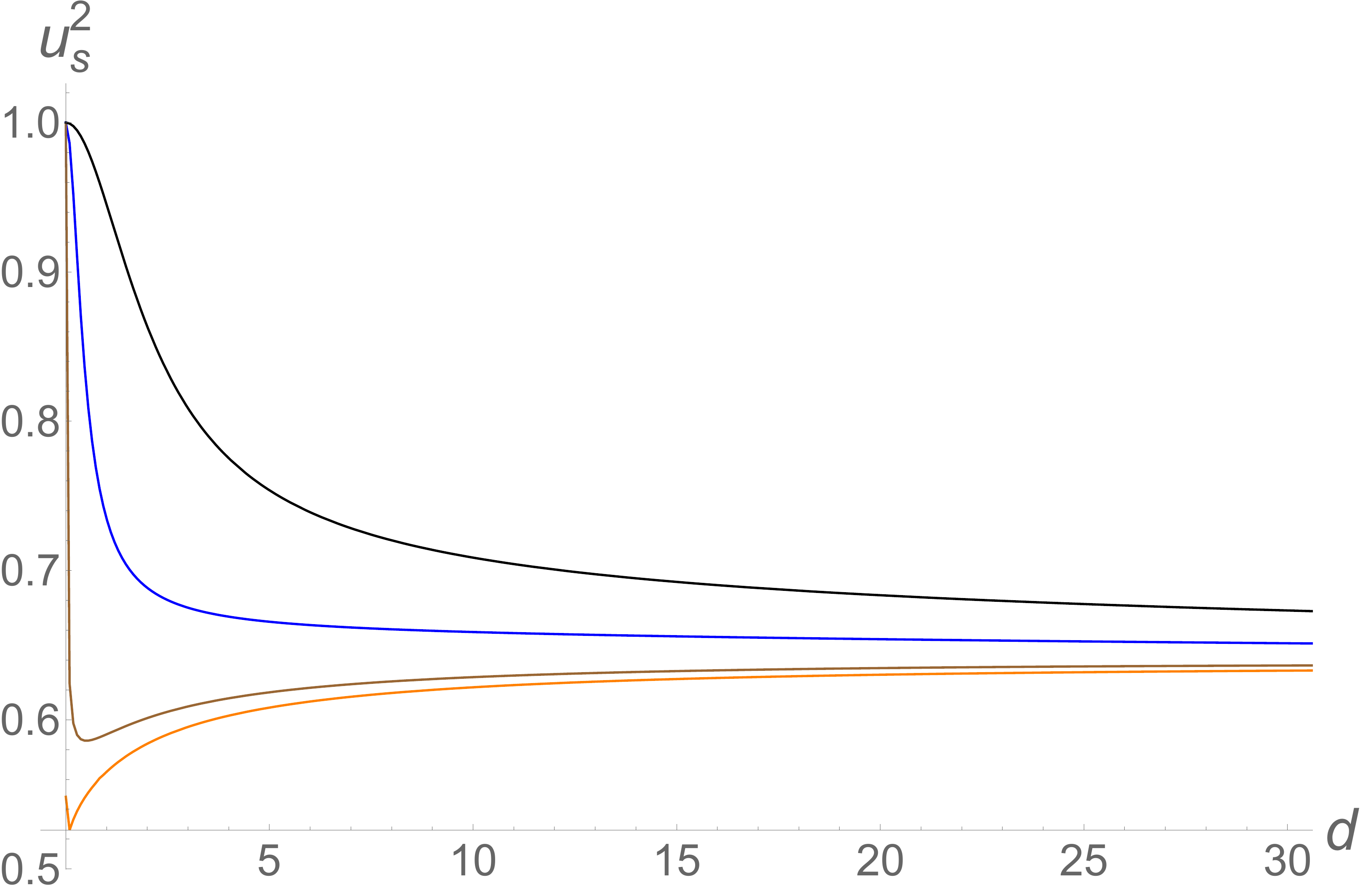}
  \caption{Speed of sound as a function of the density $\hat d$ for $x_q=0.005$ (black,top), $x_q=0.5$ (blue,second from top), $x_q=0.9$ (brown,third from top), and $x_q=0.995$ (orange,bottom).  }
\label{us_d}
\end{figure}

Let us now explore the small $\hat d$ limit. By taking $\hat d=0$ in (\ref{I1_I_2_def}) we see that the integrals $I_1$ and $I_2$ are proportional ($I_2(\hat d=0)=Z_q^4\,I_2(\hat d=0)$). 
Therefore, we get:
\beq
u_s^2 (\hat d= 0)=1\,\,.
\eeq

This result is not the one we naively expect since $\hat d\to 0$ corresponds to $m_q\to\infty$ and, as the quarks are not dynamical in this large mass limit, it would seem that we should  get the conformal result $u_s^2=1/2$ in this IR limit, at least when $x_q$ is close to one.  In order to clarify the situation, let us take $\hat d=0$ in the integrals (\ref{I1_I_2_def}) and examine the behavior of the integrand near $x=0$. 
For $x_q\not=1$ we get:
\beq
\int_{0}^{\infty} {dx\over \sqrt{\hat W}}\,{1\over (x+1-x_q)^2}\,\approx\,
\int_{0}^{\infty} dx\,\Bigg[{1\over \sqrt{6}}\,{1\over (1-x_q)^{{3\over 2}}}{1\over \sqrt{x}}+\,
{1\over  4\sqrt{6}}\,{1\over (1-x_q)^{{5\over 2}}}\,\sqrt{x}\,+\,
{\cal O}(x^{{3\over 2}})\Bigg]\,\,,
\eeq
where we have used (\ref{hat_W_inside}) to obtain the behavior of the integrand near $x=0$. This $x_q\not=1$ integral is convergent and the two integrals $I_1$ and $I_2$ are well defined. On the contrary, for $x_q=1$ we have $\hat W=1$ inside the cavity and the integrand is  divergent when $\hat d=0$ and thus we cannot take  directly $\hat d=0$ in the integrals. Actually, in the calculation of $u_s$, the limits $\hat d\to 0$ and $x_q\to 1$ do not commute. By taking $x_q\to 1$ first we indeed get the conformal result $u_s^2=1/2$ independently of the value of $\hat d$. 

The numerical values of $u_s^2$ obtained from (\ref{us_I1_I2})  as a function of $\hat d$ for different values of $\hat x_q$ have been plotted in Fig.~\ref{us_d}.  In this plot we notice that the UV asymptotic result (\ref{UV_estimate_us}) is satisfied for all values of $x_q$  and for $x_q\sim 1$ there is a minimum   at low $\hat d$,  in which $u_s^2$ approaches the conformal value $u_s^2=1/2$.  This is consistent with the behavior found in the analysis of other observables: the UV behavior is independent of $x_q$ and given by the scaling solution, whereas the IR is controlled by $x_q$. By taking $x_q$ close to one, the long distance behavior of our system becomes more isotropic.

\section{Summary and conclusions}\label{conclusions}

The goal of this paper has been the construction of a gravity dual to a system containing multiple ($2+1$)-dimensional layers in a ($3+1$)-dimensional ambient theory. 
In order to deal with this problem we adopted a top-down holographic approach and used brane engineering to generate the corresponding gravity dual.  We considered  the  setup  (\ref{D3D5intersection}),  in which D3- and D5-branes intersect along  $2+1$ dimensions and a codimension-one defect is created along the worldvolume of the D3-branes. Moreover, the D5-branes are distributed homogeneously along the gauge theory directions orthogonal to the defect, giving rise  in this way to a multilayer system. To find the corresponding supergravity solution, we regarded the D5-branes as flavor branes and used the techniques developed to find the backgrounds dual to unquenched smeared flavor. 

The background found is supersymmetric and solves the supergravity equations of motion with D5-brane sources. It is given in terms of the master function $W$, which can be obtained by solving the master equation (\ref{Master_W}). This master equation contains the profile function which can, in principle, be arbitrarily chosen and depends on the particular distribution of the  D5-brane charge  along the holographic coordinate. The solutions corresponding to a constant profile function $p$ were obtained in \cite{Conde:2016hbg}. In the flavor language, the solutions of \cite{Conde:2016hbg}  are dual to models with  massless flavors living on the defect (the corresponding black hole was constructed and analyzed in \cite{Penin:2017lqt}). The corresponding field  theories  display anisotropy in the third direction  which, by construction, is produced by the  multiple $(2+1)$-dimensional layers.

Here we generalized the scaling solution found in \cite{Conde:2016hbg} to the case in which the quarks are massive and there is a region in the bulk, which we called the cavity, in which the flavor sources vanish. The size $x_q$ of this cavity provides us with a parameter which determines the long distance behavior of the model. Indeed, as we tuned $x_q$ to its maximal value $x_q=1$, the IR behavior of several observables we analyzed approaches the one of the un-layered theory, whereas the short distance behavior is independent of $x_q$ and given by the scaling solution. Thus, as $x_q\to 1$ the theory in the IR becomes more isotropic and effectively retains its $(3+1)$-dimensional character.

We studied the running anisotropic behavior described above in several quantities, but it is clear that we have not exhausted the list of observables to analyze. Let us mention some  possible extensions of our work. In Sec.~\ref{D5_probe} we explored our background with probe D5-branes with a worldvolume gauge field dual to a chemical potential. For simplicity we considered embeddings of the probe with trivial embedding function $\chi$,  corresponding  to massless flavors. The more general case of massive embeddings can be readily obtained by considering a general function $\chi(r)$. It would be interesting to see how the quantum phase transition studied in \cite{Karch:2007br,Ammon:2012je,Itsios:2016ffv} between Minkowski and black hole embeddings is modified by the backreaction as has been seen in other $(2+1)$-dimensional systems \cite{Bea:2016fcj}.

It is clearly an interesting task to find the condensed matter system for which our holographic model could offer a framework for concrete calculations. Each layer effectively mimics graphene, when some of the D5-branes blow up in to D7'-branes \cite{Bergman:2010gm,Semenoff:2012xu,Omid:2012vy,Kristjansen:2012ny,Jokela:2012vn,Jokela:2010nu,Grau:2018keb}, so the full multilayer system could perhaps be understood as a (gapped) holographic graphite. More study is needed to make this identification precise. Nevertheless, some of the recent multilayered systems are known to have strong coupling dynamics \cite{Geisler}, so in an ideal scenario, we would love to engineer the holographic geometry to mimic the physics in these settings.
This is, however, a highly non-linear task which requires novel ideas.

Fortunately, we have a rather straightforward avenue ahead of us to find a killer app. In \cite{Hoyos:2016zke} it was demonstrated that the original D3-brane background (the dual to ${\cal N}=4$ SYM theory), supplemented with flavor D7-brane probes (introducing quenched quark matter) at finite chemical potential, provides a realistic equation of state for cold and dense quark matter. This realization paved the road for further investigations \cite{Hoyos:2016cob,Ecker:2017fyh,Annala:2017tqz,Jokela:2018ers}. 

How much then can holography help in understanding the deep cores of neutron stars where deconfined quark matter could reside? In the current context we would like to ask the following question. Black holes are known to forget about their past and are characterized by a handful of parameters (mass, rotation, charges). Neutron stars, on the other hand, seem to depend on a variety of parameters through their complicated internal composition. Surprisingly, however, certain (dimensionless) macroscopic properties (tidal deformability, quadrupole moment, moment of inertia) were found to obey universal relations to quite high degree of accuracy \cite{Yagi:2013bca} independent of the underlying equation of state.\footnote{If there is a strong first order phase transition close to the surface then the universal relations are known to be violated up to 20\% \cite{Annala:2017tqz,Sieniawska:2018zzj}.} A recent interesting paper \cite{Alexander:2018wxr} drew attention to the resemblance of the neutron star universal relations and the no-hair relations of black holes. In order to formally study the limit of strong gravity and the black hole formation, the equation of state has to be anisotropic due to Buchdahl bound \cite{Alexander:2018wxr}. We believe that the rather theoretical construction laid out in this paper (smeared D5-brane providing the necessary anisotropy) is a key step towards this direction and we hope to report progress on this front in the near future.

\vspace{2cm}
{\bf \large Acknowledgments}
We are grateful to Carlos Hoyos and Daniele Musso for discussions and comments on a draft version of this paper.
J.~M.~P. and A.~V.~R.   are funded by the Spanish grants FPA2014-52218-P and FPA2017-84436-P by Xunta de Galicia (GRC2013-024), by FEDER and by the Maria de Maeztu Unit of Excellence MDM-2016-0692.   J.~M.~P. is supported by the Spanish FPU fellowship FPU14/06300.

\appendix

\vskip 3cm
\renewcommand{\theequation}{\rm{A}.\arabic{equation}}
\setcounter{equation}{0}

\section{Details of the background}\label{Background_details}

The form of the ten-dimensional metric in Einstein frame has been written in (\ref{metric_ansatz_zeta}). In this appendix we give further details and an explicit coordinate representation. Let $\chi$ be an angular coordinate taking values in the range $0\le \chi\le \pi$ and let $\omega^{i}$ ($i=1,2,3$) be a set of left-invariant $SU(2)$ one-forms satisfying 
$d\omega^i\,=\,{1\over 2}\,\epsilon^{ijk}\omega^j\wedge \omega^k$. Then, we can write $ds^2_{10}$ as:
\bear
&&ds^2_{10}\,=\,h^{-{1\over 2}}\,\big[-(dx^0)^2+(dx^1)^2+(dx^2)^2\,+\,e^{-2\phi}\,(dx^3)^2\big]+
h^{{1\over 2}}\,\Big[dr^2\rc\rc
&&+{e^{2g}\over 4}\,\big(d\chi^2+\cos^2{\chi\over 2} ((\omega^1)^2+(\omega^2)^2)\,+\,
\cos^2{\chi\over 2} \sin^2{\chi\over 2}  (\omega^3)^2\big)+
e^{2f} \big(d\tau+{1\over 2}\,\cos^2{\chi\over 2} \omega^3\big)^2\Big]\,,\qquad\qquad
\label{10d_metric_explicit}
\eear
where the fiber $\tau$ takes values in the range $0\le \tau\le 2\pi$. Notice that we are using in (\ref{10d_metric_explicit}) a  radial variable $r$ which is different from the one in (\ref{metric_ansatz_zeta}) (see below for the relation between $r$ and $\zeta$). The one-form $A$ in (\ref{metric_ansatz_zeta}) is given by:
\beq
A\,=\,{1\over 2}\,\cos^2{\chi\over 2} \omega^3\,\,.
\eeq
The one-forms $\omega^i$ can be represented in terms of three angles $(\theta, \varphi, \psi)$ as:
\bear
&&\omega^1\,=\,\cos\psi\,d\theta\,+\,\sin\psi\,\sin\theta\,d\varphi\rc\rc
&&\omega^2\,=\,\sin\psi\,d\theta\,-\,\cos\psi\,\sin\theta\,d\varphi\rc\rc
&&\omega^3\,=\,d\psi\,+\,\cos\theta\,d\varphi\,\,,
\eear
where $0\le \theta\le \pi$, $0\le \varphi<2\pi$, and $0\le\psi<4\pi$. The coordinate representation  of the  canonical vielbein basis of  ${\mathbb C}\,{\mathbb P}^2$ is:
\bear
&&e^1\,=\,{1\over 2}\cos\big({\chi\over 2})\,\omega^1\,\,,
\qquad\qquad\qquad\,\,\,\,\,\,\,
e^2\,=\,{1\over 2}\cos\big({\chi\over 2})\,\omega^2\,\,,\rc\rc
&&e^3\,=\,{1\over 2}\cos\big({\chi\over 2})\,\sin\big({\chi\over 2})\,\omega^3\,\,,
\qquad\qquad
e^4\,=\,{1\over 2}\,d\chi\,\,.
\eear
By plugging these expressions into (\ref{hat_Omega_2}) and (\ref{F3_ansatz}) we get the value of the RR three-form $F_3$ written in (\ref{F3_ansatz}), where we already used the new radial variable $\zeta$ defined in 
(\ref{dzeta_dr}) below. 

Our solution preserves some amount of supersymmetry. Indeed, let us choose the  following vielbein basis for the 
ten-dimensional metric (\ref{10d_metric_explicit}):
\bear
&&E^{x^{\mu}}\,=\,h^{-{1\over 4}}\,dx^{\mu}\,\,,\qquad\qquad (\mu=0, 1,2),
\qquad\qquad
E^{x^{3}}\,=\,h^{-{1\over 4}}\,e^{m}\,dx^{3}\,\,,\rc\rc
&&E^{r}\,=\,h^{{1\over 4}}\,dr\,\,,
\qquad\qquad E^{i}\,=\,{1\over 2}\,h^{{1\over 4}}\,e^{g}\,\cos{\chi\over 2}\,\,\omega^{i}\,\,,
\qquad \qquad (i=1,2)\,\,,\rc\rc
&&E^3\,=\,{1\over 2}\,h^{{1\over 4}}\,e^{g}\,\cos{\chi\over 2}\,\sin{\chi\over 2}\,\,\omega^{3}\,\,,
\qquad\qquad
E^4\,=\,{1\over 2}\,h^{{1\over 4}}\,e^{g}\,d\chi\,\,,\rc\rc
&&E^5\,=\,h^{{1\over 4}}\,e^{f}\,\big(d\tau+{1\over 2}\cos^2{\chi\over 2}\,\, \omega^3\big)\,\,.
\label{10d_vielbein}
\eear
Then,  in this basis of one-forms, the Killing spinor of the background can be written as:
\beq
\epsilon\,=\,h^{-{1\over 8}}\,e^{{3\over 2}\,\Gamma_{12}\,\tau}\,\,\eta\,\,,
\label{spinor_sol}
\eeq
where $\eta$ is a constant spinor satisfying some projection conditions. If we represent the Killing spinors of type IIB supergravity as a Majorana-Weyl doublet, then $\eta$ satisfies the projections:
\bear
&&\Gamma_{rx^314}\,\sigma_1\,\eta\,=\,\eta\,\,,\rc\rc
&&\Gamma_{12}\,\eta\,=\,\Gamma_{34}\,\eta\,=\,\Gamma_{r5}\,\eta\,=\,i\sigma_2\,\eta\,\,,
\label{eta_projections}
\eear
which means that our background preserves two supercharges. Actually, the different functions of our ansatz must satisfy the following system of first-order BPS equations \cite{Conde:2016hbg}:
\bear
&&\phi'\,=\,Q_f\,p(r)\,e^{{3\,\phi\over 2}\,-2g}\rc\rc
&&g'\,=\,e^{f-2g}\rc\rc
&&f'\,=\,3\,e^{-f}\,-\,2e^{f-2g}\,+\,{Q_f\,p(r)\,\over 2}\,e^{{3\,\phi\over 2}-2g}\rc\rc
&& h'\,=\,-Q_c\,e^{-4g-f}\,-\,Q_f\,p(r)\,e^{{3\,\phi\over 2}-2g}\,h\,\,,
\label{BPS_system}
\eear
where $p(r)$ is the profile function of (\ref{F3_ansatz}),  the prime denotes derivative with respect to the radial coordinate $r$ and $Q_f$ and $Q_c$ are related to the number of flavors $N_f$ and colors $N_c$ as:
\beq
Q_f\,=\,{4\pi\,g_s\,\alpha' N_f\over 9\sqrt{3}\,L_3}\,\,,
\qquad\qquad
Q_c\,=\,16\,\pi g_s \alpha'{}^{\,2}\,N_c\,\,,
\eeq
where $L_3\,=\,\int dx^3$. Let us now write the BPS system in terms of a new radial variable $\zeta$, related to $r$ as:
\beq
{d\zeta\over dr}\,=\,e^{f-g}\,\,.
\label{dzeta_dr}
\eeq
Then, the equations for $\phi$, $g$, and $f$ take the form:
\bear
&&{d\phi\over d\zeta}\,=\,Q_f\,p(\zeta)\,e^{{3\,\phi\over 2}\,-\,f\,-g}\rc\rc
&&{d g\over d\zeta}\,=\,e^{-g}\rc\rc
&&{d f\over d\zeta}\,=\,3\,e^{g-2f}\,-\,2\,e^{-g}\,+\,{Q_f\,p(\zeta)\over 2}\,e^{{3\,\phi\over 2}\,-\,f\,-g}\,\,.
\label{BPS_system_zeta}
\eear
In this new variable $\zeta$, the BPS equation for $g$  in (\ref{BPS_system_zeta}) can be immediately integrated:
\beq
e^{g}=\zeta\,\,,
\label{zeta_g}
\eeq
and we can rewrite the remaining BPS equations as:
\bear
&&{d\phi\over d\zeta}\,=\,{Q_f\,p(\zeta)\over \zeta}\,e^{{3\,\phi\over 2}\,-\,f}\rc\rc
&&{d f\over d\zeta}\,=\,3\,\zeta\,e^{-2f}\,-\,{2\over \zeta}\,+\,{Q_f\,p(\zeta)\over 2\,\zeta}\,
e^{{3\,\phi\over 2}\,-\,f}\,\,.
\label{BPS_system_zeta_no_g}
\eear
Notice that our ansatz for the metric in the $\zeta$ variable is precisely the one written in (\ref{metric_ansatz_zeta}). 
We now introduce the new master function $W$ as:
\beq
W\,\equiv\,{e^{2f-\phi}\over \zeta^2}\,\,.
\eeq
One can  easily show that the BPS system (\ref{BPS_system_zeta_no_g})  implies that $W$ satisfies the following first-order differential equation:
\beq
{dW\over d \zeta}\,=\,{6\over \zeta}\,\big(e^{-\phi}-W\big)\,\,.
\eeq
Moreover, one can demonstrate that the equation for $\phi$ in (\ref{BPS_system_zeta_no_g})  can be written as:
\beq
{d\over d\zeta}\,e^{-\phi}\,=\,-{Q_f\,p(\zeta)\over \zeta^2\,\sqrt{W}}\,\,. 
\eeq
One can combine these last two equations in the following second-order master equation for $W$:
\beq
{d\over d\zeta}\Big(\zeta\,{d W\over d\zeta}\Big)\,+\,6\,{d W\over d\zeta}\,=\,
-{6\,Q_f\,p(\zeta)\over \zeta^2\,\sqrt{W}}\,\,,
\label{Master_W_app}
\eeq
which coincides with (\ref{Master_W}). It is also straightforward to verify that  the function $f$ and the dilaton  $\phi$ are given in terms of $W$ as in (\ref{g_f_W}) and (\ref{dilaton_W}), respectively.

As mentioned in Subsec.~\ref{massive_flavors} the master equation can be solved in powers of $\zeta_q/\zeta$ for large $\zeta$. The result for $W$ was written in (\ref{W_UV_zeta}). The  function $f$ is readily computed using (\ref{g_f_W}):
\beq
e^{f}= {3\over 2\sqrt{2}}\,\zeta\,\Big[1\,-\,{1\over 16}\,{\zeta_q^4\over \zeta^4}\,+\,{3\over 32}\,{\zeta_q^6\over \zeta^6}\,+\,
{7\over 1024}\,{\zeta_q^8\over \zeta^8}\,+\,{\cal O}\Big( {\zeta_q^{10}\over \zeta^{10}}\Big)\,\Big]\,\,.
\label{f_UV_zeta}
\eeq
The dilaton in this expansion is:
\beq
e^{\phi}= \Big({\zeta\over \sqrt{2}\,Q_f}\Big)^{{2\over 3}}\,
\Big[1\,+\,{1\over 24}{\zeta_q^4\over \zeta^4}\,+\,{1\over 48}\,{\zeta_q^6\over \zeta^6}\,+\,{43\over 4608}\,
{\zeta_q^8\over \zeta^8}\,+\,{\cal O}\Big( {\zeta_q^{10}\over \zeta^{10}}\Big)\,\Big]\,\,,
\label{dil_UV_zeta}
\eeq
and the warp factor becomes:
\bear
&&h\approx{Q_c\over \zeta^4}\,\Big[{4\over 15}\,+\,{1\over 110}\,{\zeta_q^4\over  \zeta^4}\,-\,{3\over 140}\,{\zeta_q^6\over \zeta^{6}}\,-\,
{45\over 23936}\,{\zeta_q^8\over \zeta^{8}}\,\Big]\,+\,\rc\rc
&&\qquad\qquad\qquad\qquad
+{C_1\over\zeta^{{2\over 3}}} \Big[1\,-\,{1\over 24}\,{\zeta_q^4\over \zeta^{4}}\,-\,{1\over 48}\,
{\zeta_q^{6}\over\zeta^{6}}\,-\,{35\over 4608}\,{\zeta_q^8\over \zeta^{8}}\Big]\,\,,
\label{h_UV_zeta}
\eear
where $C_1$ is a constant of integration. The squashing factor $q=e^f/\zeta$  for this solution  can be obtained from (\ref{q_W}) and takes the form:
\beq
q= {3\over 2\sqrt{2}}\,\Big[1\,-\,{1\over 16}\,{\zeta_q^4\over \zeta^4}\,+\,{3\over 32}\,{\zeta_q^6\over \zeta^6}\,+\,
{7\over 1024}\,{\zeta_q^8\over \zeta^8}\Big]\,+\,{\cal O}\Big( {\zeta_q^{10}\over \zeta^{10}}\Big)\,\,.
\label{q_UV_zeta}
\eeq

We can also solve the master equation \eqref{Master_W} perturbatively close to the cavity for $\zeta\ge \zeta_q$. In order to do that, let us substitute in the master equation an expansion of the form:
\begin{equation} \label{IRapprox}
W^{approx}_{IR} = C_0 + C_2 \left(\frac{\zeta}{\zeta_q}- 1\right) + \sum\limits_{i=4}   
\beta_i \left(\frac{\zeta}{\zeta_q} - 1\right)^{i \over 2}  \, . 
\end{equation}

Identifying the above approximate solution \eqref{IRapprox} and the analytic solution \eqref{W_inside} at the 
edge of the cavity $\zeta=\zeta_q$ we get the following two constraints
\begin{equation} \label{constraints}
C_0 \, = \, C\left(1- \frac{b^6}{\zeta_q^6} \right)
\quad , \quad
C_2 \, = \, 6 \, C \, \frac{b^6}{\zeta_q^6} \, . 
\end{equation}

The first non-vanishing coefficients are
\begin{eqnarray}
&& \beta_4 = - {7 \over 2} \, C_2 \,\,,
\qquad
\beta_5 = - {12 \, \sqrt{2} \over 5 \, \sqrt{C_0}} \, \frac{Q_f}{\zeta_q} \,\,,
\qquad
\beta_6 = {28 \over 3} \, C_2\,\,,  
\nonumber \\\\
&&
\beta_7 = \frac{3 \, \sqrt{2} \left(109 \, C_0 \,+ \, 6 \, C_2 \right)}{35 \, C_0^{3/2}} \, \frac{Q_f}{\zeta_q} \,\,,
\qquad\qquad
\beta_8 = -\, 21 \, C_2 \,\,.
\end{eqnarray}
The function $f$ corresponding to (\ref{IRapprox}) is:
\begin{eqnarray}
\frac{e^{f}}{\zeta_q \, \left(1+ \frac{1}{6} \,\frac{C_2}{C_0}\right)^{-{1 \over 2}}} & = & 
1+ \left(1+ \frac{1}{2} \,\frac{C_2}{C_0}\right) \, \left(\frac{\zeta}{\zeta_q} - 1\right) 
 + \frac{1}{ \sqrt{2} \, C_0^{3 \over 2}} \left(1+ \frac{1}{6} \,\frac{C_2}{C_0}\right)^{-1} 
\frac{Q_f}{\zeta_q} \, \left(\frac{\zeta}{\zeta_q} - 1\right)^{3 \over 2} 
\nonumber  \\
&&
- \, \frac{5}{4} \,\frac{C_2}{C_0}\, \left(1+ \frac{1}{10} \,\frac{C_2}{C_0}\right) \, 
\left(\frac{\zeta}{\zeta_q} - 1\right)^{2} +\ldots\,\,,
\end{eqnarray}
and the dilaton can be expanded as:
\begin{eqnarray}
e^{\phi} & = & C_0^{-1} \left(1+ \frac{1}{6} \,\frac{C_2}{C_0}\right)^{-1}
 + \, \frac{\sqrt{2}}{C_0^{5 \over 2}} \left(1+ \frac{1}{6} \,\frac{C_2}{C_0}\right)^{-2} 
\frac{Q_f}{\zeta_q} \, \left(\frac{\zeta}{\zeta_q} - 1\right)^{3 \over 2}
- \, \frac{41 \,C_0\, + \, 6 \, C_2}{10 \, \sqrt{2} \, C_0^{7 \over 2}} 
\nonumber  \\
&&
\times  \left(1+ \frac{1}{6} \,\frac{C_2}{C_0}\right)^{-2} 
\frac{Q_f}{\zeta_q} \, \left(\frac{\zeta}{\zeta_q} - 1\right)^{5 \over 2} 
+ \, \frac{2}{C_0^{4}} \, \left(1+ \frac{1}{6} \,\frac{C_2}{C_0}\right)^{-3} \, 
\frac{Q_f^2}{\zeta_q^2} \, \left(\frac{\zeta}{\zeta_q} - 1\right)^{3} +\ldots \ .
\qquad\qquad
\end{eqnarray}
Finally, plugging these expansions into (\ref{warp_factor_eq}) we can solve for the warp factor $h$, which is given by
\begin{eqnarray}
h & = & C_1 - \left(1+ \frac{1}{6} \,\frac{C_2}{C_0}\right) \, \frac{Q_c}{\zeta_q^4}\, 
\left(\frac{\zeta}{\zeta_q} - 1\right)
 - \frac{\sqrt{2}\, \, C_1}{C_0^{3 \over 2}} \left(1+ \frac{1}{6} \,\frac{C_2}{C_0}\right)^{-1}\,  
\frac{Q_f}{\zeta_q} \, \left(\frac{\zeta}{\zeta_q} - 1\right)^{3 \over 2} 
\nonumber  \\
&&
+ \, \frac{5}{2} \,  
\left(1+ \frac{1}{5} \,\frac{C_2}{C_0}\right) \, 
\left(1+ \frac{1}{6} \,\frac{C_2}{C_0}\right) \, 
\frac{Q_c}{\zeta_q^4} \, \left(\frac{\zeta}{\zeta_q} - 1\right)^{2}+ \ldots \ .
\end{eqnarray}
The integration constant $C_1$ is connected with the integration constant coming from solving 
the warp factor differential equation inside the cavity.

\vskip 3cm
\renewcommand{\theequation}{\rm{B}.\arabic{equation}}
\setcounter{equation}{0}

\section{Probe analysis and profile function}
\label{profile_details}

In this appendix we study the embeddings of a D5-brane probe which preserves the supersymmetry of a background  given by our ansatz. Once these embeddings are characterized we will be able to obtain the corresponding  profile function $p(\zeta)$ which, in turn, is a necessary ingredient to solve the BPS equations and determine completely the different functions of the ansatz.

The supersymmetric embeddings we are looking for must satisfy the kappa symmetry condition:
\beq
\Gamma_{\kappa}\,\epsilon\,=\,\pm\epsilon\,\,,
\label{kappa_epsilon}
\eeq
where $\epsilon$ is a Killing spinor of the background. For a D5-brane without any worldvolume gauge field, $\Gamma_{\kappa}$ is given by:
\beq
\Gamma_{\kappa}\,=\,{1\over 6!\sqrt{-g_6}}\,\,
\epsilon^{\alpha_1\,\cdots\,\alpha_6}\,\sigma_1\,\gamma_{\alpha_1\cdots\alpha_6}\,\,,
\eeq
where $g_6$ is the determinant of the induced worldvolume metric,  $\gamma_{\alpha_1\cdots\alpha_6}$ is the antisymmetrized product of induced Dirac matrices and we are 
representing the Killing spinors of type IIB supergravity as a Majorana-Weyl doublet.
We will  take $\xi^{\alpha}\,=\,(x^0, x^1, x^2, r, \theta, \psi)$ as our set of worldvolume coordinates. In this case the kappa symmetry matrix takes the form:
\beq
\Gamma_{\kappa}\,=\,{1\over \sqrt{-g_6}}\,\,
\sigma_1\,\gamma_{x^0\, x^1\, x^2\, r\,\theta\,\psi}\,\,.
\eeq
Recall that the Killing spinor of the background can be written as in (\ref{spinor_sol}) in terms of a constant spinor 
$\eta$ which satisfies the projections (\ref{eta_projections}).   We can cast the kappa symmetry condition (\ref{kappa_epsilon}) as the following condition on $\eta$:
\beq
\tilde\Gamma_{\kappa}\,\eta\,=\,\pm\eta\,\,,
\label{kappa_eta}
\eeq
where $\tilde\Gamma_{\kappa}$ is defined as the matrix:
\beq
\tilde\Gamma_{\kappa}\,\equiv\,e^{-{3\over 2}\,\Gamma_{12}\,\tau}\,
\Gamma_{\kappa}\,e^{{3\over 2}\,\Gamma_{12}\,\tau}\,\,.
\label{tilde_Gamma_kappa}
\eeq
We will consider an embedding in which $x^3$ and $\varphi$ are constant, while
\beq
\chi=\chi(r)\,\,,
\qquad\qquad
\tau=\tau (\psi)\,\,.
\label{embedding_ansatz}
\eeq
The induced $\gamma$-matrices on the worldvolume for this embedding ansatz  are:
\bear
&&\gamma_{x^\mu}\,=\,h^{-{1\over 4}}\,\Gamma_{x^{\mu}}\,\,,
\qquad\qquad (\mu=0,1,2)\rc\rc
&&\gamma_{r}\,=\,h^{{1\over 4}}\,\Gamma_{r}\,+\,{1\over 2}\,
h^{{1\over 4}}\,e^{g}\,\chi'\,\Gamma_{4}\rc\rc
&&\gamma_{\theta}\,=\,{1\over 2}\,h^{{1\over 4}}\,e^{g}\,\cos{\chi\over 2}\,\cos\psi\,
\Gamma_{1}\,+\,{1\over 2}\,h^{{1\over 4}}\,e^{g}\cos{\chi\over 2}\,\sin\psi\,\Gamma_{2}\rc\rc
&&\gamma_{\psi}\,=\,{1\over 2}\,h^{{1\over 4}}\,e^{g}\,\cos{\chi\over 2}\,\sin{\chi\over 2}\,
\Gamma_{3}\,+\,h^{{1\over 4}}\,e^{f}\,
\Big({1\over 2}\,\cos^2{\chi\over 2}\,+\,\dot\tau\Big)\,\Gamma_{5}\,\,,
\eear
where we have denoted 
\beq
\chi'\,=\,{d\chi\over dr}\,\,,
\qquad\qquad
\dot\tau\,=\,{d\tau\over d\psi}\,\,,
\eeq
and the $\Gamma$'s are constant ten-dimensional Dirac matrices. 
From these induced matrices we can compute the antisymmetrized product $\gamma_{r\theta\psi}$ and get an expression of the type:
 \beq
 \gamma_{r\theta\psi}={h^{{3\over 4}}e^{g}\over 2}\cos{\chi\over 2}\Big[
 c_1\,\Gamma_{r13}+ c_2\,\Gamma_{r23}+c_3\,\Gamma_{413}+c_4\,\Gamma_{423}+
 c_5\,\Gamma_{r15}+c_6\,\Gamma_{r25}+c_7\Gamma_{415}+c_8\Gamma_{425}\Big]\,\,,
\eeq
where the different coefficients are given by
\bear
&&c_1\,=\,{e^g\over 2}\cos{\chi\over 2}\,\sin{\chi\over 2}\,\cos\psi\,\,,
\qquad\qquad
c_2\,=\,{e^g\over 2}\cos{\chi\over 2}\,\sin{\chi\over 2}\,\sin\psi\,,\rc\rc
&&c_3\,=\,{e^{2g}\over 4}\cos{\chi\over 2}\,\sin{\chi\over 2}\,\cos\psi\,\chi'\,\,,
\qquad\qquad
c_4\,=\,{e^{2g}\over 4}\cos{\chi\over 2}\,\sin{\chi\over 2}\,\sin\psi\,\chi'\,\,,\rc\rc
&&c_5\,=\,e^f\,\Big({1\over 2}\,\cos^2{\chi\over 2}\,+\,\dot\tau\Big)\cos\psi\,\,,
\qquad\qquad
c_6\,=\,e^f\,\Big({1\over 2}\,\cos^2{\chi\over 2}\,+\,\dot\tau\Big)\sin\psi\,\,,\rc\rc
&&c_7\,=\,{e^{f+g}\over 2}\Big({1\over 2}\,\cos^2{\chi\over 2}\,+\,\dot\tau\Big)\cos\psi\,\chi'\,\,,
\qquad\qquad
c_8\,=\,{e^{f+g}\over 2}\Big({1\over 2}\,\cos^2{\chi\over 2}\,+\,\dot\tau\Big)\sin\psi\,\chi'\,\,.
\qquad\qquad
\eear
This expression can be rewritten as:
\beq
 \gamma_{r\theta\psi}={h^{{3\over 4}}e^{g}\over 2}\cos{\chi\over 2}
 \,e^{-\psi\Gamma_{12}}\,
 \Big[ d_1\,\Gamma_{r13}+d_2\,\Gamma_{413}+ d_3\,\Gamma_{r15}+d_4\Gamma_{415}\Big]\,\,,
 \eeq
where the coefficients $d_i$ are:
\bear
&&d_1\,=\,{e^g\over 2}\cos{\chi\over 2}\,\sin{\chi\over 2}\,\,,
\qquad\qquad\qquad
d_2\,=\,{e^{2g}\over 4}\cos{\chi\over 2}\,\sin{\chi\over 2}\,\chi'\,\,,\rc\rc
&&d_3\,=\,e^f\,\Big({1\over 2}\,\cos^2{\chi\over 2}\,+\,\dot\tau\Big)\,\,,
\qquad\qquad
d_4\,=\,{e^{f+g}\over 2}\Big({1\over 2}\,\cos^2{\chi\over 2}\,+\,\dot\tau\Big)\,\chi'\,\,.
\label{d_coefficients}
\eear
In order to compute the  form of $\Gamma_{\kappa}$, we use that
\beq
\gamma_{x^0\,x^1\,x^2}\,=\,h^{-{3\over 4}}\,\Gamma_{x^0\,x^1\,x^2}\,\,.
\eeq
We find
\beq
\Gamma_{\kappa}\,=\,{1\over 2}\,{e^{g}\,\cos{\chi\over 2} \over \sqrt{-g_6}}\,
 \,e^{-\psi\Gamma_{12}}\,\sigma_1\,\Gamma_{x^0\,x^1\,x^2}\,\Big[
 d_1\,\Gamma_{r13}\,+\,d_2\,\Gamma_{413}\,+\,d_3\,\Gamma_{r15}\,+\,
 d_4\,\Gamma_{415}\Big]
\eeq
Let us now calculate the matrix $\tilde \Gamma_{\kappa}$.  As
\beq
\{\Gamma_{12}, \Gamma_{r13}\}\,=\,
\{\Gamma_{12}, \Gamma_{413}\}\,=\,
\{\Gamma_{12}, \Gamma_{r15}\}\,=\,
\{\Gamma_{12}, \Gamma_{415}\}\,=\,0\,\,,
\eeq
we have that
\beq
\Gamma_{\kappa}\,e^{{3\over 2}\,\Gamma_{12}\,\tau}\,=\,
e^{-{3\over 2}\,\Gamma_{12}\,\tau}\,\Gamma_{\kappa}
\eeq
and, therefore:
\beq
\tilde\Gamma_{\kappa}\,=\,{1\over 2}\,{e^{g}\,\cos{\chi\over 2} \over \sqrt{-g_6}}\,
 \,e^{-(\psi+3\tau)\Gamma_{12}}\,\sigma_1\,\Gamma_{x^0\,x^1\,x^2}\,\Big[
 d_1\,\Gamma_{r13}\,+\,d_2\,\Gamma_{413}\,+\,d_3\,\Gamma_{r15}\,+\,
 d_4\,\Gamma_{415}\Big] \ .
 \label{Gamma_kappa_eta_1}
\eeq
We now study how the different terms of $\tilde\Gamma_{\kappa}$ act on $\eta$.  We begin by considering the action of the first term on the right-hand-side of (\ref{Gamma_kappa_eta_1}).  First of all, we notice that, using the ten-dimensional chirality condition satisfied by $\epsilon$ and the projections (\ref{eta_projections}), one can show that $\eta$ satisfies\cite{Conde:2016hbg}:
\beq
\Gamma_{x^0\,x^1\,x^2}\,\eta\,=\,i\sigma_2\,\Gamma_{x^3}\,\eta\,\,.
\label{Gamma_012_eta}
\eeq
Using (\ref{Gamma_012_eta}) and the first projection in (\ref{eta_projections}), we get:
\beq
\sigma_1\,\Gamma_{x^0\,x^1\,x^2}\,\Gamma_{r13}\,\eta\,=\,
\sigma_1\,\Gamma_{43}\,\Gamma_{14rx^3}\,\eta\,=\,i\sigma_2\,\Gamma_{34}\,\eta\,\,.
\eeq
Finally, the last projection in  (\ref{eta_projections}) allows us to write:
\beq
\sigma_1\,\Gamma_{x^0\,x^1\,x^2}\,\Gamma_{r13}\,\eta\,=\,-\eta\,\,.
\eeq
Let us next consider the second term in  $\tilde\Gamma_{\kappa}$. We  first use:
\beq
\Gamma_{413}\,=\,-\Gamma_{r4}\,\Gamma_{r13}\,\,
\eeq
from which it immediately follows that:
\beq
\sigma_1\,\Gamma_{x^0\,x^1\,x^2}\,\Gamma_{413}\,\eta\,=\,\Gamma_{r4}\,\eta\,\,.
\eeq
Similarly, since $\Gamma_{r15}=-\Gamma_{35}\,\Gamma_{r13}$, we get that the third term of 
(\ref{Gamma_kappa_eta_1}) acts on $\eta$ as:
\beq
\sigma_1\,\Gamma_{x^0\,x^1\,x^2}\,\Gamma_{r15}\,\eta\,=\,\Gamma_{35}\,\eta\,\,.
\eeq
But, according to the last set of projections in (\ref{eta_projections}), we have $\Gamma_{35}\eta\,=\,-\Gamma_{r4}\eta$. Therefore:
\beq
\sigma_1\,\Gamma_{x^0\,x^1\,x^2}\,\Gamma_{r15}\,\eta\,=\,-\Gamma_{r4}\,\eta\,\,.
\eeq
Finally, as $\Gamma_{415}=-\Gamma_{r4}\,\Gamma_{r15}$, we get:
\beq
\sigma_1\,\Gamma_{x^0\,x^1\,x^2}\,\Gamma_{415}\,\eta\,=\,-\eta\,\,.
\eeq
Thus, collecting all these results, we have:
\beq
\tilde\Gamma_{\kappa}\,\eta\,=\,{1\over 2}\,{e^{g}\,\cos{\chi\over 2} \over \sqrt{-g_6}}\,
 \,e^{-(\psi+3\tau)\Gamma_{12}}
 \Big[-d_1-d_4\,+\,(d_2-d_3)\,\Gamma_{r4}\,\Big]\,\eta\,\,.
 \label{Gamma_eta}
 \eeq
As $\tilde\Gamma_{\kappa}$  should act as (plus or minus) the identity matrix on $\eta$, we must have:
\beq
\sin(\psi+3\tau)\,=\,0\,\,,
\qquad\qquad
d_3\,=\,d_2\,\,.
\label{BPS_first}
\eeq
The first equation fixes that 
\beq
\psi+3\tau\,=\,n\,\pi\,\,,
\label{tau_psi_n}
\eeq
with $n\in {\mathbb Z}$. This  implies that 
\beq
\dot\tau\,=\,-{1\over 3}\,\,.
\label{dot_tau}
\eeq
In order to have $\tau$ having values within the range $[0,2\pi]$, we choose $n=4$ in (\ref{tau_psi_n}). Therefore,  the function $\tau=\tau(\psi)$ is:
\beq
\tau(\psi)\,=\,{4\pi-\psi\over 3}\,\,,
\label{tau_psi}
\eeq
and the values of $\tau$ for the embedding range from $\tau=0$ to $\tau={4\pi\over 3}$.  Moreover, the second equation in (\ref{BPS_first}) implies the following  first-order equation for $\chi$:
\beq
\chi'\,=\,2\,e^{f-2g}\,{\cos^2{\chi\over 2}\,+\,2\dot\tau\over 
\sin{\chi\over 2}\,\cos{\chi\over 2}}\,\,.
\eeq
By making use of (\ref{dot_tau}), this equation can be written as:
\beq
\chi'\,=\,{2\over 3}\,e^{f-2g}\,{3\,\cos\chi-1\over  \sin\chi}\,\,.
\label{chi-prime_BPSeq}
\eeq
The induced metric on the D5-brane worldvolume for our ansatz is:
\bear
&&ds^2_6\,=\,h^{-{1\over 2}}\,\big[-(dx^0)^2+(dx^1)^2+(dx^2)^2\,\big]\,+\,
h^{{1\over 2}}\,\big[1\,+\,{e^{2g}\over 4}\,(\chi')^2\,\big]\,dr^2\,+\,
{h^{{1\over 2}}\,e^{2g}\over 4}\,\cos^2{\chi\over 2}\,(d\theta)^2
\rc\rc
&&\qquad\qquad
+\,
{h^{{1\over 2}}\,e^{2g}\over 4}\,\big[\sin^2{\chi\over 2}\,\cos^2{\chi\over 2}\,+\,
e^{2f-2g}\,(\cos^2{\chi\over 2}\,+\,2\dot\tau)^2\,\big]\,(d\psi)^2\,\,,\qquad\qquad
\eear
whose determinant is equal to:
\beq
\sqrt{-g_6}\,=\,{e^{2g}\,\cos{\chi\over 2}\over 4}
\Big[1\,+\,{e^{2g}\over 4}\,(\chi')^2\,\Big]^{{1\over 2}}\,
\Big[\sin^2{\chi\over 2}\,\cos^2{\chi\over 2}\,+\,
e^{2f-2g}\,(\cos^2{\chi\over 2}\,+\,2\dot\tau)^2\,\Big]^{{1\over 2}}\,\,.
\label{sqrt_g6}
\eeq
Let us compute this determinant for the embeddings satisfying the BPS equations obtained  by imposing kappa symmetry. First of all we notice that:
\beq
\Big(1\,+\,{e^{2g}\over 4}\,(\chi')^2\Big)_{BPS}\,=\,
{\sin^2{\chi\over 2}\,\cos^2{\chi\over 2}\,+\,
e^{2f-2g}\,(\cos^2{\chi\over 2}\,+\,2\dot\tau)^2\over 
\sin^2{\chi\over 2}\,\cos^2{\chi\over 2}}\,\,,
\label{1_chi_prime_BPS}
\eeq
from which it follows that:
\beq
\sqrt{-g_6}{}_{\big|\,{BPS}}\,=\,{e^{2g}\over 4\sin{\chi\over 2}}\,
\Big[\sin^2{\chi\over 2}\,\cos^2{\chi\over 2}\,+\,
e^{2f-2g}\,(\cos^2{\chi\over 2}\,+\,2\dot\tau)^2\Big]\,\,.
\label{g_6_BPS}
\eeq
Moreover, one can easily verify that:
\beq
\sqrt{-g_6}{}_{\big|\,{BPS}}\,=\,{e^{g}\,\cos{\chi\over 2}\over 2}\,
\big(d_1+d_4\big)_{BPS}\,\,.
\eeq
Plugging this result and the BPS equations (\ref{BPS_first}) into (\ref{Gamma_eta}), we get that:
\beq
\tilde\Gamma_{\kappa}\,\eta\,=\,-\cos(\psi+3\tau)\,\eta\,=\,-\,\eta\,\,,
\eeq
which shows that our brane configuration preserves the same  amount of  SUSY as  the background. 

\subsection{General integration of the BPS equation}

Let us now integrate the BPS equation (\ref{chi-prime_BPSeq}) for $\chi$.  First of all we perform a change in the holographic variable and write (\ref{chi-prime_BPSeq}) in terms of the coordinate 
$\zeta$ defined in (\ref{zeta_g}). Using (\ref{dzeta_dr}) we get:
\beq
{\zeta\over 2}\,{d\chi\over d\zeta}\,=\,{\cos\chi\,-\,{1\over 3}\over \sin\chi}\,\,.
\label{BPS_chi_zeta}
\eeq
This equation can be integrated as:
\beq
\cos\chi\,=\,{1\over 3}\,+\,\Big({c\over \zeta}\Big)^2\,\,,
\label{chi_r_c}
\eeq
where $c$ is an integration constant.  By inspecting (\ref{chi_r_c}), it is clear that the coordinate $\zeta$ of the brane must be greater or equal than some minimal value.  Let us denote by  $\zeta_q$  this minimal value of  $\zeta$. Clearly,  $\zeta_q$ and the constant $c$ are related as:
\beq
{1\over 3}\,+\,{c^2\over \zeta_q^2}\,=\,1\,\,,
\eeq
or, equivalently:\footnote{We are implicitly assuming that $c^2$ is positive. If we take $c^2<0$ in (\ref{chi_r_c}) the minimal value  $\zeta_q$ of the coordinate $\zeta$ is achieved when $\cos\chi=-1$ and is given by $\zeta_q^2\,=\,-{3\over 4}\,c^2$.
Thus, the embedding function in this second branch can be written as:
\beq
\cos\chi\,=\,{1\over 3}\,-\,{4\over 3}\,\,\Big({\zeta_q\over \zeta}\Big)^2\,\,.
\label{chi_zeta_second}
\eeq
For these embeddings $\chi=\pi$ at the tip  of the brane and decreases when we move towards the UV region $\zeta\to\infty$, reaching the value $\chi=\chi_*$ asymptotically.
}
\beq
c^2\,=\,{2\over 3}\,\,\zeta_q^2\,\,.
\eeq
In terms of $\zeta_q$ the embedding angle $\chi$ is given by:
\beq
\cos\chi\,=\,{1\over 3}\Big[1\,+\,2\,\Big({\zeta_q\over \zeta}\Big)^2\,\Big]\,\,.
\label{chi_zeta}
\eeq
It follows from (\ref{chi_zeta}) that $\chi$ vanishes at the tip of the brane $\zeta=\zeta_q$ and grows as we move towards the UV, until it reaches an asymptotic value $\chi=\chi_*$, with $\chi_*$  given by:
\beq
\cos\chi_*\,=\,{1\over 3}\,\,.
\label{chi_star}
\eeq
Notice that $\chi=\chi_*$ is a constant angle solution of the BPS equation (\ref{BPS_chi_zeta}).  For this $\chi=\chi_*$ embedding,  the tip of the brane is at $\zeta_q=0$ (see (\ref{chi_zeta})). Therefore, the brane reaches the origin of the geometry, as it corresponds to having massless quarks.

\subsection{The profile function}

To determine the function $p(r)$ for a set of  branes ending at the same distance from the origin, \ie\ with the same mass, we compare the smeared and localized brane actions. Let us start with the WZ term, for which the smeared distribution is:
\beq
S_{WZ}^{smeared}\,= T_5 \int_{{\cal M}_{10}} C_{6}  \wedge \Xi\,\,,
\label{WZ_smeared}
\eeq
where $C_6$ is the RR six-form potential given by:
\beq
C_6\,=\,e^{{\phi\over 2}}\,{\cal K}_{(6)}\,\,,
\label{C6_K6}
\eeq
with ${\cal K}_{(6)}$  being the calibration six-form written in \cite{Conde:2016hbg}. In terms of 
the one-forms $E^{a}$ of our vielbein basis (\ref{10d_vielbein}),  ${\cal K}_{(6)}$  is given by:
\beq
{\cal K}_{(6)}\,=\,E^{x^0\,x^1\,x^2}\,\wedge \,{\rm Re}\,\Omega\,\,,
\label{K_6_expression}
\eeq
where $\Omega$ is the following complex three-form:
\beq
\Omega\,=\,e^{3i\tau}\,\big(E^1+iE^2\big)\wedge \big(E^3+iE^4\big)\wedge \big(E^r+iE^5\big)\,\,.
\eeq
In (\ref{WZ_smeared}) $\Xi$ is the smearing four-form for the massive case, $dF_3=2\kappa_{10}^2\,\Xi$, 
which, according to (\ref{dF_3}), is
\beq
2\,\kappa_{10}^2\,T_5\,\Xi\,=\,-Q_f\,\Big[\,3\,p(r)\,dx^3\wedge  {\rm Re }\,\hat\Omega_2\wedge (d\tau+A)\,+\,
p'(r)\,dx^3\wedge dr\wedge {\rm Im }\,\hat\Omega_2\,\Big]\,\,.
\label{smearing_form_massive}
\eeq
Therefore:
\beq
S_{WZ}^{smeared}\,= -{Q_f\over 2 \kappa_{10}^2} \int_{{\cal M}_{10}} e^{{\phi\over 2}}\,{\cal K}_{(6)}\,\wedge dx^3\wedge
\big[3\,p(r)\wedge  {\rm Re }\,\hat\Omega_2\wedge (d\tau+A)\,+\,
p'(r)\,\wedge dr\wedge {\rm Im }\,\hat\Omega_2\,\big]\,\,.
\label{S_WZ_smeared}
\eeq
Integrating in (\ref{S_WZ_smeared}) over the angular variables and $x^3$ for 
$-{L_3\over 2}\le x^3\le {L_3\over 2}$, we get:
\beq
S_{WZ}^{smeared}\,=\,\int d x^0 \,dx^1\,dx^2\,\,dr\,{\cal L}_{WZ}^{smeared}\,\,,
\eeq
where the smeared WZ Lagrangian density is:
\beq
{\cal L}_{WZ}^{smeared}\,=\,{\pi^3\,Q_f\,L_3\over \kappa_{10}^2}\,
e^{{\phi\over 2}+2g}\,
\big(3\,p(r)\,+\,e^f\,p'(r)\big)\,\,.
\label{smeared_WZ_massive}
\eeq
We now compare (\ref{smeared_WZ_massive}) with the WZ  action of a localized single D5-brane multiplied by $N_f$.  As all the flavor branes that form the set of our smeared distribution are identical, these two quantities should be  equal. The WZ term of the action of a D5-brane without worldvolume gauge field is:
\beq
S_{WZ}\,=\,T_5\,\int \hat C_6\,\equiv\,\int d^6\xi\,{\cal L}_{WZ}\,\,,
\label{SWZ}
\eeq
where the hat denotes pullback to the worldvolume  and $C_6$  has been written in (\ref{C6_K6}) in terms of the calibration form of (\ref{K_6_expression}). Let us suppose that we take $\zeta^{\alpha}=(x^0, x^1, x^2, r, \theta, \psi)$ as worldvolume coordinates and consider an embedding of the type (\ref{embedding_ansatz}) with $x^3$ and $\varphi$ constant. Then, the pullback of   $C_6$  is:
\beq
 \hat C_6\,=\,{e^{{\phi\over 2}+2g}\over 16}\,\cos{\chi\over 2}\,\cos(3\tau+\psi)\,
 \big(2\sin\chi+e^f\,(1+\cos\chi+4\dot\tau)\,\chi'\big)\,
 d^3x\wedge dr\wedge d\theta\wedge d\psi\,\,,
 \label{hat_C6_general}
 \eeq
where $d^3x\equiv d x^0\wedge d x^1\wedge d x^2$. Let us next assume that $\tau(\psi)$ is given by (\ref{tau_psi}) and let us integrate the action over $\theta$ and $\psi$. The resulting Lagrangian density (to be integrated over $(x^0, x^1, x^2, r)$) multiplied by $N_f$ is:
\beq
{\pi^2\,N_f\,T_5\over 4}\,e^{{\phi\over 2}+2g}\,\cos{\chi\over 2}\,
\Big[2\sin\chi\,+\,e^{f}\,\big(\cos\chi-{1\over 3}\big)\,\chi'\,\Big]\,\,.
\label{WZ_individual}
\eeq
Let us now compare the terms in (\ref{smeared_WZ_massive}) and (\ref{WZ_individual}) without $e^f$. They are equal provided the profile $p(r)$ is given by:
\beq
p(r)\,=\,{\kappa_{10}^2\,T_5\over 6\pi}\,{N_f\over Q_f\,L_3}\,\cos{\chi(r)\over 2}\,\sin\chi(r) \ .
\eeq
When $r\to\infty$ the angle $\chi\to\chi_*$, where $\chi_*$ is the value written in (\ref{chi_star}). Since  
$\cos{\chi_*\over 2}\,\sin\chi_*={4\over 3\sqrt{3}}$, we have:
\beq
p(r\to\infty)\,=\,{2\,\kappa_{10}^2\,T_5\over 9\sqrt{3}\,\pi}\,{N_f\over Q_f\,L_3}\,\,.
\eeq
As  argued on general grounds in Sec.~\ref{massive_flavors}, we should have $p(r\to\infty)=1$, \ie\ the profile
should approach the one corresponding to massless quarks in the far UV. This only occurs when $Q_f$ and $N_f$ are related as:
\beq
Q_f\,=\,{4\pi\,g_s\,\alpha'\over 9\sqrt{3}}\,{N_f\over L_3}\,\,.
\label{Q_f-N_f-L3}
\eeq
This same conclusion can also be reached by comparing (\ref{smeared_WZ_massive}) for $p=1$  and (\ref{WZ_individual}) for $\chi=\chi_*$. It follows that  the function $p(r)$ is related  to the function $\chi(r)$ for our fiducial embedding as:
\beq
p(r)\,=\,{3\sqrt{3}\over  4}\,\cos{\chi(r)\over 2}\,\sin\chi(r)\,\,.
\label{p_chi}
\eeq
Moreover, by equating the terms proportional to $e^f$ in (\ref{smeared_WZ_massive}) and (\ref{WZ_individual}) we conclude that 
$p(r)$ should satisfy the differential equation:
\beq
p'(r)\,=\,{9\sqrt{3}\over 4}\,\cos{\chi\over 2}\,
\Big(\cos^2{\chi\over 2}\,-\,{2\over 3}\Big)\,\chi'\,\,.
\label{p_prime_chi_prime}
\eeq
By computing the derivative of the right-hand side of (\ref{p_chi})  one can easily show that indeed $p'$ satisfies (\ref{p_prime_chi_prime}).

Let us now check the  equality of the DBI terms in the smeared and microscopic approach.  The smeared DBI action is:
\beq
S_{DBI}^{smeared}\,=\,- T_5 \int_{{\cal M}_{10}} e^{\phi/2} {\cal K}_{(6)}  \wedge \Xi\,=\,
\int d x^0 \,dx^1\,dx^2\,\,dr\,{\cal L}_{WZ}^{smeared}\,\,,
\label{S_DBI_smeared}
\eeq
where  the smeared DBI Lagrangian density ${\cal L}_{DBI}^{smeared}$ is:
\beq
{\cal L}_{DBI}^{smeared}\,=\,-{\pi^3\,Q_f\,L_3\over \kappa_{10}^2}\,
e^{{\phi\over 2}+2g}\,
\Big[3\,p(r)\,+\,e^{f}\,p'(r)\Big]\,=\,-{\cal L}_{WZ}^{smeared}
\,\,. 
\label{DBI_smeared_massive}
\eeq
Let us compare the action (\ref{S_DBI_smeared})  to the DBI action of the fiducial brane multiplied by $N_f$, which is equal to:
\beq
-N_f\,T_5\,\int d^6\xi\,\sqrt{g_6}\Big|_{BPS}\,\,.
\eeq
Taking into account (\ref{g_6_BPS}), we get that (\ref{DBI_smeared_massive}) should be equal to:
\beq
-\pi^2\,N_f\,T_5\,e^{{\phi\over 2}+2g}\,\,
{\sin^2{\chi\over 2}\,\cos^2{\chi\over 2}\,+\,
e^{2f-2g}\,(\cos^2{\chi\over 2}\,+\,2\dot\tau)^2\over 
\sin{\chi\over 2}}\,\,.
\label{DBI_individual}
\eeq
By identifying the terms without $e^f$ in (\ref{DBI_smeared_massive}) and (\ref{DBI_individual}) we get exactly (\ref{p_chi}), whereas the terms with $e^f$ yield:
\beq
e^{2g-f}\,p'(r)\,=\,{9\sqrt{3}\over 2}\,
{\Big(\cos^2{\chi\over 2}\,-\,{2\over 3}\Big)^2\over
\sin{\chi\over 2}}\,\,,
\label{p_prime_chi}
\eeq
which can be shown to be equivalent to (\ref{p_prime_chi_prime}) after using (\ref{chi-prime_BPSeq}). 

Let us finally obtain the explicit form of the profile in terms  of the radial variable $\zeta$. First of all, we write $p$ as:
\beq
p\,=\,{3\sqrt{3}\over 2}\,\cos^2{\chi\over 2}\,\sqrt{1-\cos^2{\chi\over 2}}\,\,.
\label{profile_chi}
\eeq
Using (\ref{chi_zeta}), we get:\footnote{
For the second branch of embeddings with fiducial embedding function given by (\ref{chi_zeta_second}), the profile $p(\zeta)$ can be obtained by substituting  (\ref{chi_zeta_second}) into (\ref{profile_chi}), with the result:
\beq
p(\zeta)\,=\,\Big[1\,-\,\Big({\zeta_q\over \zeta}\Big)^2\Big]\,
\sqrt{1\,+\,2\Big({\zeta_q\over \zeta}\Big)^2}\,\,\,
\Theta(\zeta-\zeta_q)\,\,. 
\label{p_zeta_explicit_second}
\eeq

}
\beq
p(\zeta)\,=\,\Big[1-\big({\zeta_q\over \zeta}\big)^2\Big]^{{1\over 2}}\,
\Big[1\,+\,{1\over 2}\,\big({\zeta_q\over \zeta}\big)^2\Big]\,\,\Theta(\zeta-\zeta_q)\,\,.
\label{p_zeta_explicit_app_B}
\eeq
Notice that $p(\zeta)$ is continuous at $\zeta=\zeta_q$, as it should. 

\vskip 3cm
\renewcommand{\theequation}{\rm{C}.\arabic{equation}}
\setcounter{equation}{0}

\section{Equations of motion of the probe}
\label{eom_probe}

The DBI action, in Einstein frame, of a D5-brane with a worldvolume gauge field $F$  turned on is:
\beq
S_{DBI}\,=\,-T_5\,\int d^6\xi\,e^{{\phi\over 2}}\,
\sqrt{-\det(g_{6}+e^{-{\phi\over 2}}\,F)}\,\equiv\,\int d^6\xi\, {\cal L}_{DBI}\,\,.
\label{DBI_action_At}
\eeq
We are interested in having a gauge potential $A$ dual to a $U(1)$ charge density. Therefore, we
take
\beq
A\,=\,A_0(r)\,dt\,\,,
\label{At}
\eeq
for which $F=dA$ is given by:
\beq
F\,=\,A_0'\,dr\wedge dx^0\,\,.
\label{Ftr}
\eeq
Let us choose $\xi^{\alpha}\,=\,(x^0, x^1, x^2, r, \theta, \psi)$ as our set of worldvolume coordinates and 
let us assume an embedding ansatz as in (\ref{embedding_ansatz}), \ie\ with 
$\chi=\chi(r)$ and $\tau=\tau(\psi)$. The Lagrangian density ${\cal L}_{DBI}$ corresponding to the action 
(\ref{DBI_action_At}) is:
\beq
{\cal L}_{DBI}\,=\,-{T_5\over 8}\,e^{{\phi\over 2}+2g}\,
\cos{\chi\over 2}\,\sqrt{
\sin^2\chi\,+\,e^{2f-2g}\,(1+\cos\chi+ 4\dot\tau)^2}\,
\sqrt{1\,-\,e^{-\phi}\,(A_0')^2\,
+\,{e^{2g}\over 4}\,(\chi')^2}\,\,.
\eeq
The WZ term of the action has been written in (\ref{SWZ}). Taking into account (\ref{hat_C6_general}), we can  easily demonstrate that  ${\cal L}_{WZ}$ takes the form:
\beq
{\cal L}_{WZ}\,=\,{T_5\over 16}\,e^{{\phi\over 2}+2g}\,
\cos{\chi\over 2}\, \cos(3\tau+\psi)\,\big(2\sin\chi+e^f\,(1+\cos\chi+4\dot\tau)\,\chi'\big)\,\,.
\label{WZ_lag_probe}
\eeq
Let us now analyze the equations of motion of $\chi(r)$ and $\tau(\psi)$ derived from the total Lagrangian 
${\cal L}\,=\,{\cal L}_{DBI}\,+\,{\cal L}_{WZ}$. The equation of motion  of $\chi(r)$  derived from the total action is:
\beq
{\partial\over \partial r}\Big[{\partial {\cal L}_{DBI}\over \partial \chi'}\,+\,
{\partial {\cal L}_{WZ}\over \partial \chi'}\Big]\,-\,{\partial {\cal L}_{DBI}\over \partial \chi}\,-\,
{\partial {\cal L}_{WZ}\over \partial \chi}\,=\,0\,\,.
\label{eom_chi}
\eeq
The derivatives with respect to $\chi'$ appearing  in (\ref{eom_chi}) are:
\bear
{\partial {\cal L}_{DBI}\over \partial \chi'} & = &
-{T_5\over 32}\,e^{{\phi\over 2}+4g}\,\cos{\chi\over 2}\,
{\sqrt{
\sin^2\chi\,+\,e^{2f-2g}\,(1+\cos\chi+ 4\dot\tau)^2}\over
\sqrt{1\,-\,e^{-\phi}\,(A_0')^2\,\,+\,{e^{2g}\over 4}\,(\chi')^2}}\,\,\chi'\rc\rc
{\partial {\cal L}_{WZ}\over \partial \chi'} & = &
{T_5\over 16}\,e^{{\phi\over 2}+2g+f}\,\
\cos{\chi\over 2}\,  \cos(3\tau+\psi)\,(1+\cos\chi+4\dot\tau)\,\,.
\label{partial_L_chiprime_At}
\eear
Moreover, if we define the auxiliary functions
\bear
{\cal J}_{DBI}(\chi) & \equiv & \cos{\chi\over 2}\,
\sqrt{ \sin^2\chi\,+\,e^{2f-2g}\,(1+\cos\chi+ 4\dot\tau)^2}\rc\rc
{\cal J}_{WZ}(\chi,\chi') & \equiv & \cos{\chi\over 2}\,
\Big(\,2\sin\chi+e^f\,(1+\cos\chi+4\dot\tau)\,\chi'\,\Big)\,\,,
\label{cal_Js}
\eear
the derivatives with respect to $\chi$ we need in (\ref{eom_chi})  are:
\bear
{\partial {\cal L}_{DBI}\over \partial \chi} & = & 
-{T_5\over 8}\,e^{{\phi\over 2}+2g}\,
\sqrt{1\,-\,e^{-\phi}\,(A_0')^2\,+\,{e^{2g}\over 4}\,(\chi')^2}\,\,
{\partial {\cal J}_{DBI}(\chi)\over \partial \chi}\rc\rc
{\partial {\cal L}_{WZ}\over \partial \chi} & = &
{T_5\over 16}\,e^{{\phi\over 2}+2g}\,\cos(3\tau+\psi)\,
{\partial {\cal J}_{WZ}(\chi,\chi')\over \partial \chi}\,\,.
\label{partial_L_chi_At}
\eear

Let us analyze the dependence on the angular variable $\psi$  of the different terms in (\ref{eom_chi}).
The only terms that depend on  $\psi$ in (\ref{eom_chi}) are those that contain $\dot\tau$ and $3\tau+\psi$.  By inspecting how these terms enter into (\ref{partial_L_chiprime_At}) and (\ref{partial_L_chi_At}), one easily concludes that  the equation of motion  of $\chi$ can only be satisfied if $\dot \tau$ and $3\tau+\psi$ are constant.  This last condition implies that $\dot\tau=-1/3$, as in (\ref{dot_tau}).  Let us now look at the equation of motion for $\tau(\psi)$:
\beq
{\partial\over \partial \psi}\Big[{\partial {\cal L}_{DBI}\over \partial \dot\tau}\,+\,
{\partial {\cal L}_{WZ}\over \partial \dot\tau}\Big]\,-\,{\partial {\cal L}_{DBI}\over \partial \tau}\,-\,
{\partial {\cal L}_{WZ}\over \partial \tau}\,=\,0\,\,.
\label{eom_tau}
\eeq
Let us study the different terms in (\ref{eom_tau}).  When $\dot\tau$ is constant, we have
\beq
{\partial\over \partial \psi}\Big[{\partial {\cal L}_{DBI}\over \partial \dot\tau}\Big]\,=\,0\,\,.
\eeq
Moreover
\beq
{\partial\over \partial \psi}\Big[{\partial {\cal L}_{WZ}\over \partial \dot\tau}\Big]\,=\,
-{T_5\over 4}\,e^{{\phi\over 2}+2g}\,\cos{\chi\over 2}\,
\sin(3\tau+\psi)\,(3\dot \tau+1)\,\chi'\,\,,
\eeq
which vanishes if (\ref{dot_tau}) holds. The DBI Lagrangian does not depend on $\tau$, thus:
\beq
{\partial  {\cal L}_{DBI}\over \partial \tau}\,=\,0\,\,.
 \eeq
 The remaining  $\tau$ derivative that we have to compute is:
 \beq
 {\partial  {\cal L}_{WZ}\over \partial \tau}\,=\,
 -{3T_5\over 16}\,e^{{\phi\over 2}+2g}\,\cos{\chi\over 2}\,
\sin(3\tau+\psi)\,
\Big(\,2\sin\chi+e^f\,(1+\cos\chi+4\dot\tau)\,\chi'\,\Big)\,\,,
\label{partial_L_tau_At}
\eeq
 which vanishes independent of $\chi$ if $\sin(3\tau+\psi)=0$, \ie\ when $\tau$ depends on $\psi$ as in
 (\ref{tau_psi_n}). As in the supersymmetric solution, we will take $n=4$ in this equation. 
 
 The conclusion of this  analysis is that, for the type of ansatz we are considering, the function $\tau(\psi)$ must be given by (\ref{tau_psi}) in order to satisfy the equations of motion of the probe D5-brane. We will now study these equations separately in two different cases. 

\subsection{The BPS solution}
Let us consider now the BPS configuration in which $\tau(\psi)$ is given by (\ref{tau_psi}), $A_0$ vanishes and $\chi(r)$ satisfies the first-order differential equation (\ref{chi-prime_BPSeq}) dictated by kappa symmetry. By using (\ref{1_chi_prime_BPS}) to evaluate the different derivatives of (\ref{partial_L_chiprime_At}), (\ref{partial_L_chi_At}), and (\ref{partial_L_tau_At}) one can easily show that:
\beq
{\partial {\cal L}\over \partial\chi'}\Bigg|_{BPS}\,=\,{\partial {\cal L}\over \partial\chi}\Bigg|_{BPS}
\,=\,0\,\,,
\qquad\qquad
{\partial {\cal L}\over \partial\dot\tau}\Bigg|_{BPS}\,=\,
{\partial {\cal L}\over \partial\tau}\Bigg|_{BPS}\,=\,0
\label{partial_L_BPS}\,\,,
\eeq
which implies the fullfilment of the equations of motion for the BPS configuration.

\subsection{The massless solution}
Let us consider a massless embedding of the probe brane with non-zero chemical potential. Therefore, we will try to solve  the equations of motion with $\chi$ constant. It is clear from the first equation in  (\ref{partial_L_chiprime_At})  that  $\partial {\cal L}_{DBI}/ \partial \chi'$ vanishes if $\chi'$ is zero. Moreover, from the second equation in (\ref{partial_L_chiprime_At})  we conclude that 
$ \partial {\cal L}_{WZ}/ \partial \chi'$ is zero if $\chi=\chi_*$, where $\chi_*$ is the angle defined in 
(\ref{chi_star}).  Moreover, since
\beq
{\partial {\cal J}_{DBI}\over \partial\chi}\Big|_{\chi=\chi_*}\,=\,
{\partial {\cal J}_{WZ}\over \partial\chi}\Big|_{\chi=\chi_*}\,=\,0\,\,,
\eeq
 it follows that $\chi=\chi_*$ solves the equations of motion of the probe.  Let us next define 
 ${\cal J}_*$ as:
 \beq
 {\cal J}_*\equiv  {\cal J}_{DBI}(\chi=\chi_*)\,=\,{1\over 2}\,
 {\cal J}_{WZ}(\chi=\chi_*)\,\,.
 \eeq
It follows that:
\beq
{\cal J}_*\,=\,{4\over 3\sqrt{3}}\,\,.
\eeq
It remains to satisfy the equation of motion for $A_0$.  The Lagrangian density for the gauge field is:
\beq
{\cal L}\,=\,-{\cal T}\,e^{{\phi\over 2}\,+\,2g}\,
\Big[\sqrt{1-e^{-\phi}\,A_0'^{\,2}}\,-\,1\Big]\,\,,
\label{cal_L_At}
\eeq
where the effective tension ${\cal T}$ is given by:
\beq
{\cal T}\,=\,{T_5\over 8}\,{\cal J}_*\,=\,{T_5\over 6\sqrt{3}}\,\,.
\label{cal_T}
\eeq
Since $A_0$ is a cyclic variable in the Lagrangian density (\ref{cal_L_At}), 
the equation of motion for $A_0$  can be integrated once, giving:
\beq
{e^{2g-{\phi\over 2}}\,A_0'\over 
\sqrt{1-e^{-\phi}\,A_0'^{\,2}}}\,=\,d\,\,,
\label{first_integral_At}
\eeq
where $d$ is a constant proportional to the charge density.  From this equation we get:
\beq
A_0'\,=\,{e^{{\phi\over 2}}\, d\over 
\sqrt{d^2+e^{4g}}}\,\,.
\eeq
The chemical potential $\mu$ is just  the value of $A_0$ at the boundary, and is given by the following integral:
\beq
\mu\,=\,d\,\int_0^{\infty}\,
{e^{{\phi\over 2}}\, \over 
\sqrt{d^2+e^{4g}}}\,dr\,\,.
\eeq

\vskip 3cm
\renewcommand{\theequation}{\rm{D}.\arabic{equation}}
\setcounter{equation}{0}

\section{Probes in the Higgs branch}
\label{Higgs}

We now study embeddings of a probe D5-brane in which the coordinate $x^3$ is not constant but instead bends as the holographic coordinate changes. This bending can be interpreted as  a recombination between the D3- and D5-branes, realizing the Higgs branch of the dual theory, see, {\emph{e.g.}}, \cite{Itsios:2015kja}. In order to find this configuration let us consider the following set of worldvolume coordinates $\xi^{\alpha}\,=\,(x^0, x^1, x^2, r, \theta, \psi)$ and the following embedding ansatz:
\beq
\chi=\chi(r)\,\,,
\qquad\qquad
\tau=\tau (\psi)\,\,,
\qquad\qquad
x^3=z(r)\,\,.
\label{embedding_ansatz_Higgs}
\eeq
In addition, the probe D5-brane will have a worldvolume flux,  given by:
\beq
F\,=\,Q\,d\theta\wedge d\psi\,\,,
\eeq
with $Q$ being a constant. We show below that, if certain BPS equations are satisfied, this configuration preserves the supersymmetries of the background.

The induced metric for the  ansatz (\ref{embedding_ansatz_Higgs}) is:
\bear
&&ds^2_6\,=\,h^{-{1\over 2}}\,\big[-(dx^0)^2+(dx^1)^2+(dx^2)^2\,\big]\,+\,
h^{{1\over 2}}\,\big[1\,+\,{e^{2g}\over 4}\,(\chi')^2\,+\,h^{-1}\,e^{-2\phi}\,(z')^2\big]\,dr^2
\rc\rc
&&\qquad
+\,
{h^{{1\over 2}}\,e^{2g}\over 4}\,\cos^2{\chi\over 2}\,(d\theta)^2\,
+\,
{h^{{1\over 2}}\,e^{2g}\over 4}\,\big[\sin^2{\chi\over 2}\,\cos^2{\chi\over 2}\,+\,
e^{2f-2g}\,(\cos^2{\chi\over 2}\,+\,2\dot\tau)^2\,\big]\,(d\psi)^2\,\,.\qquad\qquad
\label{induced_metric_Higgs}
\eear
The DBI action in Einstein frame  is given by (\ref{DBI_action_At}). In the present case the DBI determinant is:
\beq
-\det(g_{6}+e^{-{\phi\over 2}}\,F)={e^{4g}\over 64}
\big[{\cal J}_{DBI}^2(\chi)+64\,h^{-1}\,e^{-4g-\phi}\,Q^2\big]
\big[1+{e^{2g}\over 4}\,(\chi')^2+h^{-1}\,e^{-2\phi}\,(z')^2\big]\,\,,\qquad
\eeq
where ${\cal J}_{DBI}(\chi)$ is the function of $\chi$ defined in (\ref{cal_Js}). The DBI Lagrangian density is:
\beq
{\cal L}_{DBI}\,=\,-{T_5\over 8}\,e^{{\phi\over 2}+2g}\,
\sqrt{{\cal J}_{DBI}^2(\chi)+64\,h^{-1}\,e^{-4g-\phi}\,Q^2}
\sqrt{1+{e^{2g}\over 4}\,(\chi')^2+h^{-1}\,e^{-2\phi}\,(z')^2}\,\,.
\label{L_DBI_Higgs}
\eeq
The WZ action now takes the form:
\beq
S_{WZ}\,=\,T_5\,\int \hat C_6\,+\,T_5\int \hat C_4\wedge F
\equiv\,\int d^6\xi\,{\cal L}_{WZ}\,\,,
\label{SWZ_Higgs}
\eeq
where the WZ Lagrangian density is given by:
\beq
{\cal L}_{WZ}\,=\,{T_5\over 16}\,e^{{\phi\over 2}+2g}\,\cos(3\tau+\psi)\,{\cal J}_{WZ}(\chi,\chi')\,+\,
T_5\,Q\,e^{-\phi}\,h^{-1}\,z'\,\,,
\eeq
and  ${\cal J}_{WZ}(\chi,\chi')$ has been defined in  (\ref{cal_Js}).

Let us now study the equations of motion for $z(r)$. By inspecting ${\cal L}_{DBI}$ and 
 ${\cal L}_{WZ}$  we immediately conclude that they do not depend on $z$ (only on $z'$) and, therefore, $z(r)$ is a cyclic variable. Thus, the equation of motion for $z(r)$ can be integrated once as:
\beq
{\partial {\cal L}_{DBI}\over \partial z'}\,+\,
{\partial {\cal L}_{WZ}\over \partial z'}\,=\,{\rm constant}\,\,.
\label{first-integral_Higgs}
\eeq
We will consider the case in which the constant on the right-hand side  of (\ref{first-integral_Higgs})  is zero which, as we will verify below, is the supersymmetric configuration. Therefore, we must have:
\beq
-{\partial {\cal L}_{DBI}\over \partial z'}\,=\,
{\partial {\cal L}_{WZ}\over \partial z'}\,\,.
\label{SUSY_first_integral_Higgs}
\eeq
Moreover, the derivatives of $ {\cal L}_{DBI}$ and ${\cal L}_{WZ}$ with respect to $z'$ are:
\bear
-{\partial {\cal L}_{DBI}\over \partial z'} & = & {T_5\over 8}\,e^{-{3\phi\over 2}+2g}\,h^{-1}\,
{\sqrt{{\cal J}_{DBI}^2(\chi)+64\,h^{-1}\,e^{-4g-\phi}\,Q^2}\over
\sqrt{1+{e^{2g}\over 4}\,(\chi')^2+h^{-1}\,e^{-2\phi}\,(z')^2}}\,\,z'\rc\rc
{\partial {\cal L}_{WZ}\over \partial z'} & = & T_5\,Q\,e^{-\phi}\,h'\,\,.
\eear
Plugging these values into (\ref{SUSY_first_integral_Higgs}) and solving for $z'$, we get:
\beq
z'\,=\,8\,Q\,{e^{{\phi\over 2}-2g}\over {\cal J}_{DBI}(\chi)}\,\sqrt{1+{e^{2g}\over 4}\,(\chi')^2}\,\,.
\label{BPS_bending}
\eeq
A useful relation that can be derived from this last equation is:
\beq
\sqrt{1+{e^{2g}\over 4}\,(\chi')^2+h^{-1}\,e^{-2\phi}\,(z')^2}\,=\,
\sqrt{1+{e^{2g}\over 4}\,(\chi')^2}\,\,\,
{\sqrt{{\cal J}_{DBI}^2(\chi)+64\,h^{-1}\,e^{-4g-\phi}\,Q^2}
\over {\cal J}_{DBI}(\chi)}\,\,.
\label{useful_relation_Higgs}
\eeq

Let us now derive the equations of motion for  the embedding function $\chi(r)$. First of all, we calculate the derivatives of the DBI Lagrangian density with respect to $\chi'$ and $\chi$, which are given by:
\bear
&&{\partial {\cal L}_{DBI}\over \partial \chi'}\,=\,-{T_5\over 32}\,
e^{{\phi\over 2}+4g}\,
{\sqrt{{\cal J}_{DBI}^2(\chi)+64\,h^{-1}\,e^{-4g-\phi}\,Q^2}\over
\sqrt{1+{e^{2g}\over 4}\,(\chi')^2+h^{-1}\,e^{-2\phi}\,(z')^2}
}\,\,\chi'\rc\rc
&&{\partial {\cal L}_{DBI}\over \partial \chi}\,=\,-{T_5\over 8}\,
e^{{\phi\over 2}+2g}\,
{\sqrt{1+{e^{2g}\over 4}\,(\chi')^2+h^{-1}\,e^{-2\phi}\,(z')^2}\,
\over
\sqrt{{\cal J}_{DBI}^2(\chi)+64\,h^{-1}\,e^{-4g-\phi}\,Q^2}}\,\,\,
{\cal J}_{DBI}(\chi)
{\partial {\cal J}_{DBI}(\chi)\over \partial\chi}\,\,.
\qquad\qquad
\label{partials_DBI_Higgs}
\eear
Let us now use (\ref{useful_relation_Higgs})  to compute the square root of the right-hand-side of (\ref{partials_DBI_Higgs})  involving $\chi'$. We get:
\bear
&&{\partial {\cal L}_{DBI}\over \partial \chi'}\,=\,-{T_5\over 32}\,
e^{{\phi\over 2}+4g}\,
{{\cal J}_{DBI}(\chi)\over \sqrt{1+{e^{2g}\over 4}\,(\chi')^2}}\rc\rc
&&{\partial {\cal L}_{DBI}\over \partial \chi}\,=\,-{T_5\over 8}\,
e^{{\phi\over 2}+2g}\,\sqrt{1+{e^{2g}\over 4}\,(\chi')^2}\,
{\partial {\cal J}_{DBI}(\chi)\over \partial\chi}\,\,,
\eear
which are just the derivatives corresponding to the case $Q=z'=0$ with $A_0=0$  (see (\ref{partial_L_chiprime_At}) and (\ref{partial_L_chi_At})). As the derivatives of ${\cal L}_{WZ}$ with respect to $\chi'$ and $\chi$ are the same as in the  $Q=z'=0$ case, we immediately conclude that the equation of $\chi$ is satisfied if $\tau(\psi)$ and $\chi(r)$ fulfill (\ref{tau_psi_n}) and (\ref{chi-prime_BPSeq}) respectively. Thus, the addition of internal flux, related to the bending of the brane in the $x^3$ direction as in (\ref{BPS_bending}), does not modify the BPS solution for $\chi (r)$ and $\tau(\psi)$. 

Let us now write an explicit equation for the bending function $z(r)$. This equation can be obtained by plugging into (\ref{BPS_bending}) the relation:\footnote{
Another useful relation is
\beq
{\cal J}_{DBI}\,{\partial {\cal J}_{DBI}(\chi)\over \partial\chi}\,=\,
{1\over 2}\,\sin\chi\,\cos{\chi\over 2}\,{\partial {\cal J}_{WZ}\over \partial \chi} \ .
\eeq
}
\beq
\sqrt{1+{e^{2g}\over 4}\,(\chi')^2}\,=\,{ {\cal J}_{DBI}\over 
\sin\chi\,\cos{\chi\over 2}}\,\,.
\eeq
We get:
\beq
z'\,=\,\,{8Q\,e^{{\phi\over 2}-2g}\over \sin\chi\,\cos{\chi\over 2}}\,\,.
\label{BPS_bending_z(r)}
\eeq

\subsection{Kappa symmetry}

In this subsection we will derive the first-order equations  for $\tau(\psi)$, $\chi(r)$, and $z(r)$ from the kappa symmetry condition of the probe. We begin by computing the induced Dirac matrices for the ansatz (\ref{embedding_ansatz_Higgs}):
\bear
&&\gamma_{x^\mu}\,=\,h^{-{1\over 4}}\,\Gamma_{x^{\mu}}\,\,,
\qquad\qquad (\mu=0,1,2)\rc\rc
&&\gamma_{r}\,=\,h^{{1\over 4}}\,\Gamma_{r}\,+\,{1\over 2}\,
h^{{1\over 4}}\,e^{g}\,\chi'\,\Gamma_{4}\,+\,
h^{-{1\over 4}}\,e^{-\phi}\,z'\,\Gamma_{x^3}\rc\rc
&&\gamma_{\theta}\,=\,{1\over 2}\,h^{{1\over 4}}\,e^{g}\,\cos{\chi\over 2}\,\cos\psi\,
\Gamma_{1}\,+\,{1\over 2}\,h^{{1\over 4}}\,e^{g}\cos{\chi\over 2}\,\sin\psi\,\Gamma_{2}\rc\rc
&&\gamma_{\psi}\,=\,{1\over 2}\,h^{{1\over 4}}\,e^{g}\,\cos{\chi\over 2}\,\sin{\chi\over 2}\,
\Gamma_{3}\,+\,h^{{1\over 4}}\,e^{f}\,
\Big({1\over 2}\,\cos^2{\chi\over 2}\,+\,\dot\tau\Big)\,\Gamma_{5}\,\,.
\label{induced_gammas_Higgs}
\eear
The kappa symmetry matrix with $F_{\theta\psi}$ flux in  our conventions is:
\beq
\Gamma_{\kappa}\,=\,{1\over 
\sqrt{-\det(g_{6}+e^{-{\phi\over 2}}\,F)}}\,
\Big[\sigma_1\,\gamma_{x^0x^1x^2 r \theta\psi}\,+\,
e^{-{\phi\over 2}}\,F_{\theta\psi}\,(i\sigma_2)\,\gamma^{\theta\psi}\,
\gamma_{x^0x^1x^2 r \theta\psi}\Big]\,\,,
\label{Gamma_kappa_with_flux}
\eeq
with $F_{\theta\psi}=Q$.  In the second term in (\ref{Gamma_kappa_with_flux}) we need to compute
\beq
\gamma^{\theta\psi}\,\gamma_{\theta\psi}\,=\,g^{\theta\theta}\,g^{\psi\psi}\,
(\gamma_{\theta\psi})^2\,\,,
\eeq
with $g^{\theta\theta}$ and $g^{\psi\psi}$ being elements of the inverse induced metric (\ref{induced_metric_Higgs}). In order to calculate  the square of  $\gamma_{\theta\psi}$, let us write this matrix as:
\beq
\gamma_{\theta\psi}\,=\,A\,e^{-\psi\,\Gamma_{12}}
\Big(B\,\Gamma_{13}\,+\,C\,\Gamma_{15}\Big)\,\,,
\eeq
with  $A$, $B$, and $C$ being given by:
\beq
A\,=\,{e^g\over 2}\,h^{{1\over 2}}\,\cos{\chi\over 2}\,\,,
\qquad
B\,=\,{e^g\over 2}\,\cos{\chi\over 2}\,\sin{\chi\over 2}\,\,,
\qquad
C\,=\,{e^g\over 2}\Big(\cos^2{\chi\over 2}\,+\,2\dot\tau\Big)\,\,.
\eeq
As $\{\Gamma_{13}, \Gamma_{12}\}=\{\Gamma_{15}, \Gamma_{12}\}=0$, we can write:
\beq
(\gamma_{\theta\psi})^2\,=\,A^2\,\big(B\,\Gamma_{13}\,+\,C\,\Gamma_{15}\big)^2\,=\,
-A^2\,(B^2+C^2)\,\,,
\eeq
where, in the last step, we have used that $(\Gamma_{13})^2=(\Gamma_{15})^2\,=\,-1$ and
that $\{\Gamma_{13}, \Gamma_{15}\}=0$. Moreover, by inspecting (\ref{induced_metric_Higgs}) we can write $g^{\theta\theta}$ and $g^{\psi\psi}$ as:
\beq
g^{\theta\theta}\,=\,\big(g_{\theta\theta}\big)^{-1}\,=\, h^{{1\over 2}}\,A^{-2}\,\,,
\qquad
g^{\psi\psi}\,=\,\big(g_{\psi\psi}\big)^{-1}\,=\,h^{-{1\over 2}}\,
\big(B^2+C^2)^{-1}\,\,.
\eeq
It follows that:
\beq
\gamma^{\theta\psi}\,\gamma_{\theta\psi}\,=\,-1\,\,.
\eeq
Thus, $\Gamma_{\kappa}$ can be written as the sum of two terms:
\beq
\Gamma_{\kappa}\,=\,\Gamma_{\kappa}^{(1)}\,+\,\Gamma_{\kappa}^{(2)}\,\,,
\eeq
where $\Gamma_{\kappa}^{(1)}$ and $\Gamma_{\kappa}^{(2)}$ are given by:
\bear
&&\Gamma_{\kappa}^{(1)}\,=\,{1\over 
\sqrt{-\det(g_{6}+e^{-{\phi\over 2}}\,F)}}\,
\sigma_1\,\gamma_{x^0x^1x^2 r \theta\psi}\rc\rc
&&\Gamma_{\kappa}^{(2)}\,=\,{1\over 
\sqrt{-\det(g_{6}+e^{-{\phi\over 2}}\,F)}}\,(-i\sigma_2)\,
\gamma_{x^0x^1x^2 r}\,\,.
\eear
Let us now proceed as in App.~\ref{profile_details} and compute the antisymmetrized product 
$\gamma_{r\theta\psi}$ by using (\ref{induced_gammas_Higgs}). We get:
\beq
 \gamma_{r\theta\psi}={h^{{3\over 4}}e^{g}\over 2}\cos{\chi\over 2}
 \,e^{-\psi\Gamma_{12}}\,
 \Big[ d_1\,\Gamma_{r13}+d_2\,\Gamma_{413}+ d_3\,\Gamma_{r15}+d_4\Gamma_{415}
 +d_5\,\Gamma_{x^3 13}+d_6\,\Gamma_{x^3 15} \Big]\,\,,
 \eeq
where the coefficients $d_1$, $d_2$, $d_3$, and $d_4$ are given by the same expression as in (\ref{d_coefficients}) and the new coefficients $d_5$ and $d_6$ are:
\beq
d_5\,=\,{h^{-{1\over 2}}\,e^{g-\phi}\over 2}\,\cos{\chi\over 2}\,\sin{\chi\over 2}\,z'\,\,,
\qquad\qquad
d_6\,=\,{h^{-{1\over 2}}\,e^{f-\phi}\over 2}\,\Big(\cos^2{\chi\over 2}+2\dot \tau\Big)\,z'\,\,.
\label{d_5_d_6}
\eeq

Let us next define the rotated kappa symmetry matrix $\tilde\Gamma_{\kappa}$ as in (\ref{tilde_Gamma_kappa}). Then, the kappa symmetry condition is the one written in (\ref{kappa_eta}). Moreover,  we can write $\tilde\Gamma_{\kappa}\,=\,\tilde\Gamma_{\kappa}^{(1)}\,+\,\tilde\Gamma_{\kappa}^{(2)}$, with 
\beq
\tilde\Gamma_{\kappa}^{(i)}\,=\,
e^{-{3\over 2}\,\Gamma_{12}\,\tau}\,
\Gamma_{\kappa}^{(i)}\,e^{{3\over 2}\,\Gamma_{12}\,\tau}\,\,.
\eeq
The rotated matrix $\tilde\Gamma_{\kappa}^{(1)}$ can be written as:
\bear
&&\tilde\Gamma_{\kappa}^{(1)}\,=\,{1\over 2}\,{e^{g}\,\cos{\chi\over 2} \over
\sqrt{-\det(g_{6}+e^{-{\phi\over 2}}\,F)}}\,
 \,e^{-(\psi+3\tau)\Gamma_{12}}\,\sigma_1\,\Gamma_{x^0\,x^1\,x^2}\,\Big[
 d_1\,\Gamma_{r13}\,+\,d_2\,\Gamma_{413}\rc\rc
  &&\qquad\qquad\qquad\qquad+
 d_3\,\Gamma_{r15}\,+\,
 d_4\,\Gamma_{415}\,
 +\,d_5\,\Gamma_{x^3 1 3}\,+\,d_6\,\Gamma_{x^3 1 5}
 \Big]\,\,,
 \eear
whereas  $\tilde\Gamma_{\kappa}^{(2)}$ is given by:
\beq
\tilde\Gamma_{\kappa}^{(2)}\,=\,
{Q\,e^{-{\phi\over 2}}\,h^{-{1\over 2}}\over 
\sqrt{-\det(g_{6}+e^{-{\phi\over 2}}\,F)}}(-i\sigma_2)\,\Gamma_{x^0 x^1 x^2}
\Big(\Gamma_r\,+\,{e^g\over 2}\,\chi'\,\Gamma_4\,+\,h^{-{1\over 2}}\,e^{-\phi}\,z'\,\Gamma_{x^3}\Big)\,\,.
\eeq
Let us now calculate $\tilde\Gamma_{\kappa}^{(1)}\eta$. This calculation is similar to the one performed to derive (\ref{Gamma_eta}). In order to compute the additional terms, we need  to use the projections:
\beq
\sigma_1\,\Gamma_{x^0 x^1 x^2}\,\Gamma_{x^3 13}\eta\,=\,\sigma_3\,\Gamma_{13}\eta\,\,,
\qquad\qquad
\sigma_1\,\Gamma_{x^0 x^1 x^2}\,\Gamma_{x^3 15}\eta\,=\,-\sigma_3\,\Gamma_{r4}\,\Gamma_{13}\eta\,\,.
\eeq
Using these results, we arrive at:
\beq
\tilde\Gamma_{\kappa}^{(1)}\,\eta\,=\,{1\over 2}\,{e^{g}\,\cos{\chi\over 2} \over 
\sqrt{-\det(g_{6}+e^{-{\phi\over 2}}\,F)}}\,
 \,e^{-(\psi+3\tau)\Gamma_{12}}
 \Big[-d_1-d_4\,+\,(d_2-d_3)\,\Gamma_{r4}\,+d_5\,\sigma_3\,\Gamma_{13}\,-d_6
 \,\sigma_3\Gamma_{r4}\Gamma_{13}
  \Big]\,\eta\,\,.
 \label{Gamma_1_eta_Higgs}
 \eeq
Next, we compute $\tilde\Gamma_{\kappa}^{(2)}\,\eta$. We need the projections:
\bear
&&(-i\sigma_2)\,\Gamma_{x^0 x^1 x^2 r}\,\eta\,=\,-\sigma_3\,\Gamma_{13}\,\eta\,\,,
\qquad\qquad
(-i\sigma_2)\,\Gamma_{x^0 x^1 x^2 4}\,\eta\,=\,\sigma_3\,\Gamma_{r4}\,\Gamma_{13}\,\eta\,\,,\rc\rc
&&(-i\sigma_2)\,\Gamma_{x^0 x^1 x^2 x^3}\,\eta\,=\,-\eta\,\,,
\eear
and we obtain:
\beq
\tilde\Gamma_{\kappa}^{(2)}\eta\,=\,
{Q\,e^{-{\phi\over 2}}\,h^{-{1\over 2}}\over 
\sqrt{-\det(g_{6}+e^{-{\phi\over 2}}\,F)}}\Big[
-\sigma_3\,\Gamma_{13}\,+\,{e^g\over 2}\,\chi'\,\sigma_3\,\Gamma_{r4}\,\Gamma_{13}\,-\,
h^{-{1\over 2}}\,e^{-\phi}\,z'\Big]\,\,.
\label{Gamma_2_eta_Higgs}
\eeq
Apart  from the two  equations  written in  (\ref{BPS_first}), we get two extra conditions by imposing that $\tilde \Gamma_{\kappa}$ acts as the identity on $\eta$. From the terms containing $\sigma_3\,\Gamma_{13}$ in (\ref{Gamma_1_eta_Higgs}) and (\ref{Gamma_2_eta_Higgs}), we have:
\beq
{e^{g}\over 2}\,\cos{\chi\over 2} \,d_5\,=\,Q\,e^{-{\phi\over 2}}\,h^{-{1\over 2}}\,\,.
\label{d_5_condition}
\eeq
Moreover, the vanishing of the terms with $\sigma_3\,\Gamma_{r4}\,\Gamma_{13}$ yields:
\beq
\cos{\chi\over 2} \,d_6\,=\,Q\,e^{-{\phi\over 2}}\,h^{-{1\over 2}}\, \chi'\,\,.
\label{d_6_condition}
\eeq
Taking into account the expression of $d_5$ in  (\ref{d_5_d_6}), it straightforward to check that (\ref{d_5_condition}) is equivalent to the bending equation (\ref{BPS_bending}). Moreover, 
(\ref{d_6_condition}) is equivalent to:
\beq
Q\,\chi'\,=\,{e^{f-{\phi\over 2}}\over 2}\,\cos{\chi\over 2}\,
(\cos^2{\chi\over 2}\,+\,2\dot \tau)\,z'\,\,,
\eeq
and one can easily show that it is a consequence of the BPS equations for $\chi$ and $z$. Therefore, we can write:
\beq
\tilde\Gamma_{\kappa}\,\eta\Big|_{BPS}\,=\,-{1\over \sqrt{-\det(g_{6}+e^{-{\phi\over 2}}\,F)}}
\Big[ {e^g\over 2}\,\cos{\chi\over 2}\,(d_1+d_4)\,+\,Q\,e^{-{3\phi\over 2}}\,h^{-1}\,z'\,\Big]\,
\eta\big|_{BPS} \ .
\eeq
Let us now compute the terms containing the unit matrix in $\tilde\Gamma_{\kappa}\,\eta$  when the BPS equations are satisfied.  We get:
\bear
(d_1+d_4)\big|_{BPS} & = & {e^{g}\over 8 \sin{\chi\over 2}\,\cos^3{\chi\over 2}}\,{\cal J}_{DBI}\rc\rc
Q\,e^{-{3\phi\over 2}}\,h^{-1}\,z'\big|_{BPS}& = & {4\,Q^2\,h^{-{1}}\,e^{-2g-\phi}\over
 \sin{\chi\over 2}\,\cos^2{\chi\over 2}}\rc\rc
  \sqrt{-\det(g_{6}+e^{-{\phi\over 2}}\,F)}\Big|_{BPS}& = &
 {e^{2g}\over 8 \sin{\chi\over 2}\,\cos^2{\chi\over 2}}\,\Big({\cal J}_{DBI}^2\,+\,
 64\,h^{-1}\,e^{-4g-\phi}\,Q^2\Big)\,\,.\qquad
\eear
Using these results it is straightforward to verify that $\tilde \Gamma_{\kappa}\,\eta\,=\,-\eta$.

\end{document}